\documentclass[prd,singlecolumn,amsmath,nobibnotes,nofootinbib,superscriptaddress,preprintnumbers]{revtex4-1}

\usepackage{bm} \usepackage{graphicx}
\usepackage{color} \usepackage{amssymb}
\usepackage{epstopdf}
\usepackage{rotating}
\usepackage{url}

\def\ba{\begin{eqnarray}}
\def\ea{\end{eqnarray}}
\def\be{\begin{equation}}
\def\ee{\end{equation}}
\def\({\left(}
\def\){\right)}
\def\[{\left[}
\def\]{\right]}
\def\<{\left<}
\def\>{\right>}

\newcommand{\prob}
  {{\rm{Pr}}}
\newcommand{\ns}
  {{N_{\rm{s}}}}
\newcommand{\nblob}
  {{N_{\rm{b}}}}
\newcommand{\nsavge}
  {{\bar{N}_{\rm{s}}}}
  \newcommand{\aobs}
  {\Omega_{\rm{obs}}}
  \newcommand{\src}
  {s}
  \newcommand{\step}
  {\Theta}
\newcommand{\region}
  {r}
\newcommand{\nregion}
  {{N_{\rm{r}}}}

\newcommand{\likelihood}
  {L}
\newcommand{\npix}
  {{N_{\rm{px}}}}
\newcommand{\template}
  {\mathbf{t}}
\newcommand{\model}
  {\mathbf{m}}
  \newcommand{\fsky}
  {f_{\rm{sky}}}
  \newcommand{\data}
  {\mathbf{d}}
  \newcommand{\eg}
	{{e.g.}}
	\newcommand{\diff}
  {{\rm{d}}}
  \newcommand{\matr}[1]
        {\mbox{\bf \sf{#1}}}
  \newcommand{\vect}[1]
        {\mbox{\bf #1}}
  \newcommand{\blob}
  {b}

\newcommand{\skypos}{\hat{\mathbf{n}}}

\begin{document}

\title{First Observational Tests of Eternal Inflation: Analysis Methods and WMAP 7-Year Results}
\date{\today}

\author{Stephen M. Feeney}
\email{stephen.feeney.09@ucl.ac.uk}
\affiliation{Department of Physics and Astronomy, University College London, London WC1E 6BT, U.K.}
\author{Matthew C. Johnson}
\email{mjohnson@perimeterinstitute.ca}
\affiliation{Perimeter Institute for Theoretical Physics, Waterloo, Ontario N2L 2Y5, Canada} 
\affiliation{California Institute of Technology, Pasadena, CA 91125, USA}
\author{Daniel J. Mortlock}
\email{mortlock@ic.ac.uk}
\affiliation{Astrophysics Group, Imperial College London, Blackett Laboratory, Prince Consort Road, London, SW7 2AZ, U.K.}
\author{Hiranya V. Peiris}
\email{h.peiris@ucl.ac.uk}
\affiliation{Department of Physics and Astronomy, University College London, London WC1E 6BT, U.K.}
\affiliation{Institute of Astronomy and Kavli Institute for Cosmology, University of Cambridge, Cambridge CB3 0HA, U.K.}

\begin{abstract}
In the picture of eternal inflation, our observable universe resides inside a single bubble nucleated from an inflating false vacuum. Many of the theories giving rise to eternal inflation predict that we have causal access to collisions with other bubble universes, providing an opportunity to confront these theories with observation. We present the results from the first observational search for the effects of bubble collisions, using cosmic microwave background data from the WMAP satellite. Our search targets a generic set of properties associated with a bubble collision spacetime, which we describe in detail. We use a modular algorithm that is designed to avoid {\em a posteriori} selection effects, automatically picking out the most promising signals, performing a search for causal boundaries, and conducting a full Bayesian parameter estimation and model selection analysis. We outline each component of this algorithm, describing its response to simulated CMB skies with and without bubble collisions. Comparing the results for simulated bubble collisions to the results from an analysis of the WMAP 7-year data, we rule out bubble collisions over a range of parameter space. Our model selection results based on WMAP 7-year data do not warrant augmenting $\Lambda$CDM with bubble collisions. Data from the {\em Planck} satellite can be used to more definitively test the bubble collision hypothesis. 
\end{abstract}

\preprint{}

\maketitle

\section{Introduction}

Observations of the cosmic microwave background (CMB) radiation have ushered in a new era of precision cosmology. Full-sky temperature maps produced by the {\em Wilkinson Microwave Anisotropy Probe} (WMAP)~\cite{Bennett:2003ba} have confirmed with high precision that the observed temperature fluctuations are consistent with a nearly Gaussian and scale invariant primordial power spectrum, as predicted by inflation. The recently launched {\em Planck} satellite~\cite{Tauber2010} has a resolution three times better than that of WMAP, with an order of magnitude greater sensitivity, and significantly wider frequency coverage (allowing for far more robust foreground removal, and therefore reduced systematics). These high quality data sets allow for the possibility of observing deviations from the standard inflationary paradigm, some of which could have drastic consequences for our understanding of the universe and its origins. 

Perhaps the largest gap in our description of the early universe lies in an understanding of its initial conditions. One possibility, motivated by the proliferation of vacua in compactifications of string theory (known as the string theory landscape~\cite{Susskind:2003kw}), is that our observable universe is only a tiny piece of a vast multiverse, the majority of which is still inflating. This picture of {\em eternal inflation} (for a review, see, e.g., Ref.~\cite{Aguirre:2007gy}) arises when the rate at which local regions exit an inflating phase is outpaced by the accelerated expansion of the inflating background. Eternal inflation is a fairly generic consequence of any theory containing positive vacuum energy and multiple vacua, highlighting the importance of understanding how this scenario might be confronted with observational tests.

The first attempts to embed our cosmology inside an eternally inflating universe led to ``open inflation"\cite{Bucher:1994gb,Gott:1982zf}; see Ref.~\cite{GarciaBellido:1997uh} for a review. In this scenario, a scalar field (or set of scalar fields) has a potential with a high energy metastable minimum that drives the eternally inflating phase. Transitions out of this vacuum proceed via the Coleman-de Luccia (CDL) instanton~\cite{Coleman:1977py,Coleman:1980aw}, resulting in expanding bubbles inside which the scalar field rests on an inflationary plateau. The symmetries of the CDL instanton ensure that there is a very nearly homogeneous and isotropic open universe inside the bubble; inflation, reheating, and standard cosmological evolution follow. 

In any given bubble, the future light cone of the nucleation event forms the ``Big Bang'' (where the scale factor vanishes) of an open FRW universe. The eternally inflating phase outside our bubble can therefore be thought of as a pre-Big Bang epoch, and one might expect inflation to erase any of the scant observational evidence of our parent vacuum. In single bubble open inflation, various anomalies are induced in the CMB temperature power spectrum (see Ref.~\cite{GarciaBellido:1997uh} and references therein), but unfortunately, the size of these effects decreases with the present energy density in curvature (related to the number of inflationary $e$-folds), rendering them negligible at all but the lowest multipoles where cosmic variance dominates. However, our bubble does not evolve in isolation. There are other nucleation events from the false vacuum, containing a phase that might be identical to ours, or perhaps very different. If one of these secondary nucleation events occurs close enough to our bubble wall, then a collision inevitably results. In fact, since our bubble grows to reach infinite size, there are an infinite number of collisions~\cite{Guth:1981uk,Guth:1982pn,Gott:1984ps,Garriga:2006hw} (a finite subset of which are causally accessible to any one observer). This raises the possibility that if such collisions are both survivable and only small perturbations on top of standard cosmology, they might leave observable signatures of eternal inflation~\cite{Aguirre:2007an}; it is these signatures which our analysis targets. 

If we are to detect such bubble collisions, their predicted signatures must be consistent with our observed cosmology, but sufficiently distinct to be differentiated from other possible signals in the CMB. In addition, the theory must predict that we expect to have causal access to bubble collisions. While these criteria are not met in every model of eternal inflation, recent work~\cite{Guth:1981uk,Guth:1982pn,Hawking:1982ga,Wu:1984eda,Gott:1984ps,Garriga:2006hw,Aguirre:2007an,Aguirre:2007wm,Aguirre:2008wy,Chang:2007eq,Chang:2008gj,Dahlen:2008rd,Freivogel:2009it,Easther:2009ft,Larjo:2009mt,Zhang:2010qg,Czech:2010rg} (for a review, see Ref.~\cite{Aguirre:2009ug}) has established that bubble collisions could in some theories be both expected and detectable. Bubble collisions produce a fairly characteristic set of inhomogeneities in the very early universe, which are processed into temperature anisotropies in the CMB. From the spherical symmetry of the colliding bubbles, the collision possess azimuthal symmetry, and by causality must be confined to a disc on the sky. The CMB temperature and its derivatives need not be constant across the causal boundary. Therefore, the signals we are searching for are localized, and because they are primordial, consist of a long-wavelength modulation of the standard inflationary density fluctuations inside the affected region~\cite{Chang:2008gj}. The amplitude and angular scale of the signal is dependent upon the underlying model and kinematics of the collision.

These general features suggest a set of strategies for data analysis. The localization of the collision implies that wavelet analysis could be a sensitive tool for picking out both the location and angular scale of a candidate signal. The causal boundary, across which the temperature and its derivatives need not be constant, suggests the use of edge detection algorithms similar to those used in searches for cosmic strings \cite{Kaiser:1984iv,Lo:2005xt,Danos:2008fq,Amsel:2007ki}. Finally, the prediction that the temperature modulation induced by the collision is rather long-wavelength yields a sufficiently generic template to perform a full Bayesian parameter estimation and model selection analysis. 

In this paper, we describe a modular analysis algorithm designed to look for the signatures of eternal inflation, and apply it to the WMAP 7-year data~\cite{Jarosik:2010iu}. This algorithm can easily be adapted to test any model that predicts a population of spatially localized sources in addition to the standard fluctuations predicted by $\Lambda$CDM. A summary of our results was presented in Ref.~\cite{Feeney:2010jj}; in this paper we describe our analysis in detail. Currently available full sky CMB data are rather limited in their sensitivity to the signatures of bubble collisions listed above; the main current limitation is the low resolution. Therefore, we apply our algorithm to current data mainly as a validation exercise; to exploit its full power would require future high resolution data, e.g., from {\em Planck}.  

The individual steps of our analysis pipeline are calibrated using realistic simulations of the WMAP experiment with and without bubble collisions. The calibrated pipeline applied to data is fully automated, identifying the candidate signals and processing them without any human intervention. This removes any {\em a posteriori} choices from our analysis, which must be carefully avoided in any analysis of a large data-set such as the WMAP 7-year data~\cite{Bennett:2010jb}.

The plan of the paper is as follows. In Sec.~\ref{sec:collisionintro}, we review some of the background on bubble collisions in eternal inflation, and outline the predicted observable signatures. Our analysis pipeline is summarized in Sec.~\ref{sec:summaryofpipeline}. We describe some properties of the WMAP experiment and our simulations in Sec.~\ref{sec:simulatedmaps}, and detail our analysis tools in Sec.~\ref{sec:analysistools}. Sec.~\ref{sec:WMAP7} summarizes the results of our analysis of the WMAP 7-year data, and we conclude in Sec.~\ref{sec:conclusions}.

\section{The observable effects of bubble collisions}\label{sec:collisionintro}

The simplest model of eternal inflation involves a single scalar field in four dimensions, with a double-well potential. In many models (as long as the average curvature of the potential between the minima is small compared to the Planck scale), the Coleman-de Luccia (CDL) instanton~\cite{Coleman:1977py,Coleman:1980aw} mediates a transition from the false (higher energy) to the true (lower energy) vacuum. This tunneling event corresponds to the appearance of an expanding bubble of the true vacuum embedded in the false. As long as the probability that a bubble nucleates in each horizon volume of the false vacuum during a Hubble time is less than one (so that the background expansion of the false vacuum on average prevents bubbles from merging), the phase transition never completes and inflation is eternal~\cite{Guth:1981uk} (see Ref.~\cite{Sekino:2010vc} for a modern treatment of the percolation problem in eternal inflation). The O(4)-invariance of the instanton guarantees that the bubble interior possesses SO(3,1) symmetry, and therefore contains an infinite open Friedman Robertson Walker (FRW) universe. Although homogeneity is ensured by the symmetries of the instanton, if the interior of a bubble is to resemble our own universe, a second epoch of inflation inside the bubble is necessary to dilute the negative curvature and provide the correct spectrum of primordial density perturbations to seed structure. Models of this type are known in the literature as open inflation, and have been explored in detail (see Ref.~\cite{GarciaBellido:1997uh}).

The signatures of single-bubble open inflation include negative curvature and modifications to the power spectrum. These modifications are most important at large angular scales (see Ref.~\cite{GarciaBellido:1997uh} and references therein) where cosmic variance is dominant, and would be very difficult to detect. Since curvature alone would not be a very distinguishing prediction, we do not consider these signals further.

A less ambiguous signature of eternal inflation would be the visible remnants of collisions between bubbles. Although the bubbles formed during eternal inflation do not percolate, there are many (in fact, an infinite number of) collisions. These collisions lead to inhomogeneities in the inner-bubble cosmology, perhaps leaving observable signatures in the CMB.  Assessing the observational consequences of bubble collisions in an eternally inflating universe has been an active area of research~\cite{Garriga:2006hw,Aguirre:2007an,Aguirre:2007wm,Aguirre:2008wy,Chang:2007eq,Chang:2008gj,Dahlen:2008rd,Freivogel:2009it,Easther:2009ft,Larjo:2009mt,Zhang:2010qg,Czech:2010rg} (for a review, see Ref.~\cite{Aguirre:2009ug}). These studies have established that a number of criteria necessary for the observation of bubble collisions~\cite{Aguirre:2007an} can be satisfied, at least in some models:

{\bf Compatibility:} In order to satisfy this criterion, there must be a bubble we can collide with that only minimally disturbs the homogeneity of the observable portion of the surface of last scattering. Such collisions do seem to exist, as evidenced by thin-wall junction condition analysis~\cite{Aguirre:2007wm,Chang:2007eq} as well as numerical simulations~\cite{Aguirre:2008wy} and a study of the inflaton field in a background thin-wall collision geometry~\cite{Chang:2008gj}.

{\bf Probability:} We should expect to have a collision in our causal past. The number of collisions in our past is $\bar{N} = \lambda V_4^{F}$, where $\lambda$ is the bubble nucleation probability per unit four-volume and $V_4^{F}$ is the four-volume outside the bubble to which we have causal access. The expected number (assuming the original open FRW foliation) is formally infinite~\cite{Aguirre:2007an}; however, collisions that contribute to this divergence only produce very long wavelength fluctuations at last scattering, and so would not be observable~\cite{Aguirre:2008wy,Freivogel:2009it} (this is similar to the infrared divergence found in models of slow-roll inflation).  The average number of collisions that affect the {\em observable} portion of the surface of last scattering is finite~\cite{Freivogel:2009it,Aguirre:2009ug}, and is given by
\begin{equation}\label{eq:collnum}
\bar{N} \simeq \frac{16 \pi}{3} \lambda H_F^{-4} \left( \frac{H_F}{H_I} \right)^2 \Omega_k^{1/2} \, , 
\end{equation}
where $H_F$ is the false vacuum Hubble constant, $H_I$ is the inflationary Hubble constant inside our bubble, and $\Omega_k$ is the current component of energy in curvature. For the expected number of observable collisions to be one or larger, the separation of scales between $H_F$ and $H_I$ must be large enough to compensate for the low probability $\lambda$ (which is exponentially suppressed because this is a tunneling process) and the observational constraint $\Omega_k \alt .0084$~\cite{Komatsu:2010fb}. Given a particular scalar potential underlying eternal inflation, $\bar{N}$ for each possible type of collision is fixed. However, in a theory with a complicated potential landscape for the scalar field(s), it makes sense to think of $\bar{N}$ as a continuous parameter with some prior probability distribution\footnote{We return to this point in Sec.~\ref{sec:bayes} when discussing the Bayesian framework for testing bubble collision models.}. Without detailed knowledge of the theory underlying eternal inflation and an associated measure, it is difficult to assess how likely it is to have $\bar{N} > 1$, but see Refs.~\cite{Freivogel:2009it,Aguirre:2009ug} for some speculative comments. There is also an exponential pressure from the nucleation rates towards $\bar{N} \gg 1$ or $\bar{N} \ll 1$. In the following, we assume $\bar{N}$ can be order one.

{\bf Observability:} Since the effects of a collision must pass through the entire inner-bubble cosmology, they can be thought of as perturbations of the Big Bang in an FRW cosmology. As such, they are stretched by inflation, and we expect the strength of most signatures to scale with (some power of) $\Omega_k$. We therefore must require that there are not too many more $e$-folds than required to satisfy the observational bound on curvature. Given a field theory model, the number of $e$-folds of inflation inside the bubble is uniquely determined by the instanton. However, if we consider a landscape of scalar potentials, then it is necessary to find a measure over the number of inflationary $e$-folds (or equivalently $\Omega_k$). For some work in this direction, see, e.g., Refs~\cite{Freivogel:2005vv,DeSimone:2009dq}.

\begin{figure}[tb]
   \includegraphics[width=6 cm]{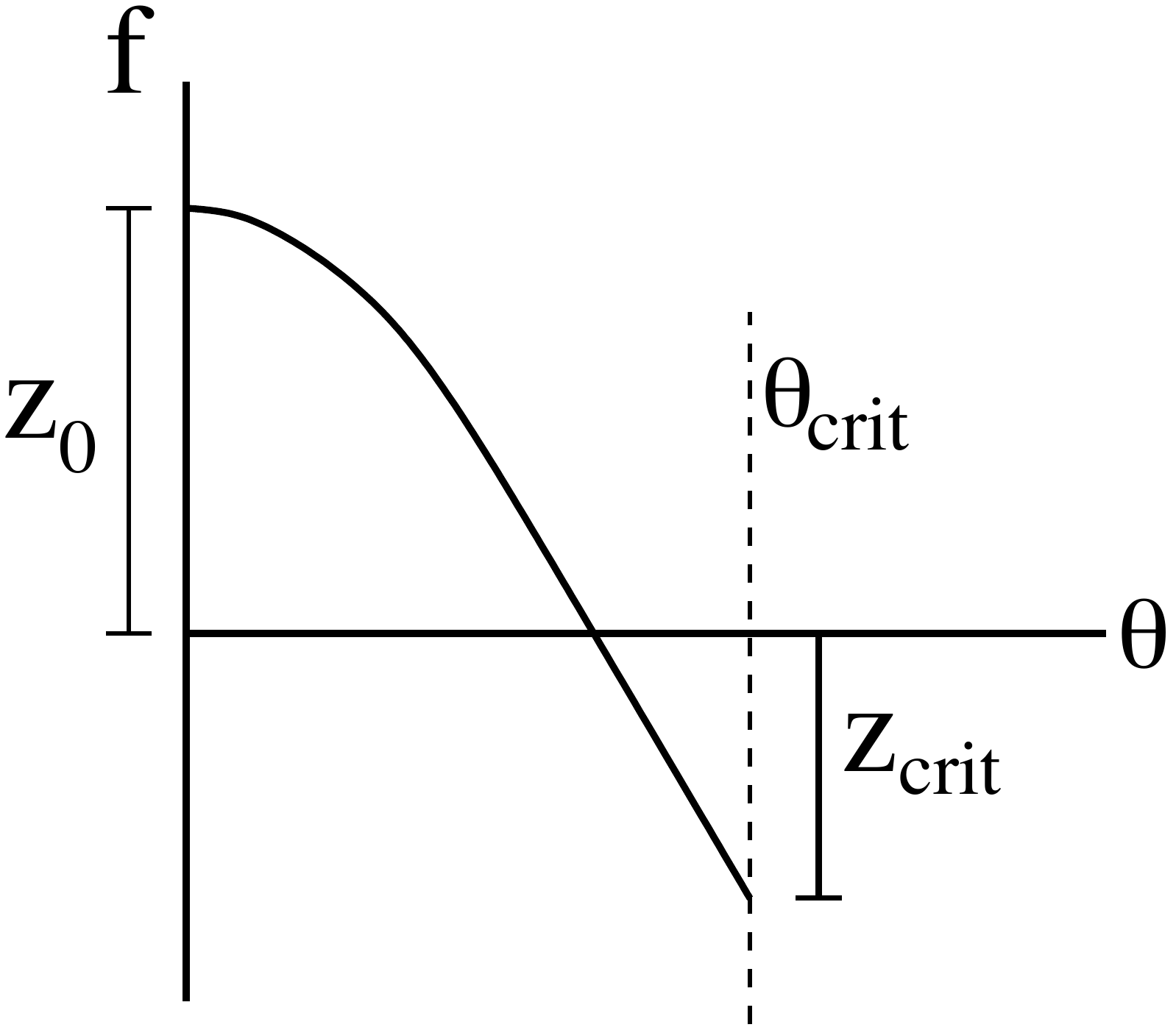}
\caption{The radial temperature modulation Eq.~\ref{eq:collfluct} induced by a bubble collision centered on the the north pole ($\theta = 0$). }
\label{fig-template}
\end{figure}

Much remains to be learned about the full spectrum of possible outcomes of a bubble collision and the exact details of the associated observational signatures. Nevertheless, all potentially observable bubble collisions involving two bubbles share a sufficiently general set of properties to allow for a meaningful observational search even in the absence of a detailed model. In summary, we expect all such observable bubble collisions to possess:

\begin{itemize}
\item Azimuthal symmetry: A collision leaves an imprint on the CMB sky that has azimuthal symmetry. This is a consequence of the SO(2,1) symmetry of the spacetime describing the collision of two vacuum bubbles~\cite{Wu:1984eda,Garriga:2006hw,Aguirre:2007an}.
\item A causal boundary: The surface of last scattering can only be affected inside the future light cone of a collision event. The intersection of  our past light cone, the future light cone of a collision, and the surface of last scattering is a ring. This is the causal boundary of the collision on the CMB sky. The temperature and its derivatives need not be continuous across this boundary. Neglecting the backreaction of the collision on the geometry of the bubble interior, the distribution of ring sizes was found in Ref.~\cite{Freivogel:2009it} to be
\begin{equation}
\label{eq-angdist}
\frac{d \bar{N}}{d\theta_{\rm crit}} \sim 4 \pi \lambda H_F^{-4} \left( \frac{H_F}{H_I} \right)^2 \Omega_k^{1/2} \sin \left( \theta_{\rm crit} \right),
\end{equation}
where $\theta_{\rm crit}$ is the angular radius measured from the center of the disc to the causal boundary and the other quantities are as defined in Eq.~\ref{eq:collnum}.
\item An overall modulation of the background fluctuations: We assume that the temperature fluctuations, including the effects of the collision, at a location on the sky $\hat{\mathbf{n}}$ can be written as~\cite{PhysRevD.72.103002,Chang:2008gj}
\begin{equation} \label{eq:tempmod}
\frac{\delta T(\hat{\mathbf{n}})}{T_{0}} =  (1+ f(\hat{\mathbf{n}})) (1 + \delta(\hat{\mathbf{n}})) - 1,
\end{equation}
where $ f(\hat{\mathbf{n}})$ is the modulation induced by the collision and $\delta(\hat{\mathbf{n}})$ are the temperature fluctuations induced by modes set down during inflation. This is motivated by the observation that the main effect of the bubble collision is to slightly advance or retard the inflaton inside our bubble. The modulation is multiplicative under the assumption that the normal inflationary density fluctuations simply "paint" the perturbed surface of last scattering and have identical statistical properties in both the regions affected and unaffected by the collision.
\item Long-wavelength modulation: A collision is a pre-inflationary relic. The effects of a collision inside the causal boundary are stretched by inflation, and so we can expect that the relevant fluctuations are large-scale. As we describe below, this implies that the temperature modulation due to a collision centered on the north pole ($\theta = 0$) has the form
\begin{equation}\label{eq:collfluct}
f(\hat{\mathbf{n}}) = (c_0 + c_1 \cos \theta + \mathcal{O} (\cos^2 \theta)) \Theta (\theta_{\rm crit} - \theta) \, ,
\end{equation}
where the $c_i$ are constants related to the properties of the collision, $\theta$ is the angle measured from the center of the affected disc, and $\Theta (\theta_{\rm crit} - \theta)$ is a step function at the causal boundary $\theta_{\rm crit}$. Truncating the sum at $\mathcal{O}(\cos \theta)$, the constants $c_0$ and $c_1$ can be expressed in terms of a central amplitude $z_0$ and edge discontinuity $z_{\rm crit}$:
\begin{equation}
c_0 =  \frac{z_{\rm crit} - z_0 \cos \theta_{\rm crit}}{1 - \cos \theta_{\rm crit}}, \ \ c_1 = \frac{z_0 - z_{\rm crit}}{1 - \cos \theta_{\rm crit}} \, ,
\end{equation}
as shown in Fig.~\ref{fig-template}. Allowing the collision to be centered on an arbitrary location $\{ \theta_0, \phi_0 \}$ on the celestial sphere, the induced temperature modulation can be expressed as a function of five parameters: $\{z_0, z_{\rm crit}, \theta_{\rm crit}, \theta_0, \phi_0\}$. A modulation of this form was first derived in Ref.~\cite{Chang:2008gj}, where it was obtained from the observed modulation of a field representing the inflaton inside our bubble, numerically evolved in a background thin-wall bubble collision geometry. These authors did not predict the existence of a temperature discontinuity $z_{\rm crit}$. While further work is needed to better predict the precise form of the template, in our analysis we allow bubble collisions to produce modulations with and without discontinuities.
\end{itemize}

\begin{figure}
\includegraphics[width=7.5cm]{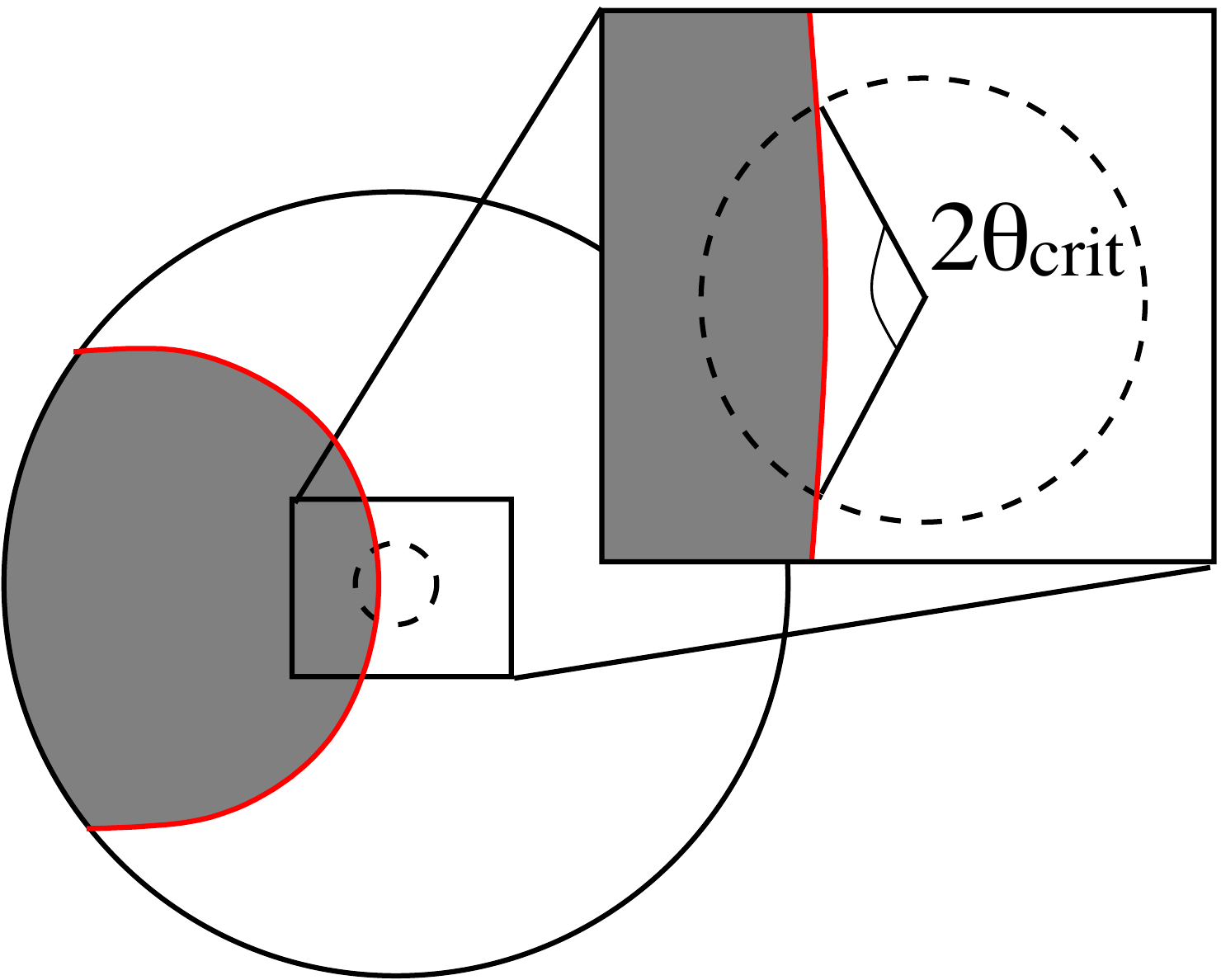}
\caption{A Poincare-disc representation of the surface of last scattering inside our parent bubble, with one dimension suppressed. The future light cone of the collision at this time is denoted by the dark red line, with the shaded region representing the portions of the surface of last scattering that are to the future of the collision. Our past light cone at last scattering is represented by the dashed circle. From the present bounds on curvature, the size of our past light cone must be much smaller than one curvature radius. Zooming in on the portion of the surface of last scattering that we have causal access to (inset), the universe is very close to flat, and the region affected by the collision has approximate planar symmetry. The region affected by the collision appears as a disc of angular radius $\theta_{\rm crit}$ on the CMB sky. 
  \label{fig-poincarefig}
}
\end{figure}

In Fig.~\ref{fig-poincarefig}, we show the Poincare-disc representation (see Ref.~\cite{Aguirre:2007an} for the details of this construction) of the surface of last scattering inside our parent bubble. The collision affects the shaded portion of this surface. The observed CMB is formed at the intersection of our past light cone (dashed circle) with the surface of last scattering, which in this case includes regions both affected and unaffected by the collision. From the underlying azimuthal symmetry, the collision appears as a disc on the observer's CMB sky. Zooming in on the neighborhood of our past light cone (inset), we can treat the universe as being flat. In addition, because we have causal access to much less than one curvature radius at last scattering (again from the observational bound on $\Omega_k$), the collision has an approximate planar symmetry.

The collision introduces pre-inflationary inhomogeneities into our bubble. The exact nature of these inhomogeneities depends on the specific model underlying the formation of our bubble and the subsequent epoch of slow-roll inflation, as well as the specifics of the collision. In dramatic cases, the collision ends slow-roll inflation everywhere within its future light cone~\cite{Aguirre:2008wy}, induces the transition to another vacuum state~\cite{Easther:2009ft,Giblin:2010bd,Johnson:2010bn}, or produces a post-collision domain wall that eats into our bubble interior~\cite{Aguirre:2007wm,Chang:2007eq}. These scenarios are obviously in conflict with observation, and we do not consider them further. In mild cases, which will be our focus in the remainder of this paper, collisions satisfy the ``compatibility" criterion defined above: the observable portion of the surface of last scattering is only minimally disturbed by the collision. Thin-wall analysis~\cite{Aguirre:2007wm} and numerical simulations~\cite{Aguirre:2008wy,Chang:2008gj} indicate that it is indeed possible to find situations where the effects of a collision are compatible with our observed cosmology. 

The disturbance caused by a collision is a pre-inflationary relic and thus is stretched by the period of inflation inside the bubble. From the current bound on curvature~\cite{Komatsu:2010fb}, we can infer that our past light cone encompasses less than one horizon volume at the onset of inflation. This implies that the initial disturbances caused by a collision, which is smeared out on the scale of the inflationary horizon after a few $e$-folds of inflation, has a wavelength today that is larger than the current horizon size. Together with the planar symmetry of the collision at last-scattering (by convention along the $y$-$z$ plane), this implies that we can Taylor-expand the Newtonian potential (see Ref.~\cite{Czech:2010rg} for a translation between the Newtonian potential and the originally postulated temperature modulation presented in Ref.~\cite{Chang:2008gj}) about the causal boundary of the collision at $x=x_{\rm crit}$ as
\begin{equation}\label{eq:newtonian}
\Phi_{\rm coll} = \Phi (a) \left( \bar{c}_0 + \bar{c}_1 (x-x_{\rm crit}) + \mathcal{O}( (x - x_{\rm crit})^2) \right) \Theta(x - x_{\rm crit}),
\end{equation}
where $\Phi (a)$ encodes the evolution of the potential with scale factor $a$ and the $\bar{c}_i$ are model-dependent constants.~\footnote{We are modeling the collision as a collection of super-modes truncated at the causal boundary, and our treatment is therefore very similar to the so-called ``tilted universe" scenario \cite{Turner:1991dn,Erickcek:2008jp}. The important distinction in the case of bubble collisions is that the perturbation vanishes at the causal boundary $x_{\rm crit}$. Because the collision entered our past light cone only relatively recently, we are still comoving with respect to the undisturbed FRW foliation, and the cancellation of the dipolar temperature modulation seen in Ref.~\cite{Turner:1991dn,Erickcek:2008jp,Zibin:2008fe} does not occur.} 

There are contributions to the observed temperature modulation from the Sachs-Wolfe effect, the integrated Sachs-Wolfe effect, and a Doppler effect (coming from the induced bulk peculiar velocity ${\bf v}$ of the fluid in the region affected by the collision): 
\begin{equation}
\frac{\delta T}{T} \simeq \frac{\Phi_{\rm coll} (a_{\rm ls})}{3} + 2  \int_{a_{\rm ls}}^{1} da \frac{d \Phi_{\rm coll}}{da} + \left( {\bf v} \cdot {\hat{\mathbf{n}}} + \mathcal{O}(v^2)\right),
\end{equation}
where $a_{\rm ls}$ is the scale factor at last scattering, $a=1$ today, and
\begin{equation}
{\bf v} \propto \nabla \Phi_{\rm coll} + a \frac{d}{da} \nabla \Phi_{\rm coll}.
\end{equation}
To leading order in ${\bf v}$, the temperature induced by the collision is linear in $\Phi_{\rm coll}$ and its derivatives. Therefore, since $x = x_{\rm ls} \cos \theta$ (where $x_{\rm ls}$ is the comoving distance out to which we can see on the surface of last scattering), the temperature fluctuations induced by a collision are generally of the form Eq.~\ref{eq:collfluct}. Further, even if the Newtonian potential is continuous across $x=x_{\rm crit}$, the resulting temperature fluctuations need not be continuous across the causal boundary at $\theta_{\rm crit}$. This discontinuity arises from the ISW and Doppler contributions to the observed temperature fluctuation. Effects that we have neglected, including the finite thickness of the surface of last scattering and uncertainties about how the perturbations caused by a bubble collision propagate through our bubble interior, are encapsulated by the higher order terms in Eq.~\ref{eq:collfluct}. These effects could smear out the causal boundary enclosing the collision on sub-degree scales. These corrections could be incorporated into our analysis as theoretical understanding improves.

Given a specific model for the scalar fields making up the bubbles
and driving eternal inflation, the kinematics of a
particular collision, and our position inside our bubble, it is in principle possible to determine the
free parameters in Eq.~\ref{eq:collfluct}. Treating the colliding bubbles in the
thin-wall approximation, some measure of the strength of a collision
can be specified in terms of the vacuum energies in the bubbles, wall
tensions, and kinematics as in Ref.~\cite{Aguirre:2007wm}. The kinematics will induce a probability distribution for the free parameters in Eq.~\ref{eq:collfluct}. 
However, an accurate treatment requires a calculation of the back-reaction of the collision
on the behaviour of the inflaton inside our bubble. Preliminary
work in this direction has been done~\cite{Chang:2008gj,Aguirre:2008wy,Easther:2009ft,Giblin:2010bd}, providing a handful of
examples. However, a systematic investigation has not yet been
performed. This is distinct from the case where an ensemble of field theory
models is considered, representing the string theory landscape. In
this case, the fundamental parameters governing the structure of the
colliding bubbles (wall tensions and vacuum energies) and the
properties of the inner-bubble cosmology (including the number of
inflationary $e$-folds etc) are drawn from some probability
distribution. This again will induce a probability distribution for the 
free parameters in Eq.~\ref{eq:collfluct}, whose nature is presently poorly understood.

\begin{figure}[tb]
   \includegraphics[width=8 cm]{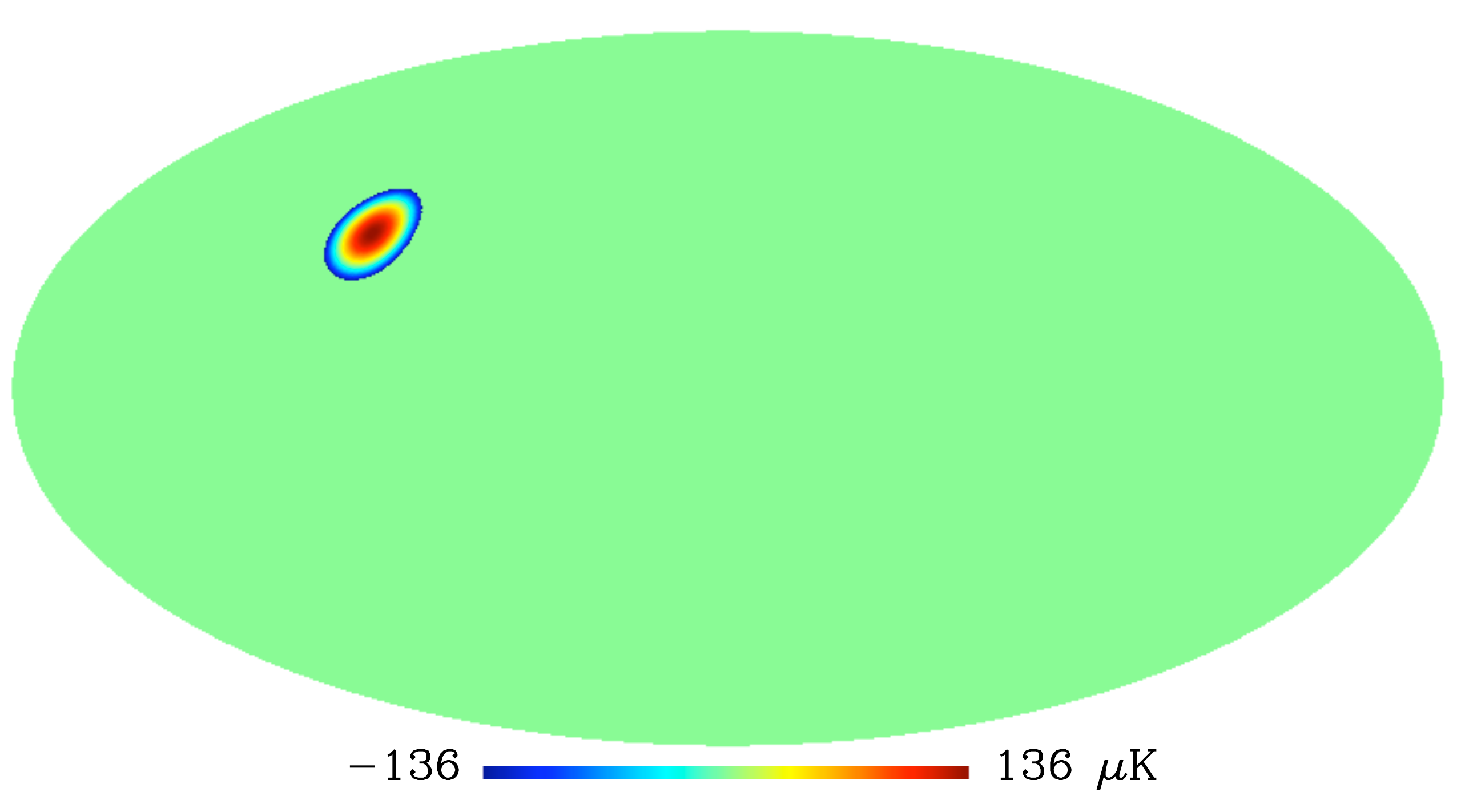}
   \includegraphics[width=8 cm]{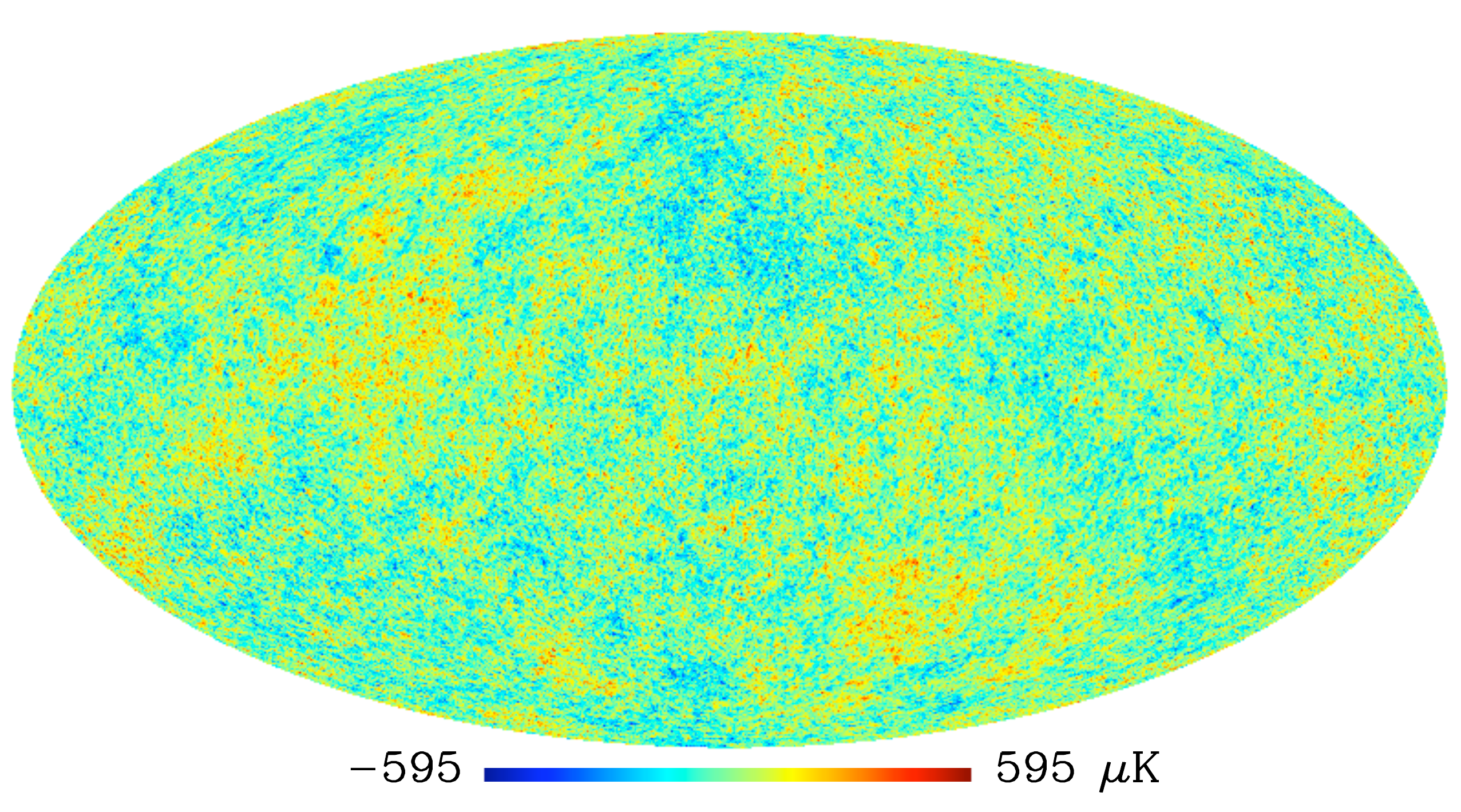}
\caption{On the left, we show a bubble collision template with $\{ z_0 = 5.0 \times 10^{-5}, z_{\rm crit} = -5.0 \times 10^{-5}, \theta_{\rm crit} = 10.0^{\circ}, \theta_0 = 57.7^{\circ}, \phi_0 = 99.2^{\circ}\}$. On the right we add simulated background fluctuations, smoothing, and instrumental noise.}
\label{fig-benchmarkexample}
\end{figure}

What would a bubble collision embedded in a CMB temperature map look like? In Fig.~\ref{fig-benchmarkexample} we show a large-amplitude collision with and without background CMB fluctuations. In the following sections, we apply the various stages of our analysis pipeline to this example to illustrate the algorithm. We make extensive use of such simulations in calibrating our analysis pipeline, and the details of their construction are presented in Sec.~\ref{sec:simulatedmaps}. Although there could conceivably be many overlapping collisions, the predicted observational signatures of this scenario have yet to be explored, and we focus on simulations of distinct individual bubble collisions. Again, as theoretical understanding improves, our analysis could be extended to include the possibility of overlapping collisions.

What would the detection, or absence, of a bubble collision tell us about the underlying theory of eternal inflation? To examine what the answer to this question might be, let us make some further assumptions about the temperature modulations caused by a bubble collision. First, assume that the potential induced by the collision (Eq.~\ref{eq:newtonian}) is composed mostly of a single long-wavelength mode of physical wavenumber $k$. Second, assume that the Sachs-Wolfe effect is the dominant contribution to the observed temperature modulation. Under these assumptions, the amplitude of an observed temperature modulation is:
\begin{equation}
z_0 \simeq \frac{2}{3} \frac{k}{H_0} \Phi (a_{\rm ls}) \left( 1 - \cos \theta_{ \rm crit} \right),
\end{equation}
where $a_{\rm ls}$ is the scale factor at last scattering. If the initial wavelength of the disturbance was of order one inflationary Hubble length $k \sim H_I$ (since any fine-structure in the collision would be smeared within the first few $e$-folds of inflation), then $\Phi (a_{\rm ls}) = \Phi (a=0)$, and the physical size of such a mode at last scattering is given by
\begin{equation}
k \simeq \Omega_k^{1/2} H_0.
\end{equation}
In this case, we have
\begin{equation}\label{eq:z0bound}
z_0 \simeq \Phi (0) \Omega_k^{1/2}  \left( 1 - \cos \theta_{\rm crit} \right).
\end{equation}
If a bubble collision is detected, and a similar set of assumptions is proven correct in a specific model, the measured values of $z_0$ and $\theta_{\rm crit}$ allow one to infer the value of $\Omega_k$.~\footnote{The values of $\Omega_k$ that one might be able to infer are conceivably below both the observational bound $\Omega_k \leq .0084$ and the theoretical observational bound $\Omega_k \leq 10^{-5}$. For example, assuming $z_0 \sim 10^{-5}$ and $\Phi (0) \sim 1$ (since the collision involves a relatively large release of energy), if a collision were observed at large angular scale (where $\cos \theta_{\rm crit} \sim 0$), we can infer that $\Omega_k \sim 10^{-10}$. This implies that a collision is in principle observable even when curvature is not. We
thank Lam Hui for elucidating this point.} In the absence of a detected collision, Eq.~\ref{eq:z0bound} can be turned into a bound on a combination of $\Omega_k^{1/2}$ and $\Phi (0)$:
\begin{equation}\label{eq:griszeleffect}
\Omega_k^{1/2} \Phi (0) < \left[ z_0 /  \left( 1 - \cos \theta_{\rm crit} \right) \right]_{\rm observational \ upper \ bound}
\end{equation}
This analysis can be recognized as an example of the Grishchuk-Zel'dovich effect in an open universe~\cite{GarciaBellido:1995wz,Turner:1991dn}. 

Determining the detailed properties of the theory underlying eternal inflation through the observation of bubble collisions is likely to be a messy business. However, any model will predict an expectation value for the number of observable bubble collisions $\bar{N}$, making this a very useful phenomenological parameter. Any constraints on $\bar{N}$ from data will also yield interesting information about our parent vacuum through Eq.~\ref{eq:collnum}. The most naive application of such a constraint, where we have evidence that $\bar{N} >1$ or $\bar{N} < 1$, would yield the inequalities
\begin{equation}\label{eq:collbound}
\lambda H_F^{-2} < \frac{3 H_I^2}{16 \pi \Omega_k^{1/2}} \ \ {\rm (no \ detected \ collisions)},  \ \ \ \lambda H_F^{-2} > \frac{3 H_I^2}{16 \pi \Omega_k^{1/2}}  \ \ {\rm (collision \ detection)},
\end{equation}
These bounds would be most useful if we detect $\Omega_k$ and/or $B$-mode polarization (the amplitude of which can be related to $H_I$) in future data. In the most optimistic scenario~\footnote{In models where a collision is expected to be in our past, there might be good reason to expect a correlation between observed $B$-modes and the observation of a bubble collision~\cite{Aguirre:2008wy}. This is because large-field models of inflation, which generically predict a larger value for the tensor to scalar ratio, are much more robust in the presence of a  bubble collision. In models of small-field inflation, a bubble collision can end inflation everywhere in its future light cone, implying that collisions in such models are not compatible with our observed cosmology.}, if primordial $B$-mode polarization is detected by the {\em Planck} satellite, we can infer that $H_I \sim 10^{11} - 10^{13}$ GeV. Further, if curvature is detected at the level $\Omega_k \sim 10^{-3}$, then in Eq.~\ref{eq:collbound}, $\lambda H_F^{-2} $ would be bounded from above or below by $\sim 10^{26}\, {\rm GeV}^2$. The condition for eternal inflation is $\lambda H_F^{-4} < 1$. Any application of Eq.~\ref{eq:collbound} must be consistent with this inequality. For example, assuming a Planckian false vacuum energy ($H_F \sim 10^{19}\, {\rm GeV}$), the nucleation probability $\lambda H_F^{-4}$ could be bounded from above or below by $\sim 10^{-4}$, remaining consistent with the condition for eternal inflation.

\section{Summary of the analysis pipeline}\label{sec:summaryofpipeline}

Before providing a detailed description of our analysis pipeline, we motivate and summarize its various components. Eternal inflation can arise from a wide range of inflationary potentials, each producing a different expected number of detectable collisions on the CMB sky, $\nsavge$. We will therefore use $\nsavge$ as a continuous parameter that characterizes particular models of eternal inflation. The standard cosmological model is given by the special case in which $\nsavge = 0$. Our primary goal is to determine, given the WMAP 7-year data, what constraints can be placed on $\nsavge$ and whether models predicting $\nsavge > 0$ should be be preferred over models predicting $\nsavge = 0$.

The optimal approach to achieving this goal would be to construct the full posterior for $\nsavge$ from Bayes' theorem given full-resolution CMB data on the whole sky. Unfortunately, this would require inverting the full-sky full-resolution CMB covariance matrix as well as integrating the bubble-collision likelihood over a many-dimensional parameter space. These tasks are  computationally intractable. However, taking advantage of the fact that bubble collisions produce discrete localized effects on the CMB sky, it is possible to approximate the full-sky Bayesian analysis by a patch-wise analysis if the most promising candidate signatures can be identified in advance. The implementation of such an approximation scheme requires two assumptions. First, we assume that the likelihood of models predicting $\nsavge > 0$ is peaked in the regions of the sky containing the candidate collisions, and that the integral over the likelihood can therefore be estimated by concentrating on these regions, which make the largest contribution to the full integral. Second, we assume that these regions are separated widely enough to be uncorrelated with each other, so much smaller local covariance matrices can be used. These assumptions allow the results of a small number of localized (and therefore computationally-feasible) Bayesian model selection tests to be combined into estimates of the required full-sky statistics. Put simply, our algorithm implements a conservative approximation to the required numerical integral.  A complete treatment of the full-sky analysis and the assumptions on which it is formed can be found in Appendix~\ref{section:method}. In addition, once a set of candidates have been identified, it is possible to apply further tests of the data in parallel. 

The full-sky approximation necessitates the development of an algorithm that identifies the most promising regions of the CMB sky and then processes them individually. Upon segmenting the full data set, it is important to avoid biasing oneself with {\em a posteriori} selection effects~\cite{Bennett:2010jb}, and it is therefore critical to minimize human intervention in choosing what portions of the sky to analyze. Thus our analysis pipeline is fully automated, tested and calibrated on realistic simulations of the data, and frozen before being applied to the real data. The final pipeline contains no algorithmic choices tunable via human intervention. As discussed in Appendix~\ref{section:method}, missing a bubble collision candidate which makes a significant contribution to the full integral leads to a {\em conservative} bias towards models predicting $\nsavge = 0$. This alleviates the worry that selection effects might lead to a spurious detection. 

Our analysis pipeline consists of a candidate identification step, followed by two parallel verification procedures:

\begin{itemize}
\item {\bf Blob detection}: To begin, we attempt to locate the most promising candidate signals using wavelets. Wavelet analysis is a compromise between working purely in position or harmonic space, and therefore yields information both about the location and angular scale of particular features in the temperature map. Specifically, we employ standard~\cite{Marinucci:2007aj,Pietrobon:2006gh,Pietrobon:2008rf,2006math......6599B,Guilloux:2007gh} and Mexican~\cite{Scodeller:2010mp} spherical needlets, two classes of wavelets defined on the sphere. The statistics of the needlet representation of a purely Gaussian CMB temperature map (expected in the absence of a bubble collision at large scales where WMAP is cosmic-variance-dominated), combined with simulations of a bubble-free masked CMB sky, can be used to quantify the significance of various features. A set of significance thresholds are then defined to ensure a manageable number of ``false detections'' in the end-to-end simulation of the WMAP experiment (see Sec.~\ref{sec:endtoend}). Regions of the sky passing these thresholds are sewn into ``blobs," whose size and location is determined by the needlet responses, and passed onto the next stages of the pipeline.
\item {\bf Edge detection}: Once a set of candidate signatures is found, we look for circular edges across which the temperature is discontinuous. As discussed above, such causal edges are expected to be a generic signature of bubble collisions. We use the Canny algorithm~\cite{11275}, adding an adaptation of the Circular Hough Transform (CHT)~\cite{360677}, to focus our search on circular edges. The algorithm consists of identifying the most likely centre for a noisy circular edge. The significance of this response is calibrated from a detailed analysis of bubble collision simulations including cosmic variance, spatially-varying WMAP instrumental noise, and smoothing due to the instrumental beam.  We verify that this step produces no false detections in the WMAP end-to-end simulation.
\item {\bf Bayesian parameter estimation and model selection}: The regions highlighted by the blob detection step can be used to construct an approximation to the full-sky posterior probability distribution for $\nsavge$ using the methods outlined in Appendix~\ref{section:method}. We first perform a pixel-based evaluation of the likelihood and Bayesian evidence in each blob for bubble collision templates of the form given in Eq.~\ref{eq:collfluct}, sampling the parameter space using the nested sampler Multinest~\cite{Feroz:2008xx}. The likelihood analysis includes cosmic variance, spatially varying WMAP instrumental noise, and the smoothing due to the instrumental beam. Combining the evidences from each blob we obtain the posterior probability distribution for $\nsavge$, which is used to derive constraints on $\nsavge$ and perform model selection to determine if a theory with $\nsavge \neq 0$ is preferred over a theory with no predicted collisions. The significance of a detection is again calibrated from an analysis of simulated collisions and an end-to-end collision-free simulation of the experiment.
\end{itemize}

The most important output of our pipeline is the approximation to the full-sky posterior probability distribution for $\nsavge$. This allows us to derive marginalized constraints on $\nsavge$, and perform model selection between theories with $\nsavge = 0$ and $\nsavge \neq 0$. In addition, for each blob identified by the first set of the pipeline, we obtain a set of marginalized posterior constraints on the model parameters $\{ z_0, z_{\rm crit}, \theta_{\rm crit}, \theta_0, \phi_0 \}$,  a maximum needlet significance, CHT score, and a local Bayesian evidence ratio with respect to the no-bubble-collision model.

\section{Simulations}\label{sec:simulatedmaps}

Our analysis pipeline is general, but each step must be calibrated using simulations of the particular data-set under consideration, in this case the WMAP 7-year data release~\cite{Jarosik:2010iu}. WMAP has measured the intensity and polarization of the microwave sky in five frequency bands. The resolution of the instrument in each band is limited by the detectors' beams, and is highest at $0.22^\circ$ in the 94 GHz W band. We perform our analysis on the foreground-subtracted W-band WMAP temperature map, as this combines the highest resolution full-sky data currently available with the least foreground contamination. To minimize the effects of the residual foregrounds we cut the sky with the conservative KQ75 mask, leaving 70.6\% of the sky unmasked. 

We carry out extensive simulations to quantify the thresholds at which areas of the sky are passed from one step to another. To find the best approximation to the full-sky Bayesian analysis, we process as much of the sky as is computationally feasible.

To determine the response of our pipeline to bubble collisions over the range of possible parameters, we generate simulations containing a variety of bubble collisions plus CMB, realistic noise and Gaussian beam smoothing. However, we also wish to ensure that we have a method to guard against systematic effects (e.g., foreground residuals and any map-making artifacts that may be present) that we do not have capability to simulate. These effects might lead to false detections in the ``blob detection'' stage, or critically, the edge detection and Bayesian analysis stages. It is impossible to claim a detection without first ensuring that there are no such false detections due to systematics. 

\subsection{WMAP end-to-end simulation}\label{sec:endtoend}

A realistic simulation of a WMAP-quality data-set that does not contain a bubble collision is an important tool for calibrating and quantifying the expected false detection rate of our analysis pipeline when applied to data which may include systematics (such as foreground residuals) that are not captured in our simulations or likelihood function. For this purpose we use a complete end-to-end simulation of the WMAP experiment provided by the WMAP Science Team~\footnote{\url{http://lambda.gsfc.nasa.gov/product/map/dr4/sim_maps_info.cfm}}. The temperature maps in this simulation are produced from a simulated time-ordered data stream, which is processed using the same algorithm as the actual data. The data for each frequency band is obtained separately from simulated sources including diffuse Galactic foregrounds, CMB fluctuations, realistic noise, smearing from finite integration time, finite beam size, and other instrumental effects. In our analysis, we utilize the foreground-reduced W-band simulation.

\subsection{Simulated bubble collisions}

The temperature fluctuations observed in the CMB, including the effects of a bubble collision (originally found in Refs.~\cite{PhysRevD.72.103002,Chang:2008gj}), can be written as
\begin{equation}
\delta T(\hat{\mathbf{n}})=  \left[ T_0' (1+f(\hat{\mathbf{n}})) (1+ \delta(\hat{\mathbf{n}})) - T_0 \right]_{\rm smoothed} + \delta T_{\rm noise}(\hat{\mathbf{n}}) \, ,
\end{equation}
where $\skypos = \{\theta, \phi\}$ is the position on the sky\footnote{These angular positions can be expressed in terms of Galactic coordinates through longitude $l = \phi$ and latitude $b =90^{\circ} - \theta$.}, $T_0$ is the average temperature of the map including the modulation, $T_0'$ is the average temperature without the modulation, $\delta T_{\rm noise}$ is the contribution from instrumental noise, $f(\hat{\mathbf{n}})$ and $\delta(\hat{\mathbf{n}})$ are defined as in Eq.~\ref{eq:tempmod}. The quantities in the brackets are smoothed with a Gaussian beam of $0.22^\circ$ (approximating the beam size of the WMAP experiment in the W band). We use the WMAP best-fit 7-year power spectrum~\cite{Larson:2010gs} in the multipole range $2 \leq \ell \leq 1024$ to generate fluctuation maps $\delta T_{\rm syn} (\hat{\mathbf{n}}) = T^{\prime}_0\delta(\hat{\mathbf{n}})$ at the full WMAP resolution of $N_{\rm side} = 512$ (with 3,145,728 pixels). The noise term $\delta T_{\rm noise}$ is generated from WMAP 7-year noise variances at the same resolution. Since the templates we consider add a relatively small temperature excess/deficit in one location, the features do not cause the power spectrum to deviate from that measured by WMAP~\cite{Chang:2008gj}. Additionally, we can replace $T_{0} \approx T^{\prime}_{0}$, which gives
\begin{equation}
\delta T(\hat{\mathbf{n}}) = \left[ (1 + f(\hat{\mathbf{n}}))(T_{0} + \delta T_{\rm syn}(\hat{\mathbf{n}})) - T_{0} \right]_{\rm smoothed} + \delta T_{\rm noise}(\hat{\mathbf{n}})
\label{eq:final_dT}.
\end{equation}

\begin{figure}[tb]
   \includegraphics[width=10 cm]{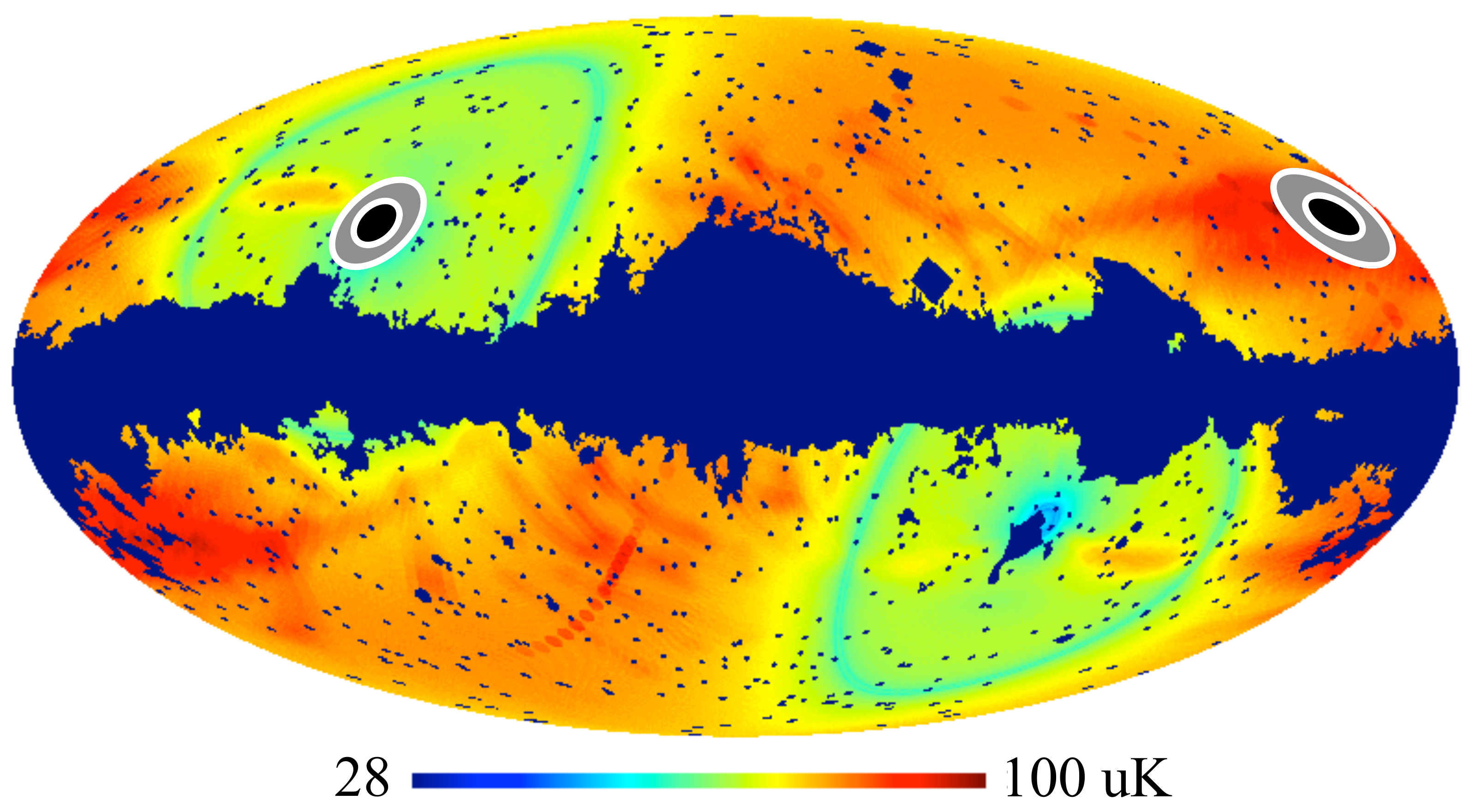}
\caption{The two locations chosen for our simulated bubble collisions are over-plotted on the WMAP 7-year noise variances with the KQ75 7-year mask applied. The regions encompassed by the $5^{\circ}$ and $10^{\circ}$ simulated collisions are shaded black and grey respectively.  Bubbles are centered in an unmasked high-noise $\{ \theta_0 = 56.6^{\circ}, \phi_0 = 193.0^{\circ} \}$ (left) and low-noise $\{ \theta_0 = 57.7^{\circ}, \phi_0 = 99.2^{\circ}\}$ region (right).}
\label{fig-bubble_locations}
\end{figure}

We consider collisions with $\theta_{\rm crit} = 5^{\circ}, 10^{\circ}, 25^{\circ}$ and choose centers in a high-noise region ($ \theta_0 = 56.6^{\circ}, \phi_0 = 193.0^{\circ} $) and a low-noise region $( \theta_0 = 57.7^{\circ}, \phi_0 = 99.2^{\circ})$ of the sky that remain significantly outside of the main body of the WMAP KQ75 7-year mask. The regions of the sky affected by $5^{\circ}$ and $10^{\circ}$ collisions are over-plotted in Fig.~\ref{fig-bubble_locations} on a masked map of the instrumental noise variance. For each $\theta_{\rm crit}$ and location, we generate $35$ simulated collisions with parameter values logarithmically spaced in the ranges $10^{-6} \leq z_{\rm 0} \leq 10^{-4}$ and $-10^{-4} \leq z_{\rm crit} \leq -10^{-6}$. The response of our pipeline depends only on the absolute value of $z_0$ and $z_{\rm crit}$, so the choice of sign for $z_0$ and $z_{\rm crit}$ is arbitrary. We repeat this for three realizations of the background CMB fluctuations, yielding a total of $210$ simulated sky maps for each of the three collision sizes.

\section{Analysis tools}\label{sec:analysistools}

We now describe in detail the analysis tools which make up our pipeline and how they are calibrated with simulations before being applied to the data. Readers wishing to skip these details may wish to study Figs.~\ref{fig-needletexclusion} and \ref{fig-chtexclusion} and turn to Sec.~\ref{sec:det_conditions} for a summary of the outputs of the pipeline at each stage, and the conditions under which a detection can be claimed. 

\subsection{Needlets}

Wavelet analysis is a powerful tool for identifying features localized on the sky, the type of signal expected for bubble collisions. There exist families of wavelets that are defined on the sphere known as standard~\cite{Marinucci:2007aj,Pietrobon:2006gh,Pietrobon:2008rf,2006math......6599B,Guilloux:2007gh} and Mexican~\cite{Scodeller:2010mp} spherical needlets. These functions form what is known as a ``tight frame," allowing for a well-defined forward and reverse needlet transform. As in other forms of wavelet analysis, decomposing the temperature map into a sum over such functions yields information both about the location and angular scale of  specific features. For a purely Gaussian temperature field, the statistical properties of the needlet transform can be straightforwardly related to the power spectrum, allowing a quantification of the significance of a possible detection. In addition, the spatial localization properties of the standard and Mexican needlets make it possible to avoid many of the problems associated with working on a cut sky. In this section we outline the properties of needlet transforms and analyze their utility in searching for bubble collisions.

\subsubsection{Definition of the spherical needlet transform}

The needlet transform is defined as:
 \begin{equation}
T (\skypos) = \sum_{j, k} \beta_{jk} \psi_{jk} (\skypos),
\end{equation}
where $\skypos$ denotes a direction $\{\theta, \ \phi \}$ on the sky, $\beta_{jk}$ are constant needlet coefficients, and $\psi_{jk} (\skypos)$ are the needlet functions. The members of this family of functions are labeled by the index $k$ of the pixel at which they are centered, and their ``frequency" $j$, which is related to the spatial extent of the needlet profile in real space. The sum in the needlet transform is over all pixels $k$, and all frequencies $j=0, 1, 2, \ldots, \infty$. For fixed $j$, there is one needlet coefficient $\beta_{jk}$ for each pixel $k$, allowing us to represent the needlet coefficients at fixed $j$ as a map on the pixelated sky. The needlet functions are defined in terms of spherical harmonics $Y_{\ell m}(\skypos)$ as
\begin{equation}
\psi_{j k} (\skypos) = \sqrt{\lambda_{jk} } \sum_\ell b \left( \ell, B, j \right) \sum_{m=-\ell}^{\ell} Y_{\ell m}^* (\skypos ) Y_{\ell m} (\skypos_{k} ).
\end{equation}
Here, $\lambda_{jk}$ are the cubature weights, which are related to the area of each pixel. In the equal-area HEALPix pixelization~\cite{Gorski:2004by} we employ, all cubature weights are equal to $\lambda_{jk} = 4 \pi / N_{\rm pix}$, where $N_{\rm pix}$ is the number of pixels, and we absorb this constant into the needlet coefficients $\beta_{jk}$. The function $b(\ell, B, j)$ acts as a filter in harmonic space, where $B$ is a constant bandwidth parameter. It is chosen such that the family of functions $\psi_{jk}(\skypos)$ form a tight frame (see e.g., Ref.~\cite{Marinucci:2007aj}), which guarantees the existence of an inverse needlet transform given by
\begin{equation}\label{eq:betas}
\beta_{j k} = \int T(\skypos) \psi_{j k} (\skypos) d\Omega.
\end{equation}

\begin{figure}[tb]
   \includegraphics[width=6 cm]{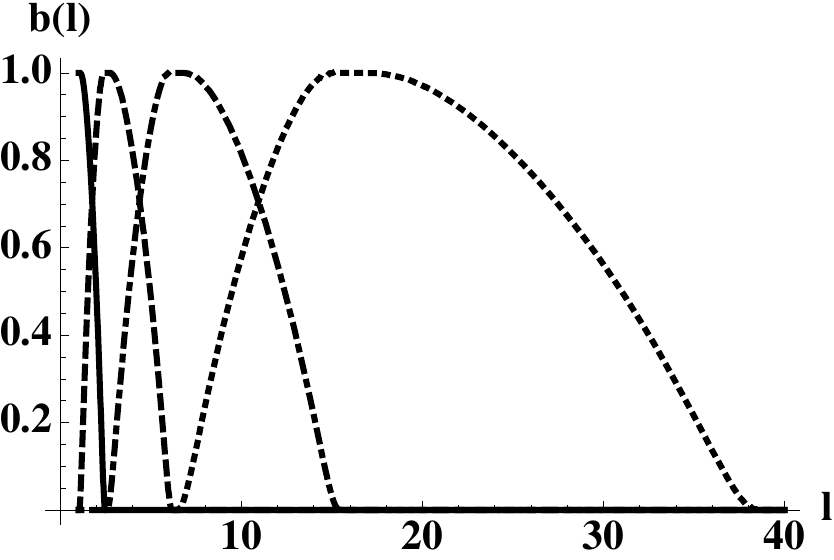}
   \includegraphics[width=6 cm]{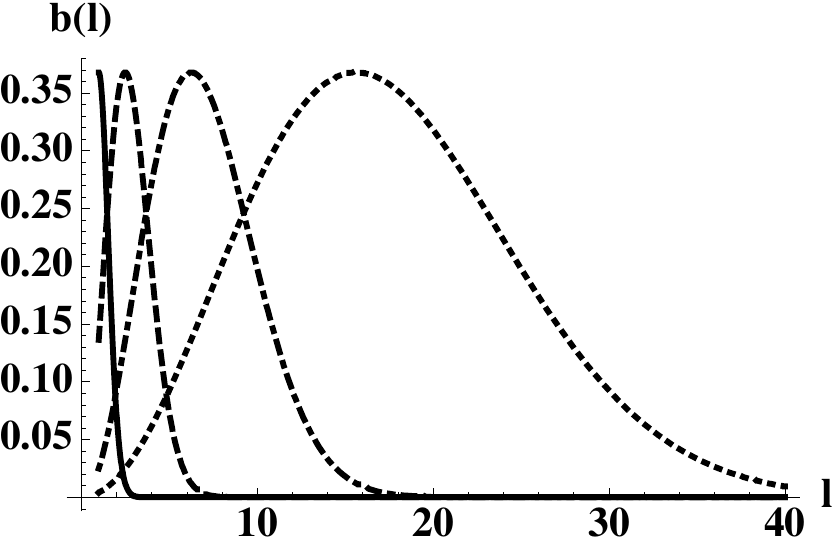}
\caption{The filter function $b(\ell, B, j)$ for standard (left) and Mexican (right) needlets with $B=2.5$ for $j=0,1,2,3$ (solid, dashed, dot-dashed, dotted).}
\label{fig-filterfunctions}
\end{figure}

There are a number of possible choices for the function $b(\ell, B, j)$, which distinguish the standard and Mexican needlets. A description of the explicit form of the function $b(\ell, B, j)$ can be found for standard needlets in Ref.~\cite{Marinucci:2007aj} and Mexican needlets in Ref.~\cite{Scodeller:2010mp}. We plot $b$ as a function of the multipole moments $\ell$ in Fig.~\ref{fig-filterfunctions}. For standard needlets, $b$ only has support for values of $\ell$ between $B^{j-1} < \ell < B^{j+1}$. The bandwidth parameter $B$ controls the width of each window function in harmonic space. Mexican needlets have support over all $\ell$ at each frequency $j$, and again have a bandwidth parameter $B$ which controls the localization properties of the functions in harmonic space.

\begin{figure}[tb]
   \includegraphics[width=6 cm]{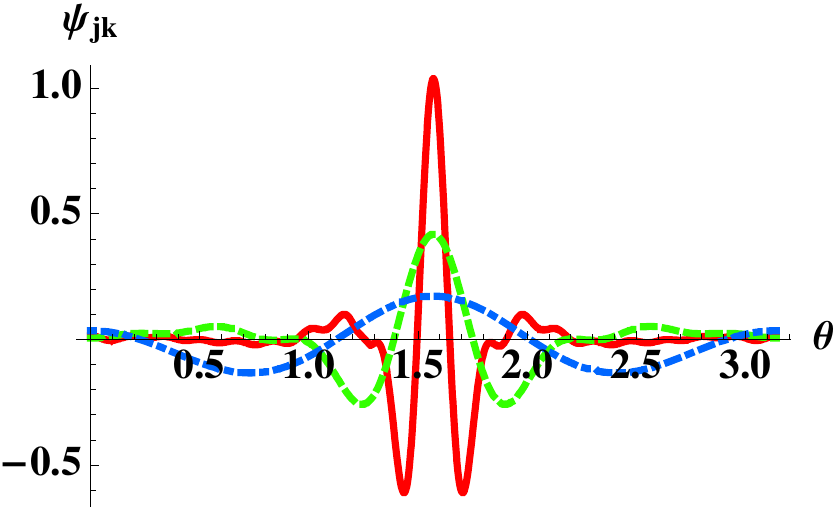}
   \includegraphics[width=6 cm]{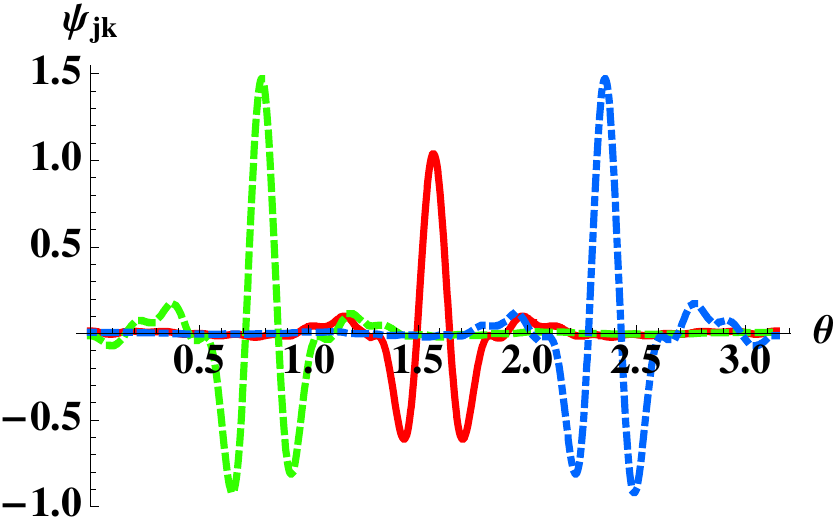}
\caption{Standard needlets in pixel space. On the left, we show standard needlets $\psi_{jk}$ with $B=2.5$ for $j=1,2,3$ (dot-dashed, dashed, solid) at fixed $k$ as a function of the polar angle $\theta$. On the right, we show standard needlets $\psi_{jk}$ for fixed $j=3$ at three pixels $k$ (dashed, solid, dot-dashed) as a function of the polar angle $\theta$ (note: since we are projecting onto $\theta$, the needlets appear asymmetric).}
\label{fig-needletsreal}
\end{figure}

In Fig.~\ref{fig-needletsreal} we plot the wavelet functions in pixel space. As is to be expected, increasing the width of the function $b(\ell, B, j)$ in harmonic space corresponds to improved localization in pixel space. In the limit of large $j$, there is an extremely small overlap of the needlet functions at nearby pixels. The compact support of $b(\ell, B, j)$ in harmonic space for standard needlets leads to slightly poorer localization in pixel space than is enjoyed by the Mexican needlets.

If we decompose the temperature map into spherical harmonics:
\begin{equation}
T (\skypos) = \sum_{\ell,m} a_{\ell m} Y_{\ell m} (\skypos),
\end{equation}
then  Eq.~\ref{eq:betas}, together with the inverse transform
\begin{equation}
a_{\ell m} = \int T(\skypos) Y_{\ell m}^* d\Omega
\end{equation}
leads to:
\begin{equation}\label{eq:finalbetas}
\beta_{j k} = \sqrt{\lambda_{jk} } \sum_\ell b \left( \ell, B, j \right) \sum_{m=-\ell}^{\ell} a_{\ell m} Y_{\ell m} (\skypos_{k} ),
\end{equation}
where $a_{\ell m}$ are the spherical harmonic coefficients. This formula allows us to transform directly from the $a_{\ell m}$s to the spherical needlet coefficients $\beta_{j k}$. In our analysis, the needlet transforms are accelerated by generating the $a_{\ell m}$s at full WMAP resolution ($N_{\mathrm{side }} = 512$) but limiting the reconstruction multipoles to $2 \leq \ell \leq 256$, and needlet positions $k$ to the pixels at $N_{\mathrm{side}} = 128$. This retains the resolution required to reconstruct features from half-sky to half-degree scales, encompassing the range of all detectable collisions.

\subsubsection{Needlet response to the bubble collision templates}

\begin{figure}[tb]
   \includegraphics[width=6 cm]{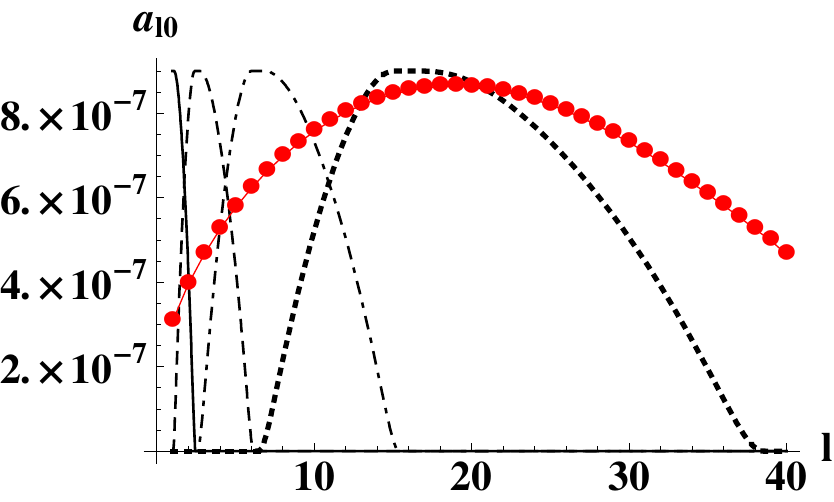}
   \includegraphics[width=6 cm]{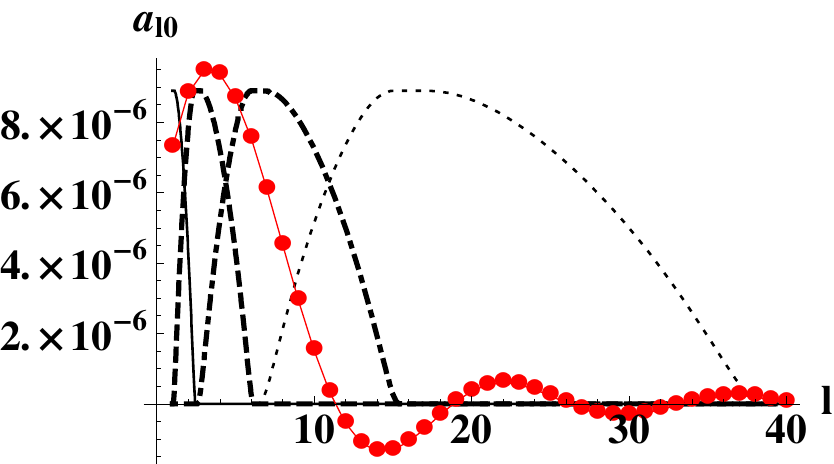}
\caption{The spherical harmonic transform (connected dots) of a $\theta_{\rm crit} = 5^{\circ}$ (left) and $\theta_{\rm crit} = 25^{\circ}$ (right) bare collision template centered on the north pole  on top of the filter function $b(\ell, B, j)$ for standard needlets with $B=2.5$ for $j=0,1,2,3$ (solid, dashed, dot-dashed, dotted). The overlap of the filter function $b(\ell, B, j)$ with the spherical harmonic transform of the bubble template (see Eq.~\ref{eq:finalbetas}) determines for which value of $j$ the needlet transform yields the largest signal. In these examples, the $5^{\circ}$ collision has the largest needlet response at $j=3$ and the $25^{\circ}$ collision at $j=2$. The needlet response as a function of angle for the $25^{\circ}$ collision is plotted in Fig.~\ref{fig-betaresponse}.}
\label{fig-aandneedlets}
\end{figure}

We now quantify the sensitivity of the needlet transform to the presence of a collision. In the absence of the background Gaussian fluctuations, we perform the needlet transform on a set of bare collision templates. As an illustration, in Fig.~\ref{fig-aandneedlets} we plot the spherical harmonic coefficients as a function of $\ell$ (all coefficients for $m \neq 0$ vanish by symmetry if we center the template on the north pole) for $25^{\circ}$ and $5^{\circ}$ collision templates, overlaid on top of the rescaled filter function $b(\ell, B, j)$ for spherical needlets. The spherical harmonic coefficients for the collision templates peak at a value of $\ell$ related to the angular scale of the causal boundary. Therefore, the needlet coefficients are largest at a frequency $j$ that is directly related to the angular scale of the collision. This can be seen in Fig.~\ref{fig-aandneedlets}, where the $5^{\circ}$ collision has a maximum response at $j=3$ and the $25^{\circ}$ collision has a maximum response at $j=2$. 

\begin{figure}[tb]
   \includegraphics[width=6 cm]{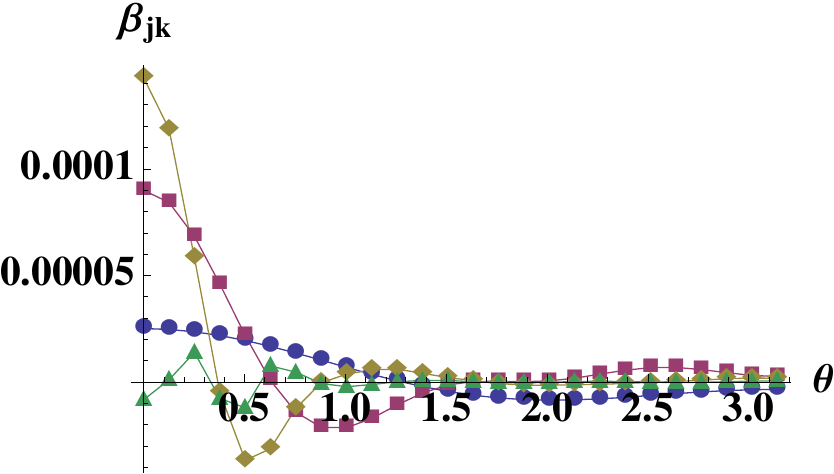}
\caption{$B=2.5$ for standard needlets and a $25^{\circ}$ collision. $j=0$ (circles), $j=1$ (squares), $j=2$ (diamonds), and $j=3$ (triangles).}
\label{fig-betaresponse}
\end{figure}

In Fig.~\ref{fig-betaresponse}, we plot the needlet coefficients for the $25^{\circ}$ template at a variety of polar angles, for $0 \le j \le 3$. The needlet coefficients are largest in the center of the region affected by the collision (here, the template is centered on $\theta = 0$), and at a frequency $j$ correlated with the angular scale of the collision ($j=2$). As expected, the needlet response is sensitive to both the location and angular scale of the collision.

By studying the needlet response to a variety of bare collision templates given by Eq.~\ref{eq:collfluct}, we can find optimal values of the bandwidth parameter $B$ for each needlet type. Larger-bandwidth needlets produce stronger signals, but also respond to a greater range of bubble sizes. We have found that the values $B=1.8$ and $B=2.5$ for the standard and $B=1.4$ for the Mexican needlets are a good compromise between signal strength and angular localization of response. In our analysis, we use this suite of three needlet transforms, ensuring that we are sensitive to temperature modulations with a variety of profiles, and allowing us to cross-check any candidate signals.

As an important step in our analysis pipeline, we build lookup tables containing the possible range of bubble collision scales, $\theta_{\rm crit, \mathrm{min}} \le \theta_{\rm crit} \le \theta_{\rm crit, \mathrm{max}}$, to which each needlet type and frequency is sensitive. We first generate a set of 100 templates at each integer $\theta_{\rm crit}$ between $1^{\circ}$ and $89^{\circ}$ by randomly sampling $z_{\rm crit}$ between $-5 \times 10^{-5} \leq z_{\rm crit} \leq 5 \times 10^{-5}$ with $z_0 = 5 \times 10^{-5}$ for $z_{\rm crit} > 0$, and $z_0 = z_{\rm crit} + 5 \times 10^{-5}$ for $z_{\rm crit} < 0$. This creates a set of templates with uniform total amplitude (i.e., constant $f(\theta, \phi) - f(\theta_{\rm crit}, \phi)$) but a variety of profiles at each angular scale. Next, we calculate the needlet coefficients for each of the three members of our needlet suite, and record the frequency generating the maximum central needlet response for each template. The range in $\theta_{\rm crit}$ recorded at each frequency is used to generate the lookup tables in Table~\ref{tab:anglelookup}.

\begin{table*}
\begin{tabular}[t]{c c c}
\hline
\hline
 $j$ \ & \  $\theta_{\rm crit, min}$ \ & \ $\theta_{\rm crit, max}$ \ \\
\hline
0 & $60\,^{\circ}$ & $90\,^{\circ}$ \\
1 & $33\,^{\circ}$ & $71\,^{\circ}$ \\
2 & $12\,^{\circ}$ & $32\,^{\circ}$ \\
3 & $5\,^{\circ}$ & $14\,^{\circ}$ \\
4 & $2\,^{\circ}$ & $5\,^{\circ}$ \\
5 & $1\,^{\circ}$ & $2\,^{\circ}$ \\
\hline
\hline
 \end{tabular} 
 \
 \
 \
 \begin{tabular}[t]{c c c}
\hline
\hline
 $j$ \ & \  $\theta_{\rm crit, min}$ \ & \ $\theta_{\rm crit, max}$ \ \\
\hline
1 & $56\,^{\circ}$ & $90\,^{\circ}$ \\
2 & $28\,^{\circ}$ & $64\,^{\circ}$ \\
3 & $17\,^{\circ}$ & $38\,^{\circ}$ \\
4 & $10\,^{\circ}$ & $21\,^{\circ}$ \\
5 & $6\,^{\circ}$ & $12\,^{\circ}$ \\
6 & $3\,^{\circ}$ & $7\,^{\circ}$ \\
7 & $2\,^{\circ}$ & $4\,^{\circ}$ \\
8 & $1\,^{\circ}$ & $2\,^{\circ}$ \\
\hline
\hline
 \end{tabular} 
 \
 \
 \
  \begin{tabular}[t]{c c c}
\hline
\hline
 $j$ \ & \  $\theta_{\rm crit, min}$ \ & \ $\theta_{\rm crit, max}$ \ \\
\hline
0 & $86\,^{\circ}$ & $90\,^{\circ}$ \\
1 & $78\,^{\circ}$ & $90\,^{\circ}$ \\
2 & $55\,^{\circ}$ & $90\,^{\circ}$ \\
3 & $36\,^{\circ}$ & $71\,^{\circ}$ \\
4 & $27\,^{\circ}$ & $48\,^{\circ}$ \\
5 & $19\,^{\circ}$ & $37\,^{\circ}$ \\
6 & $14\,^{\circ}$ & $27\,^{\circ}$ \\
7 & $10\,^{\circ}$ & $20\,^{\circ}$ \\
8 & $8\,^{\circ}$ & $16\,^{\circ}$ \\
9 & $5\,^{\circ}$ & $12\,^{\circ}$ \\
10 & $4\,^{\circ}$ & $8\,^{\circ}$ \\
11 & $3\,^{\circ}$ & $6\,^{\circ}$ \\
12 & $2\,^{\circ}$ & $4\,^{\circ}$ \\
13 & $1\,^{\circ}$ & $2\,^{\circ}$ \\
\hline
\hline
 \end{tabular} 
  \begin{center}
 \caption{Angular scale lookup tables for standard needlets with $B=2.5$ (left), standard needlets with $B=1.8$ (center), and Mexican needlets with $B=1.4$ (right). For a needlet frequency $j$, the needlet transform is sensitive to bubble collisions on scales $\theta_{\rm crit, min} \leq \theta_{\rm crit} \leq \theta_{\rm crit, max}$. No results are shown for the standard needlets with $B = 1.8$, $j = 0$ as they have no support over the range of angular scales considered.
   \label{tab:anglelookup}}
 \end{center}
\end{table*}

\subsubsection{Needlet coefficients on a cut sky}

The CMB is completely dominated by foreground emission in the region of the Galactic plane, and is also affected by bright point-sources. These issues are typically handled by applying a mask which covers the Galactic plane and known point-sources. The needlet transform can be applied directly on the masked temperature maps, and because the needlet functions are localized in pixel space, needlet coefficients far from the mask for sufficiently high frequency $j$ are not significantly affected. These high-frequency needlets are mainly composed of high-$\ell$ spherical harmonics, and so cut-sky $a_{\ell m}$s can safely be used to calculate the needlet coefficients through Eq.~\ref{eq:finalbetas}. Unfortunately, the low-frequency needlets are quite sensitive to the presence of the mask. To partially mitigate this sensitivity, we calculate the optimal unbiased maximum-likelihood estimators of the $a_{\ell m}$s~\cite{deOliveiraCosta:2006zj} at low $\ell$. 

Such maximum-likelihood estimators are computationally very expensive, and we must balance accuracy against limited computational resources. Another minor complication arises from the smoothing that is necessary to band-limit the data when performing the maximum-likelihood reconstruction algorithm of Ref.~\cite{deOliveiraCosta:2006zj}: information leaks from inside the mask. Comparing the reconstruction on masked and unmasked simulated temperature maps using $10^{\circ}$-FWHM Gaussian smoothing, we have determined that a reasonably small bias is obtained when maximum-likelihood $a_{\ell m}$s are used for $\ell < 10$.

This set of  hybrid $a_{\ell m}$s -- maximum-likelihood reconstructed $a_{\ell m}$s for $\ell \leq 10$ and cut-sky $a_{\ell m}$s for $\ell > 10$ -- is used in Eq.~\ref{eq:finalbetas} to calculate the needlet coefficients in the analysis that follows.

\subsubsection{Statistical properties of needlet coefficients}

For a Gaussian CMB without sky cuts, the statistical properties of the spherical harmonic coefficients~\cite{Marinucci:2007aj} are
\begin{equation}
\langle a_{\ell m}\rangle = 0, \ \ \ \langle |a_{\ell m}|^2 \rangle = C_{\ell}.
\end{equation}
These are related to the statistical properties of the $\beta_{jk}$ in a straightforward way by
\begin{equation}\label{eq:fullskyneedletstats}
\langle \beta_{jk} \rangle = 0, \ \ \ \langle \beta_{jk}^2 \rangle = \sum_{\ell} b \left( \ell, B, j \right) \frac{2 \ell+1}{4 \pi} C_{\ell},
\end{equation}
which are identical at each pixel $k$. Thus, in a full-sky analysis, comparison with the Gaussian variance yields a measure of how likely it is to find a particular needlet coefficient in a purely Gaussian realization of the CMB sky. 

In the presence of foregrounds, however, it is necessary to work on a cut sky, which introduces a $j$- and $k$-dependent bias. Following Ref.~\cite{Pietrobon:2008rf}, we determine the significance of a needlet coefficient on the cut sky by~\footnote{One can also define composite significances, involving needlet coefficients at multiple frequencies. An example is
\begin{equation}
S_{jj'} = \frac{|\beta_{jk} \beta_{j'k} - \langle \beta_{jk} \beta_{j'k}\rangle |}{\sqrt{\langle (\beta_{jk} \beta_{j'k})^2 \rangle}}.
\end{equation}
We have evaluated this statistic on a variety of collision templates modulating Gaussian realizations of the background CMB fluctuations, and found that it returns about half the significance given by Eq.~\ref{eq:needletsignificance}.} 
\begin{equation}\label{eq:needletsignificance}
S_{jk} = \frac{|\beta_{jk} - \langle\beta_{jk}\rangle_{\mathrm{gauss, cut}}|}{\sqrt{\langle\beta^{2}_{jk}\rangle_{\mathrm{gauss, cut}}}},
\end{equation} 
where the average, $\langle\beta_{jk}\rangle_{\mathrm{gauss, cut}}$, and variance, $\langle\beta^{2}_{jk}\rangle_{\mathrm{gauss, cut}}$, are calculated at each pixel from the needlet coefficients of 3000 collision-free Gaussian CMB realisations with the WMAP 7-year KQ75 sky cut applied. Simulating only cosmic variance is sufficient here because the measurements made by WMAP are cosmic-variance-limited on the scales of interest, $\theta_{\rm crit} \agt 5^{\circ}$. 

\begin{figure}[tb]
   \includegraphics[width=8 cm]{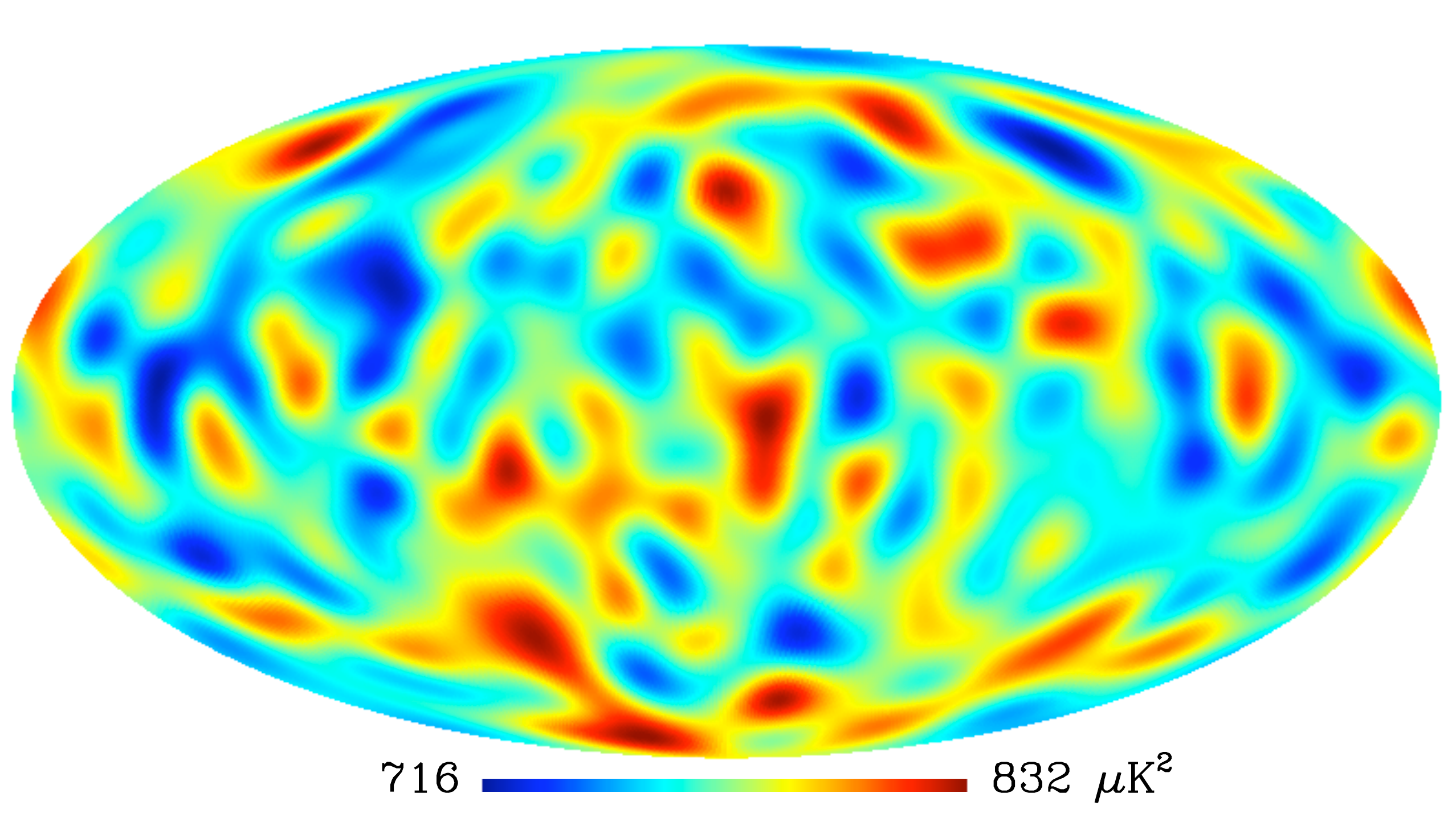}
   \includegraphics[width=8 cm]{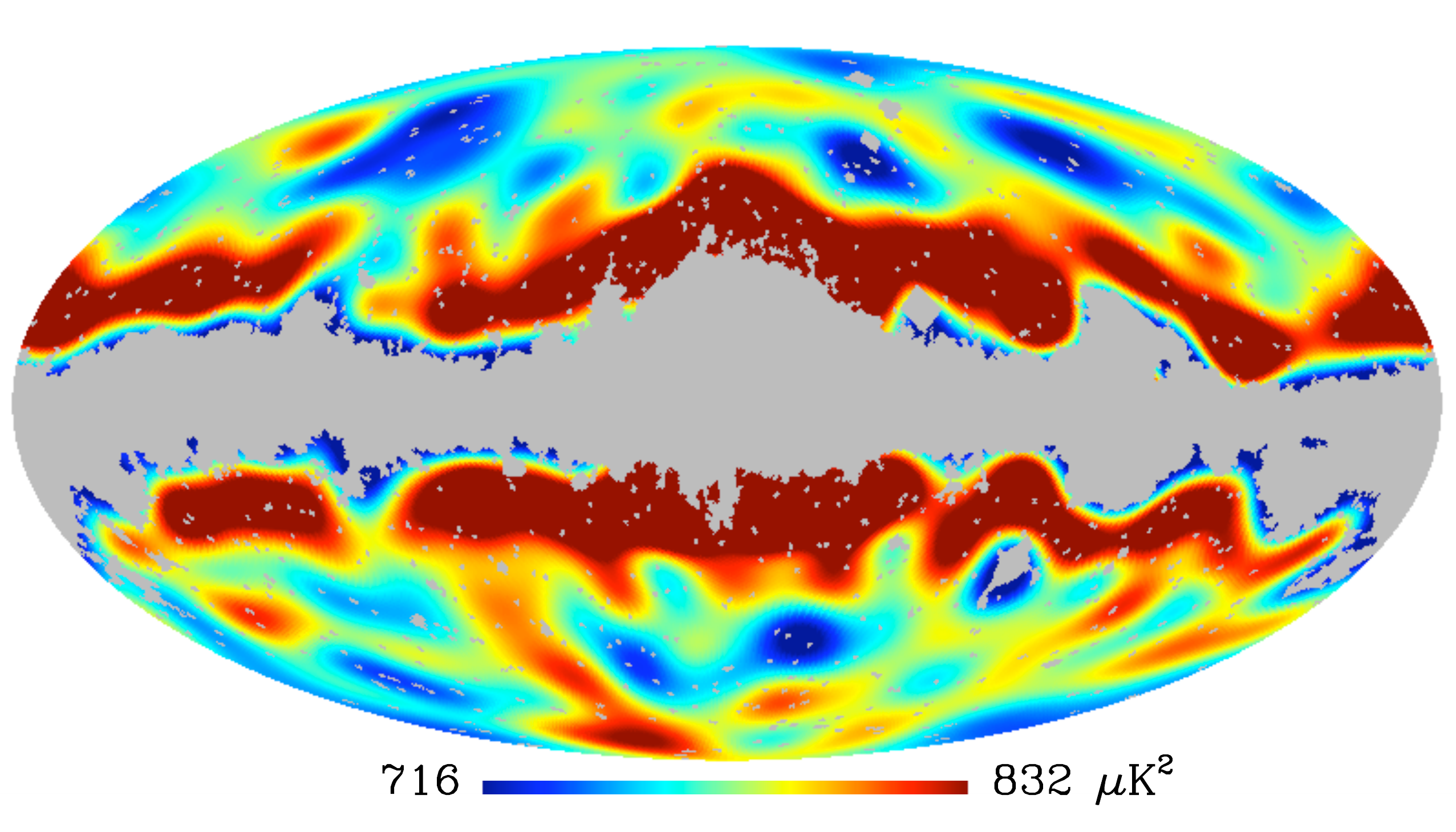}
   \includegraphics[width=8 cm]{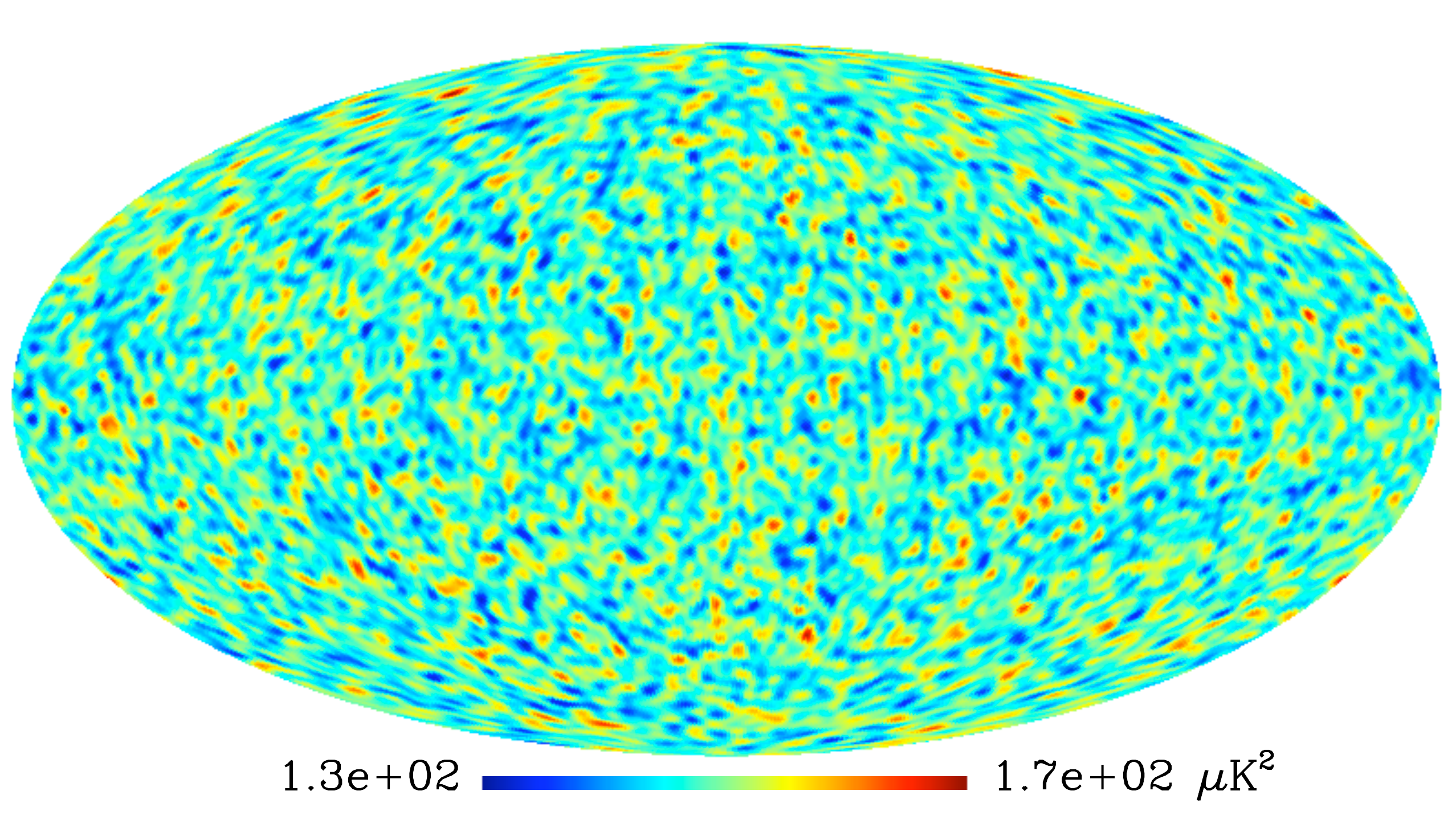}
   \includegraphics[width=8 cm]{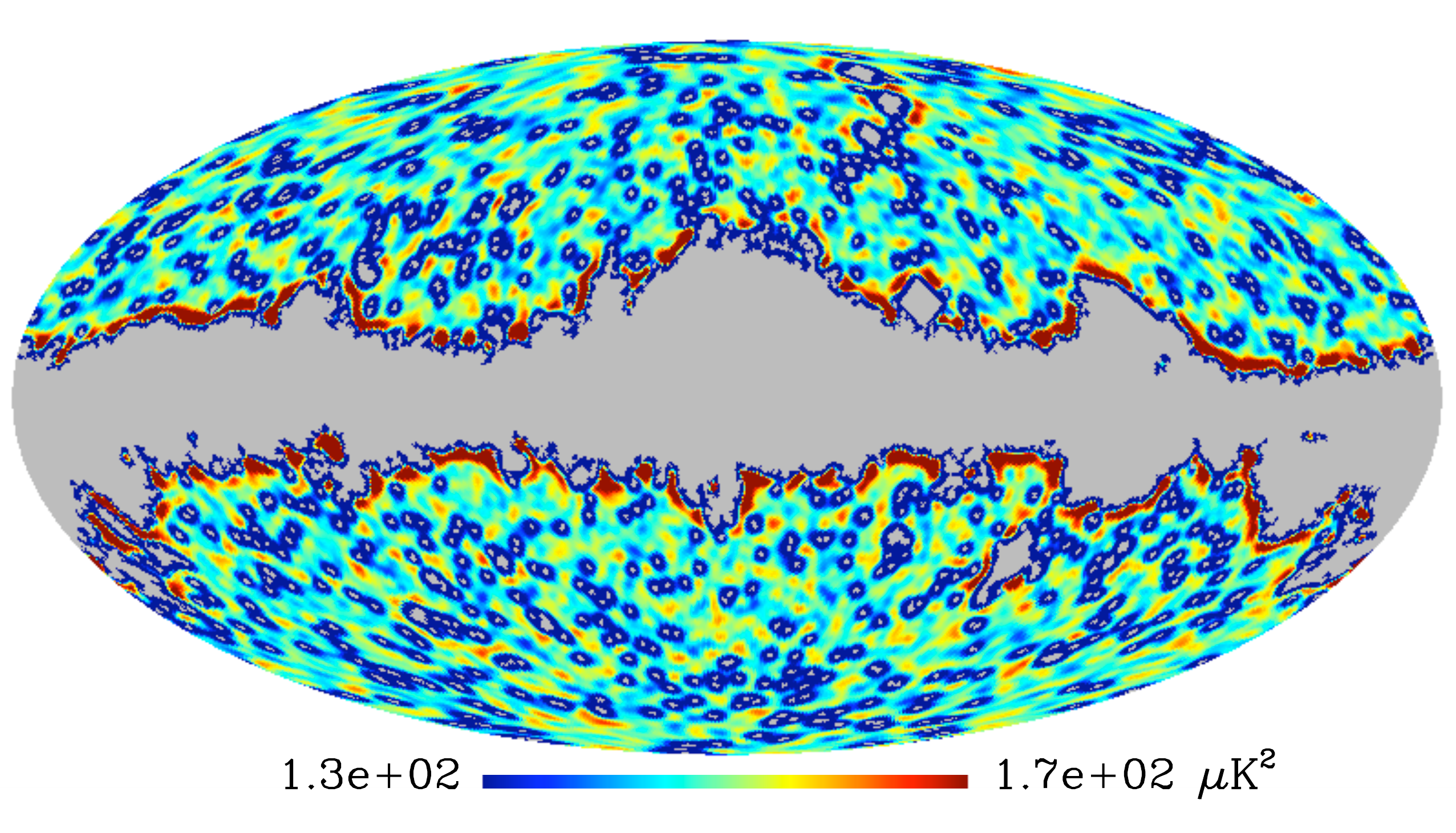}
\caption{Needlet-coefficient variance maps for standard needlets with $B = 2.5$, $j = 2$ (top row) and Mexican needlets with $B = 1.4$, $j = 11$, generated using an ensemble of 3000 Gaussian CMB realizations. The map-averaged variances of the full-sky maps (left column) agree well with expectations from theory (Eq.~\ref{eq:fullskyneedletstats} predicts $774\ \mu$K$^{2}$ and $150\ \mu$K$^{2}$ respectively), as do regions of sky sufficiently distant from the 7-year KQ75 mask (right column). The low-frequency needlets are affected predominantly by the Galactic cut, whereas the high-frequency needlets are affected by point-source masking.}
\label{fig-needletvarmaps}
\end{figure}

Maps of the needlet variances obtained from simulations are shown in Fig.~\ref{fig-needletvarmaps} for an example with low (top) and high needlet (bottom) frequency. On the left are the needlet variances calculated without a mask, which agree at the $5 \%$ level with the expected variances from Eq.~\ref{eq:fullskyneedletstats}. On the right are the masked needlet variances, which are clearly biased within a certain distance from the mask in both cases. Variances in the low-frequency example are affected predominantly by the Galactic cut. Variances in the high-frequency example are affected in a much smaller region of the Galactic cut (reflecting the increased spatial localization of needlets at high frequencies), but are much more significantly affected by the point source masks.

\begin{figure}[tb]
   \includegraphics[width=14 cm]{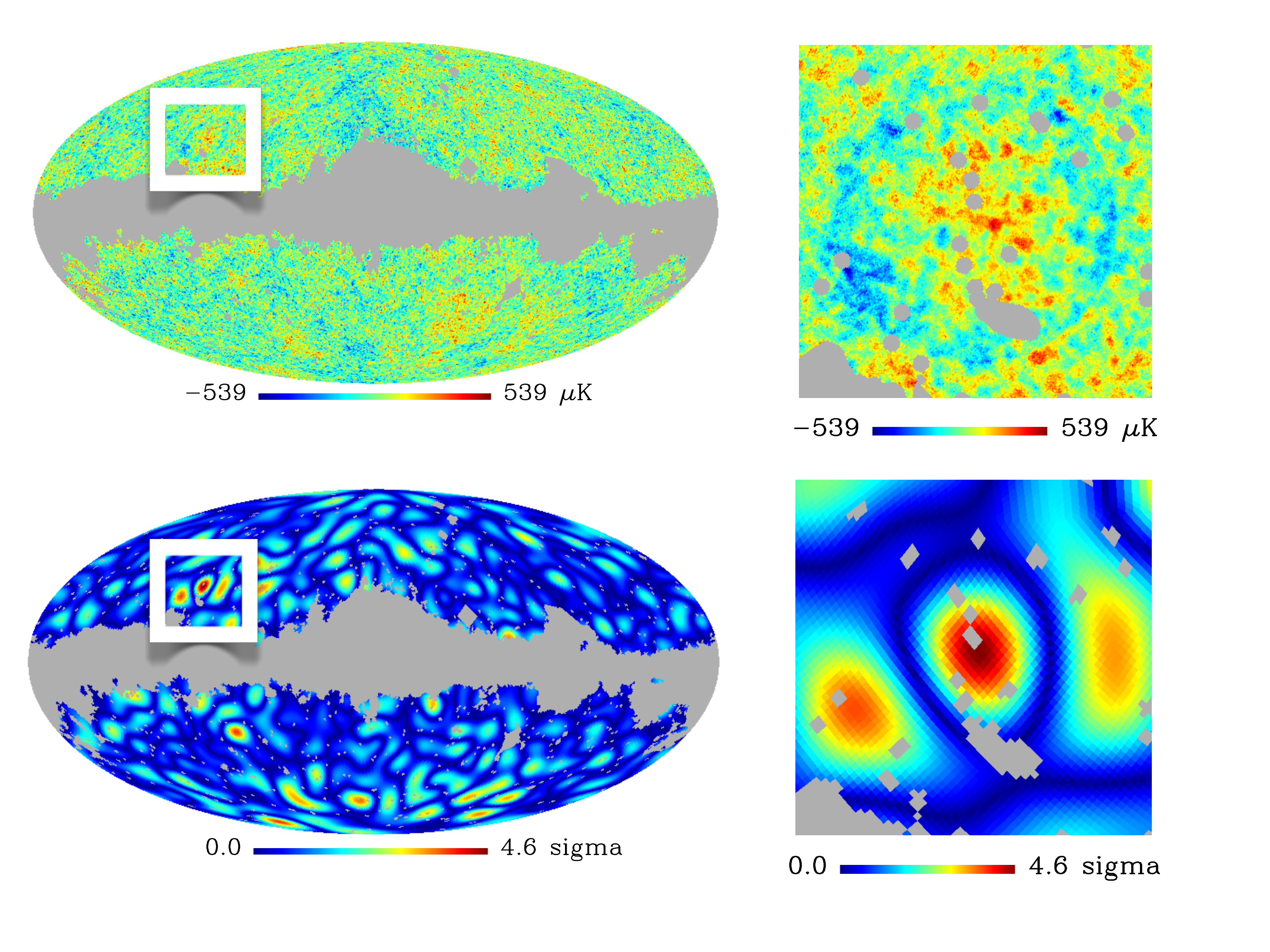}
\caption{The temperature (top) and needlet coefficient significance (bottom) maps for a simulated bubble collision with $z_0 = 5 \times 10^{-5}$, $z_{\rm crit} = -5 \times 10^{-5}$, $\theta_{\rm crit} = 10^{\circ}$ on the CMB sky with the WMAP 7-year KQ75 mask applied. We show the map of needlet coefficients which gives the largest needlet response, in this case $j=3$ for standard needlets with $B=2.5$. The right-hand panels provide close-ups of the collision region.} 
\label{fig-needletbubble}
\end{figure}

To illustrate the expected response to a bubble collision, in Fig.~\ref{fig-needletbubble} we show the temperature map of our illustrative example of a simulated bubble collision on the cut sky, and the significances of its needlet coefficients calculated from Eq.~\ref{eq:needletsignificance} at $j=3$ using standard needlets with $B=2.5$. The location of the collision is clearly highlighted in the map of needlet coefficients. The significance of the needlet coefficients in pixels in the center of the collision form a global maximum on the entire map.

In order to identify a set of needlet coefficients with a particular feature, we sew regions with 5 or more pixels whose needlet coefficients exceed a frequency-dependent threshold into ``blobs" (we discuss in more detail below how these thresholds are set). The core of each blob contains all adjacent pixels that pass the significance threshold. This core is then extended by first finding the average position $\skypos_0$ of the pixels in a blob and modeling it as a disc of radius $\Delta \theta / 2$ (where $\Delta \theta$ is the maximum separation between any two pixels in the blob) centred on $\skypos_0$. The blob is then extended to a radius of $1.1 \times ( \theta_{\rm crit, max} + \Delta \theta /2)$, where $\theta_{\rm crit, max}$ is found from Table~\ref{tab:anglelookup} (which is dependent upon the needlet type and frequency at which a feature is found) to ensure we include all related pixels. All pixels not contained in a blob are masked, and this new temperature map is passed to subsequent steps in the analysis pipeline. Eliminating irrelevant pixels allows us to drastically reduce the computational effort needed in the subsequent analysis.

\subsubsection{Analysis of the WMAP end-to-end simulation}

As there are many independent needlet coefficients over the sky, it is inevitable that highly significant features will be detected in even a purely Gaussian CMB temperature map. In addition, residual foregrounds and instrumental artifacts may give rise to features which are mis-classified as candidate bubble collisions by the pipeline. To understand how these effects might contribute to the needlet significances, we ran the suite of needlet transforms on the end-to-end simulation of the WMAP experiment (described in Section~\ref{sec:simulatedmaps}) with the 7-year KQ75 mask applied. As an illustration of our results, in Table~\ref{tab:bobcounts} we give the number of blobs of varying significance found in the masked end-to-end simulation using standard needlets with $B=2.5$. At increasingly high frequency, for which there are more independent needlet coefficients, more and more blobs are found that pass relatively large significance thresholds. 

\begin{table*}
\begin{tabular}{c c c c c c}
\hline
\hline
$ j$ & \  $S = 3$ \ & \ 3.25 \ & \ 3.5 \ & \ 3.75 \ & \ 4 \ \\
\hline
0 & 0 & 0 & 0 & 0 & 0 \\
1 & 0 & 0 & 0 & 0 & 0 \\
2 & 0 & 0 & 0 & 0 & 0 \\
3 & 10 & 4 & 2 & 1 & 0 \\
4 &  23 & 10 & 4 & 0 & 0 \\
\hline
\hline
 \end{tabular}
 \begin{center}
 \caption{The number of blobs found in the masked WMAP end-to-end simulation above significances ranging from $3 \leq S \leq 4$ for standard needlets with $B=2.5$. 
   \label{tab:bobcounts}}
 \end{center}
\end{table*}

We use the results of Table~\ref{tab:bobcounts} (and similar tables for other members of the needlet suite) to define a set of needlet frequency-dependent significance thresholds that allow a manageable number of false-positives, while retaining sensitivity to a fairly large range of collision template parameters. The significance thresholds we use in our analysis are shown in Table~\ref{tab:thresholds}. There are a total of 17 blobs in the masked end-to-end simulation that pass these thresholds. Comparing their locations on the sky, we can associate these blobs with 13 features (if a feature is picked up by multiple needlet types or frequencies, it can have multiple blobs associated with it). For three of these features, the set of pixels that pass the needlet threshold intersect the Galactic cut. We associate these with a response to the mask, and do not consider them further. For the other ten features, the needlet type and frequency which yielded the maximum significance is recorded in Table~\ref{tab:endtoendfeatures}.

\begin{table*}
\begin{tabular}[t]{c c c}
\hline
\hline
 $j$ & \  $S_{\rm min}$ \ & $N_{\rm blobs}$ \ \\
\hline
0 & 3 & 0 \\
1 & 3 & 0  \\
2 & 3.5 & 0 \\
3 & 3.5 & 1 \\
4 &  3.75 & 0 \\
\hline
\hline
 \end{tabular}
\begin{tabular}[t]{c c c}
\hline
\hline
 $j$ & \  $S_{\rm min}$ \ & $N_{\rm blobs}$ \ \\
\hline
1 & 3 & 0  \\
2 & 3 & 0 \\
3 & 3.25 & 0 \\
4 &  3.25 & 1 \\
5 &  3.25 & 3 \\
6 &  3.5 & 1 \\
7 &  3.5 & 3 \\
\hline
\hline
 \end{tabular}
\begin{tabular}[t]{c c c}
\hline
\hline
 $j$ & \  $S_{\rm min}$ \ & $N_{\rm blobs}$ \ \\
\hline
0 & 3 & 0  \\
1 & 3 & 0  \\
2 & 3 & 0 \\
3 & 3 & 0 \\
4 &  3 & 0 \\
5 &  3 & 0 \\
6 &  3.5 & 0 \\
7 &  3.5 & 0 \\
8 &  3.5 & 0 \\
9 &  3.5 & 0 \\
10 &  3.5 & 1 \\
11 &  3.5 & 1 \\
12 &  3.75 & 1 \\
\hline
\hline
 \end{tabular}
 \begin{center}
 \caption{Sensitivity thresholds $S_{\rm min}$ and the number of significant blobs detected in the end-to-end simulations $N_{\rm blobs}$ for standard needlets with $B=2.5$ (left), standard needlets with $B=1.8$ (center), and Mexican needlets with $B=1.4$ (right). Only blobs that do not intersect the galactic cut are reported. No results are shown for the standard needlets with $B = 1.8$, $j = 0$ as they have no support over the range of angular scales considered.
   \label{tab:thresholds}}
 \end{center}
\end{table*}

\begin{table*}
\begin{tabular}{c c c c c}
\hline
\hline
 feature & \ \ blob & \ \ $B$ \ \ & \ \ $j$ \ \ &  \ \ $S$ \ \ \\
\hline
1 & 1 & 2.5 & 3 & 3.83 \\
1 & 2 & 1.8 & 5 & 3.55 \\
2 & 1 & 1.8 & 4 & 3.99 \\
3 & 1 & 1.8 & 5 & 3.28 \\
4 & 1 & 1.8 & 5 & 3.33 \\
4 & 2 & 1.4 & 10 & 3.77 \\
5 & 1 & 1.8 & 6 & 3.96 \\
6 & 1 & 1.8 & 7 & 4.13 \\
7 & 1 & 1.8 & 7 & 3.97 \\
8 & 1 & 1.8 & 7 & 4.34 \\
9 & 1 & 1.4 & 11 & 3.71 \\
10 & 1 & 1.4 & 12 & 4.14 \\
\hline
\hline
 \end{tabular}
 \begin{center}
 \caption{Blobs found by the needlet transform in the WMAP end-to-end simulations with the 7-year KQ75 mask.}
   \label{tab:endtoendfeatures}
 \end{center}
\end{table*}

\subsubsection{Analysis of bubble collision simulations}

To assess how robustly the needlet transform can pick out a collision in the temperature map, we have performed an analysis of the simulated collisions described in Section~\ref{sec:simulatedmaps}. If the later steps of the analysis are to have a chance at detecting a simulated collision, it must be contained within the set of pixels defined by the blob. To determine if the needlet analysis has detected a bubble collision, we therefore require that a blob exists which fully contains the region affected by the collision, and that the true center of the collision lies within the set of pixels passing the significance threshold.

\begin{figure}[tb]
   \includegraphics[width=6 cm]{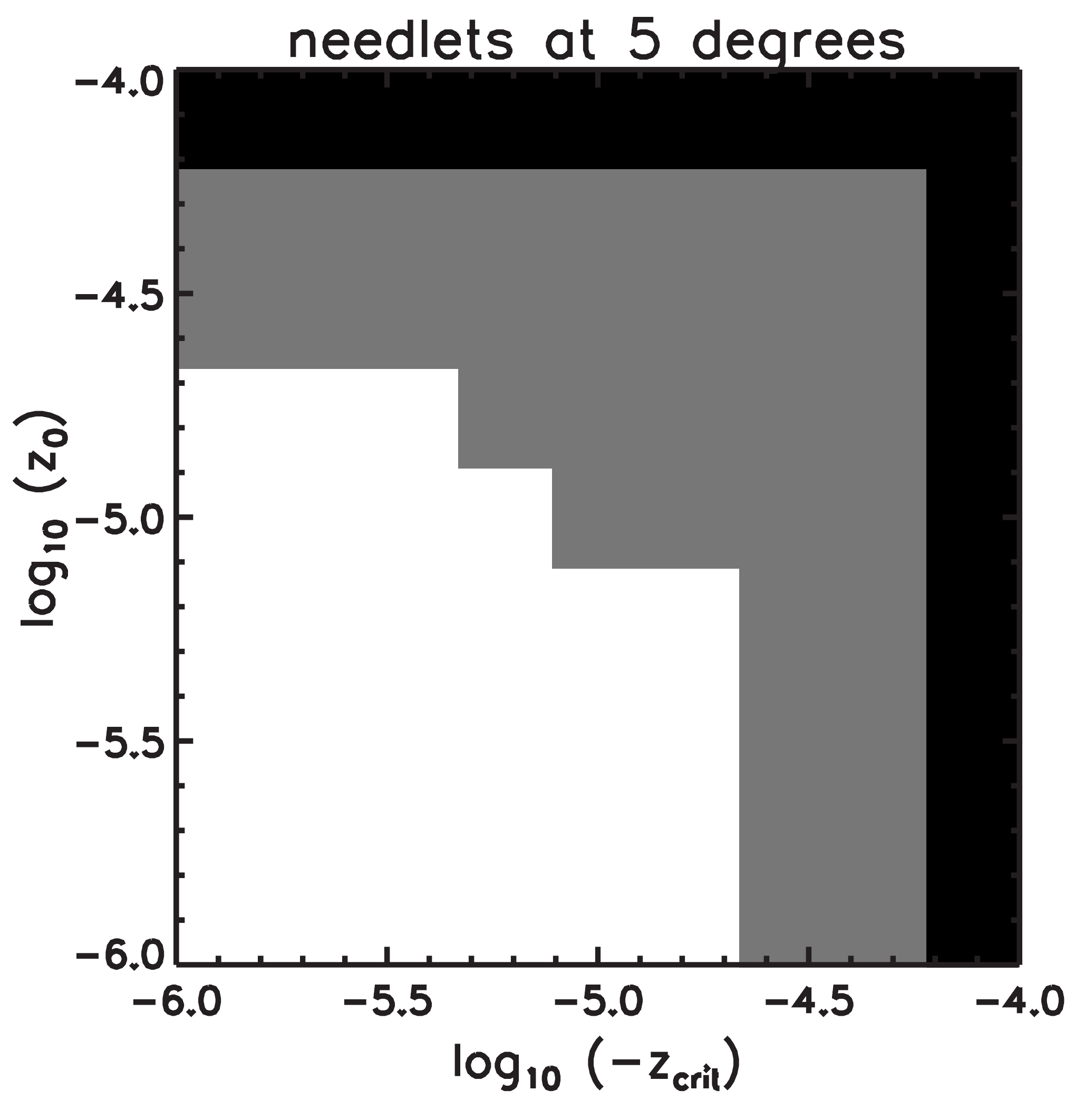}
    \includegraphics[width=6 cm]{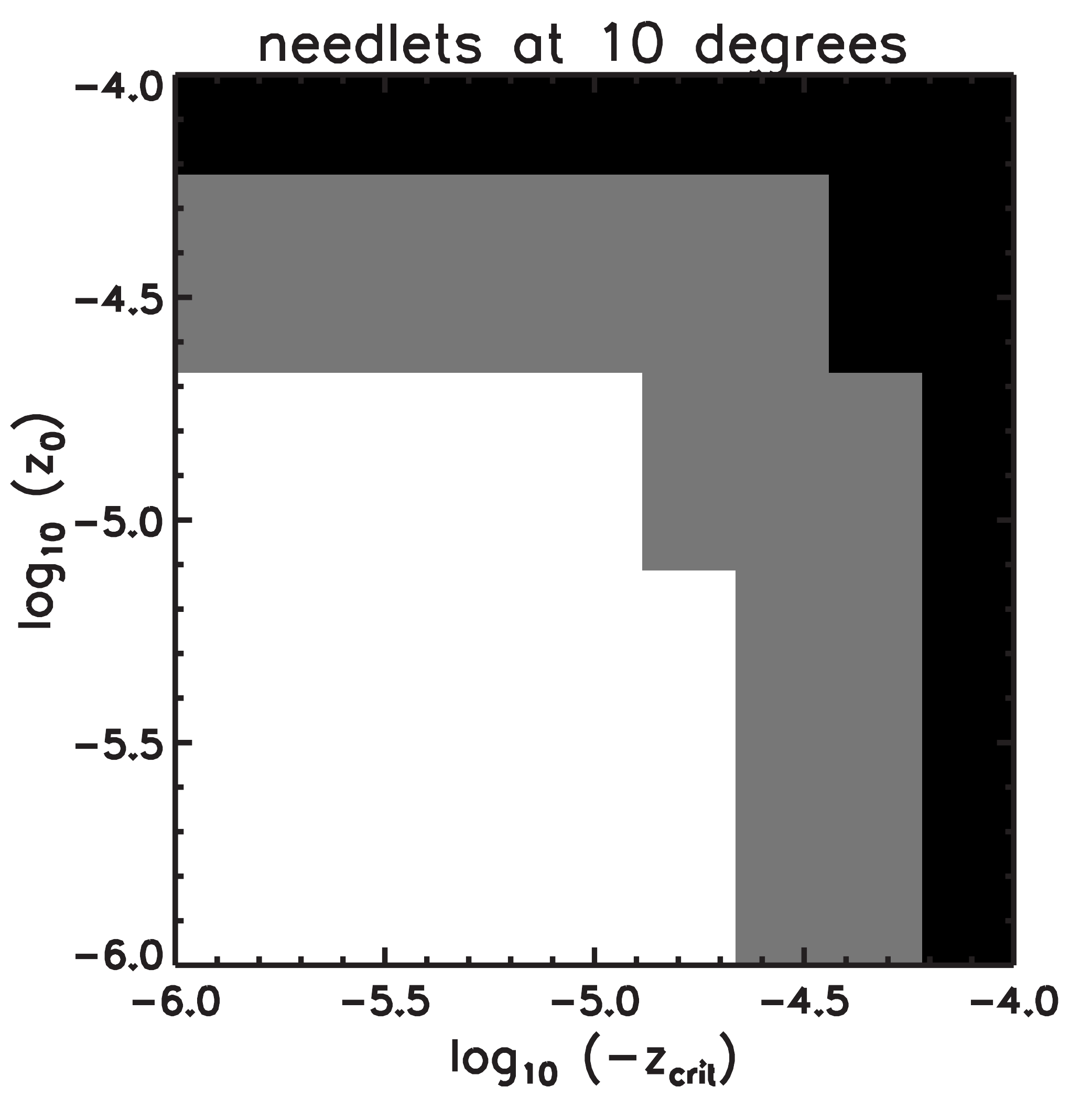}   
\caption{Exclusion (black) and sensitivity (grey) regions for the needlet step of the analysis pipeline applied to a set of $5^{\circ}$ (left) and $10^{\circ}$ (right) simulated bubble collisions. If all realizations at the high and low noise locations yield a detection, we include the collision in the exclusion region; such collisions would certainly be detected as long as they were not significantly masked. If a detection is made in any realization/location, we include the collision in the sensitivity region; such collisions could be found if they were in a favorable location of the sky (i.e., low noise, or a region with a specific realization of background CMB fluctuations).}
\label{fig-needletexclusion}
\end{figure}

The results of this analysis for the $5^{\circ}$ and $10^{\circ}$ collisions are shown in Fig.~\ref{fig-needletexclusion}. We define the ``exclusion region" of these plots as the part of parameter space for which all six realizations/locations yield a detection. If there were a bubble collision in the WMAP 7-year data with these parameters, it would be detected with high significance regardless of its location on the sky (as long as the collision was not significantly masked). The ``sensitivity region" is defined as the part of parameter space for which {\em any} of the six realizations/locations yields a detection. A bubble collision in this range of parameter space would be detected only for a favorable location or realization of the background fluctuations. The exclusion and sensitivity regions for the $25^{\circ}$ collisions are identical to those for the $10^{\circ}$ collisions. 

Looking at the simulations in detail, there are a few general trends. First, for the needlets to pick out a collision, it is sufficient to have either a relatively large central amplitude $z_0$ or a relatively large temperature discontinuity $z_{\rm crit}$ at the causal edge. This is clear from the shape of the exclusion region in Fig.~\ref{fig-needletexclusion}. From the size of the sensitivity region in this plot, one can also see that the amount of instrumental noise and particular realization of the background fluctuations can greatly affect the significance of the needlet coefficients in the vicinity of a collision. Many more collisions were detected in the low-noise region than the high noise region of the sky. For collisions in the exclusion region, there is a significant needlet response for all three needlet types over a range of frequencies, with an average significance exceeding $S>5$. Collisions in the sensitivity region are typically detected by only one needlet type and frequency, with an average significance near $S\simeq 4$. As our ability to detect $5^{\circ}$, $10^{\circ}$, and $25^{\circ}$ collisions is nearly the same, we conclude that these results are fairly representative of our detection limits over all angular scales $\gtrsim 5^{\circ}$.

These general trends can be contrasted with the response to features found in the end-to-end simulations. Here, blobs are typically detected with a single needlet type and are near the significance threshold (not surprisingly, as the threshold was chosen to have this property). A feature detected in the data by multiple needlet types and/or frequencies at a significance of $S \geq 4$ would be a good bubble collision candidate. However, we stress that many different underlying features could give rise to such a signal. The following steps in the analysis pipeline, which we now describe, are designed to verify if these candidates are in fact bubble collisions. 

\subsection{Edge detection}

The first of the two parallel verification steps of the pipeline tests whether features highlighted by the blob detection stage have circular temperature discontinuities. The unambiguous detection of a circular temperature discontinuity would strongly suggest that a particular feature highlighted by the needlets is in fact a bubble collision. We employ the Canny edge detection algorithm~\cite{11275}, whereby the gradient of an image is generated and thinned into single-pixel proto-edges, the best of which are stitched together into ``true'' edges. We also use an adaptation of the Circular Hough Transform (CHT) algorithm~\cite{360677}, which assigns a ``score" dependent on how many edge pixels lie on circles of varying centers and radii. In this section, we describe our edge detection algorithm, and study its performance on an end-to-end simulation of the experiment, as well as on simulated bubble collisions.

\subsubsection{The Canny edge detection algorithm}

\begin{figure*}[tb]
   \includegraphics[width=16 cm]{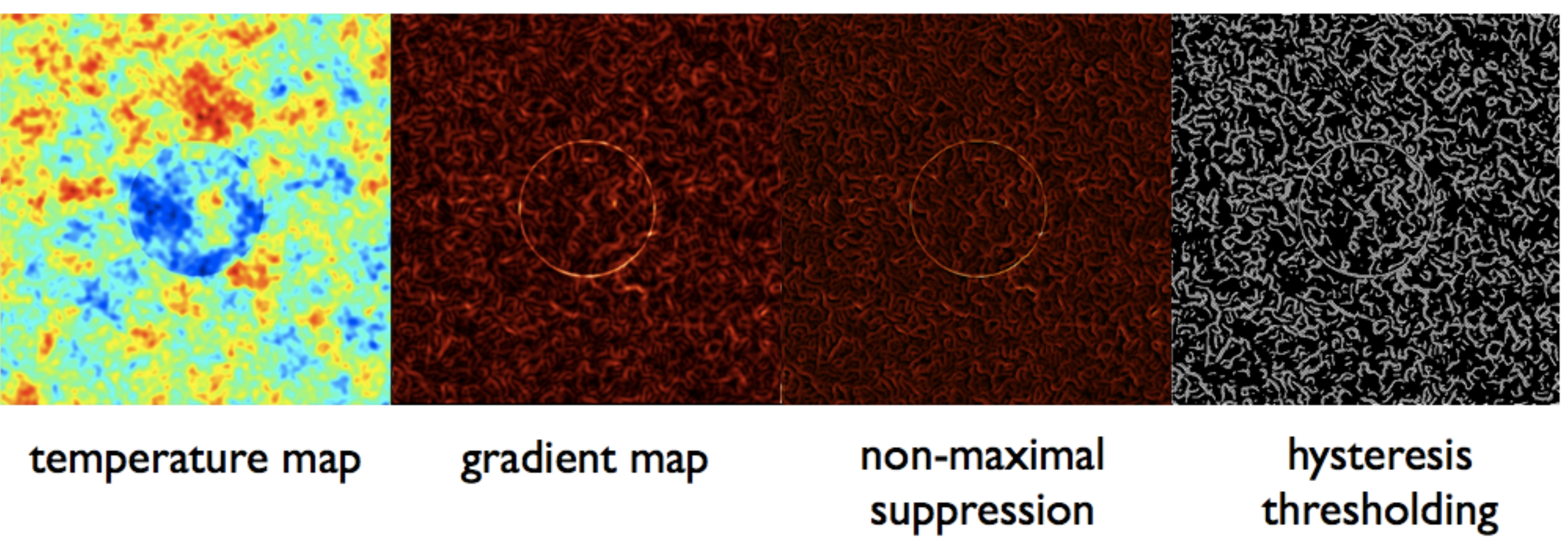}
\caption{An illustration of the Canny algorithm for edge detection. The input temperature map contains a circular discontinuity, which can be seen in a  map of the gradient magnitude as a local maximum. Non-maximal suppression selects for local maxima in the gradient map. The hysteresis thresholding step finds stitched edges by comparing the local direction of the gradient at adjacent pixels.}
\label{fig-cannyfig}
\end{figure*}

The Canny edge detector is {\em the} standard edge detection algorithm in image-processing software, and has recently been used to search for cosmic strings \cite{Danos:2008fq,Amsel:2007ki}. Designed as the optimal algorithm for localized, duplicate-free detection of edges within a noisy image, it uses three steps -- smoothed gradient generation, non-maximal suppression and hysteresis thresholding -- to extract contiguous edge sections. In Fig.~\ref{fig-cannyfig}, we depict each of these three steps as applied to a temperature map containing a circular discontinuity; each of these steps are in turn described below.
\begin{enumerate}

\item {\bf Smoothed gradient generation:} The gradient of a Cartesian image is traditionally generated by convolving the image with two small symmetric filters, each determining one orthogonal component of the gradient. A number of simple filters -- typically $3 \times 3$ pixels -- perform the job adequately, but the optimally adaptable filters are the first partial derivatives of the two-dimensional Gaussian~\cite{11275}. Using these filters is equivalent to first convolving the image with a small Gaussian filter (and thus smoothing out the effects of small-scale noise on the gradient calculation, an important step given the small number of pixels involved in the calculation) and then finding its gradient components.

Unfortunately it is impossible to construct symmetric pixel-based gradient filters that cover the whole sphere. We therefore carry out both the Gaussian smoothing and gradient generation steps in harmonic space, making use of HEALPix's in-built {\tt alm2map\_der} subroutine to calculate the magnitude and direction of the gradient at each pixel. We smooth with a Gaussian filter of FWHM $0.22^{\circ}$ -- approximately two pixels' width at the resolution of our input maps -- to minimize features due to pixel noise while retaining longer edges.

The gradient maps are generated before masking to reduce ``ringing'' from the sharp sky cut back into the map. The smoothing step ensures that any leakage from masked features is restricted to areas a few pixels within the sky cut, and affects only areas a few pixels outside of the cut. Nevertheless, any features found close to the mask should be carefully examined to check if they are generated from within the mask.

\item {\bf Non-maximal suppression:} The second step of the Canny algorithm reduces the smoothed gradients produced by the first step into local maxima. At this stage, all pixels are assumed to belong to a local edge, whose direction is defined to be perpendicular to the local gradient direction. Taking each pixel at a time, the two direct neighbors which lie closest to the local edge are found. The gradient magnitudes of the three pixels are compared, and the central pixel's gradient magnitude is set to zero unless it is the largest of the three. Processing each pixel in turn reduces the gradient magnitude map to only the local maxima (see the central panels of Fig.~\ref{fig-cannyfig}).

As an example, consider the simplest case of crossing a sharp discontinuity along a perpendicular path. The gradient direction is constant at each pixel, whereas the smoothed gradient magnitudes increase until the edge is crossed, when they start to decrease. A non-maximal suppression algorithm tracks along the path setting all of the gradient magnitudes to zero apart from the pixel on (or closest to) the edge.

\item {\bf Hysteresis thresholding:} At this stage of the Canny algorithm, we have gradient magnitudes and directions for a set of local maxima of varying amplitude, some corresponding to true edges (which may have been disrupted by noise) and others to runs of noisy pixels or to more slowly-varying boundaries of CMB patches. To filter out true edges from the noise, and stitch together any edges that have been split, the final step of the algorithm takes advantage of the fact that, unlike randomly-oriented noise, true edges conserve their gradient magnitude and direction (to an extent affected by the shape of the edge, the pixelization scheme, and the noise level) over their path.

Hysteresis thresholding involves first setting an upper threshold for the gradient magnitude: any pixels surpassing this threshold are considered to be part of true edges, and a new ``true edge'' map is created with these pixels' positions marked. A second, lower threshold is then set. Hysteresis thresholding then proceeds as follows:

\begin{enumerate}
\item The map is scanned until a true edge pixel is found.
\item The next potential edge pixel is defined to be the direct neighbor closest to $90^{\circ}$ clockwise from the local gradient direction. The gradient direction of this pixel is compared to the current pixel's. To do so, the local phase angles to the current pixel's nearest neighbors are found, and used to define the neighbor closest to the current gradient direction. The neighbors adjacent to this pixel are then determined. The gradient direction at the next potential edge pixel is required to lie between the phase angles of these neighbors. This rather loose requirement allows the algorithm to step along pixelated curved edges.
\begin{enumerate}
\item If the two pixels' gradient directions match within the tolerance,
\begin{enumerate}
\item If the neighbor's gradient magnitude passes the low threshold but not the high, it is considered to be part of a potential true edge. Its position is marked in a history array; then the algorithm ``steps onto'' this pixel and the process is repeated from step (b) until one of the other conditions is met.
\item If the neighbor's gradient magnitude passes the high threshold, all pixels found on the way from the source pixel are confirmed as a true edge. Their positions are marked on the true edge map, and the algorithm returns to step (a).
\item If the neighbor's gradient magnitude fails both thresholds, the edge is considered to be 
false: the history of potential edge pixels found on the way from the source pixel is erased, and the 
algorithm returns to step (a).
\end{enumerate}
\item If the two pixels' gradient directions do not match within the required tolerance, 
\begin{enumerate}
\item If the neighbor's gradient magnitude passes the high threshold or the neighbor is already marked in the history array, all pixels found on the way from the source pixel are confirmed as a true edge. Their positions are marked on the true edge map, and the algorithm returns to step (a). This ensures simple branched and looped edges can be reconstructed.
\item If the neighbor's magnitude does not pass the high threshold and the pixel is not already marked in the history array, the edge is considered to be false, the history of potential edge pixels found on the way from the source pixel is erased, and the algorithm returns to step (a).
\end{enumerate}
\end{enumerate}
\end{enumerate}

The entire process is then repeated, choosing the neighbor closest to $90^{\circ}$ {\em counter-clockwise} to the local gradient direction in step (b).

\end{enumerate}

\noindent The end product of the Canny algorithm is a Boolean map of stitched true edges (see right-hand panel of Fig.~\ref{fig-cannyfig}). To reduce computation time, the pipeline from hysteresis thresholding onwards is restricted to the blobs produced by extending the regions passing the needlet significance test to ensure any discontinuity is fully contained.

Care must be taken when setting the thresholds used in the hysteresis thresholding step. If either threshold is set too high, very few edges are confirmed. If the low threshold is set too low, a huge amount of potential edges are considered, and the algorithm runs extremely slowly. As the edges we could potentially find must be comparable in amplitude to the CMB signal and detector noise (as they have not yet been discovered by eye) and have been smoothed by the WMAP beam, we set low thresholds to err on the side of caution. Low and high thresholds of $30\%$ and $40\%$ of the maximum gradient magnitude found in each search region are found empirically to confirm edges generated in simulations in feasible computation timescales. This means that the gradients associated with the strongest CMB features -- typically $\sim 1^{\circ}$ in scale -- are classified as edges, as is shown in Fig.~\ref{fig-cannyfig}.

\subsubsection{Circular Hough Transform}

\begin{figure*}[tb]
   \includegraphics[width=13 cm]{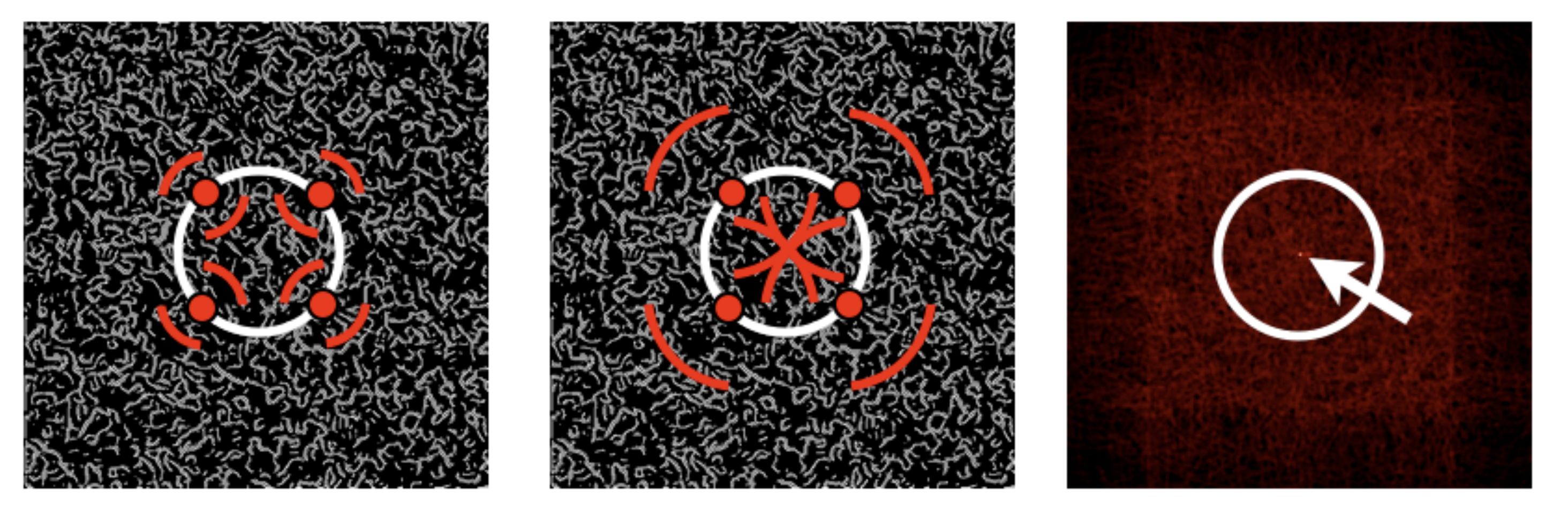}
\caption{A depiction of the Circular Hough Transform (CHT). On the left is a Boolean map of edge pixels, as output by the Canny algorithm. Centering a pair of arcs oriented in the direction of the local gradient about each edge pixel, the CHT counts the number of times each pixel is intersected. The presence of a circular edge is indicated by a maximum in the CHT score -- the hit count divided by the arc radius -- as the arc radii are varied. On the left, we show a set of arcs, centered on four pixels on the circular edge we wish to detect; there will be no clear peak in the CHT score for this radius. Increasing the arc radius to match that of the circular edge (center), there will be a large number of hits at the true center of the edge. On the right, we show the actual map of the CHT score at each pixel for this example. As this has been scanned at the correct angular scale, there is a large peak at the center of the circular edge.}
\label{fig-CHTfig}
\end{figure*}

The maps of stitched candidate edges found using the Canny algorithm are processed using the Circular Hough Transform to search for the presence of circular edges. The basic idea, as shown in Fig.~\ref{fig-CHTfig},  is to count the number of intersections between circular arcs of varying radii centered on each of the candidate edge pixels and oriented along the local gradient direction. If there is a circular edge in the map, the number of intersections will be maximized at the center of the circular edge when the radius of the circular arc matches that of the edge. 

Assuming an edge pixel forms part of a circular edge of angular radius $\theta_{\rm crit}$, one can define a prescription for the set of pixels that are the potential centers of the edge. The two most likely candidates in this set are the pixel $\theta_{\rm crit}$ away in the direction of the local gradient, and the pixel the same distance in the opposite direction; the edge could be a step up or a step down. Building in flexibility to cope with pixelation and noise effects, this set is expanded to two annular arcs of radius $\theta_{\rm crit}$, oriented in the direction of the local gradient and centred on the edge pixel.

The CHT works by assuming that {\em all} edge pixels are part of circular edges. The most likely circle center at a given radius is defined to be the pixel that is included in the greatest number of these arcs, counted using an accumulator array. If the search radius matches the radius of a circular edge within the map, we expect all of the circular edge pixels' arcs to include the true center, and the CHT accumulator will show a single clear peak. If the search and true radii are discrepant, fewer of the circular edge pixels' arcs will intersect, and this peak will appear as a ring with decreased amplitude (see Fig.~\ref{fig-CHTfig}). When the search and true radii are very discrepant, any rings will disappear beneath the background due to randomly-oriented noise and other non-circular edges. Note that ``non-circular edges'' will include the $\sim 1^{\circ}$ CMB fluctuations that are qualified as edges by the hysteresis thresholding step.

To compare the CHT results at different search radii, one must divide out the approximately linear growth with angular radius of the number of pixels in each annular arc. We call this normalized quantity the ``CHT score''. The most likely center and radius of a circular edge within a map can therefore be found by scanning the map with the CHT at a range of radii and determining the maximum CHT score.

The blob detection step provides the range of scales $\theta_{\rm crit, min} \le \theta_{{\rm crit}, i} \le \theta_{\rm crit, max}$ of potential circular edges present in each blob. To determine whether a blob contains a circular edge, we compare the CHT scores obtained by scanning at every $0.25^{\circ}$ increment within this range, using annular arcs that are $0.25^{\circ}$ thick and which cover $45^{\circ}$ of phase about each edge pixel. The annular arcs are therefore approximately two pixels thick, and are fairly wide to account for the effects of pixelation on the gradient direction. The thickness of the CHT annular arcs leads to an uncertainty in the CHT radius of $0.25^{\circ}$ and position of $0.50^{\circ}$. If a circular edge is detected, we expect a clear peak in the CHT results for a particular blob.

\begin{figure}[tb]
   \includegraphics[width=14 cm]{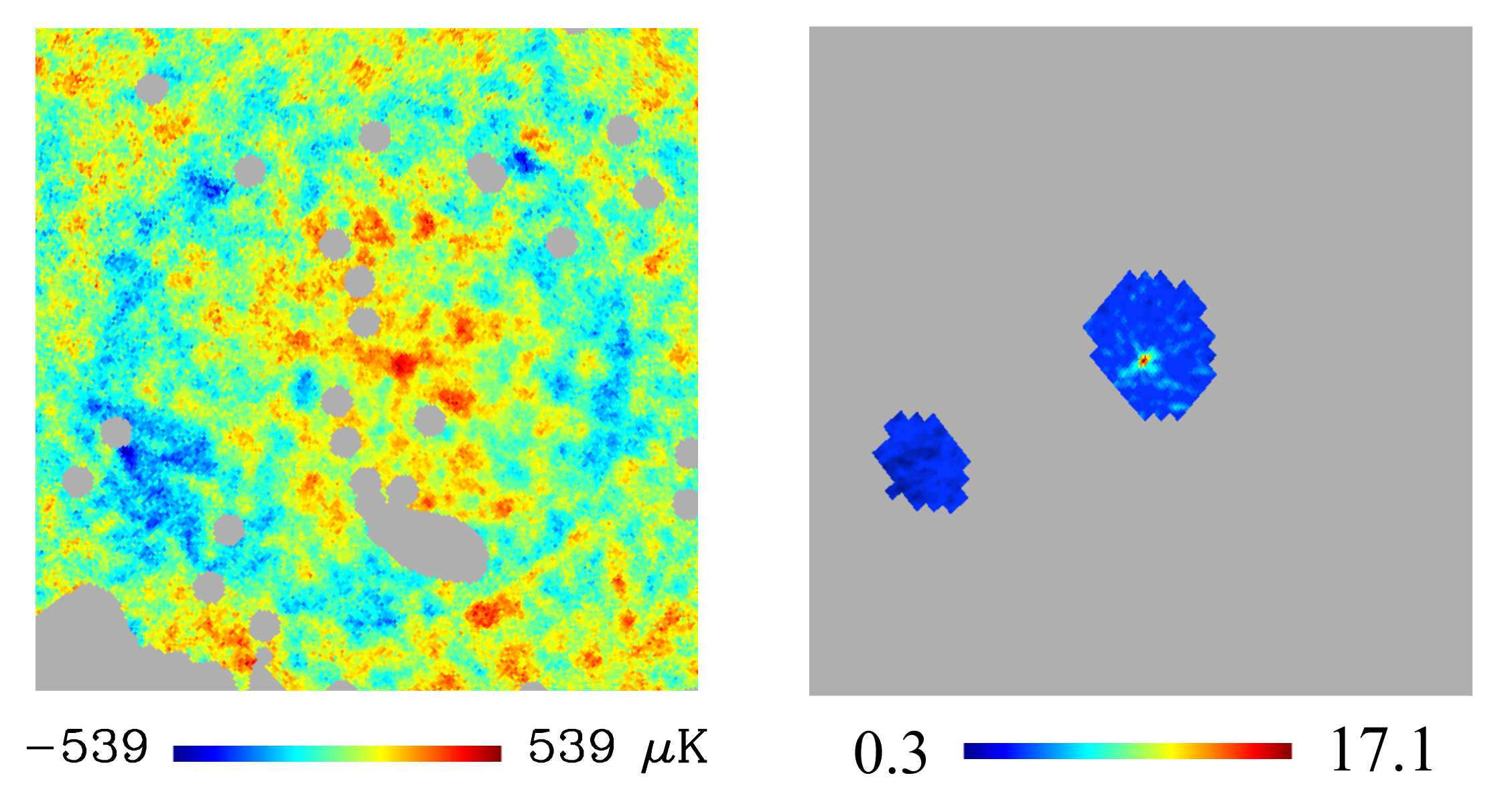}
\caption{The temperature map (left) and CHT score (right) for our illustrative example of a $10^{\circ}$ bubble collision simulation. The CHT score is recorded at each pixel passing the needlet significance threshold. For a search radius of $10^{\circ}$ there is a clear peak in the CHT score at the center of the simulated collision.}
\label{fig-bubble_and_CHT}
\end{figure}

\begin{figure}[tb]
   \includegraphics[width=9 cm]{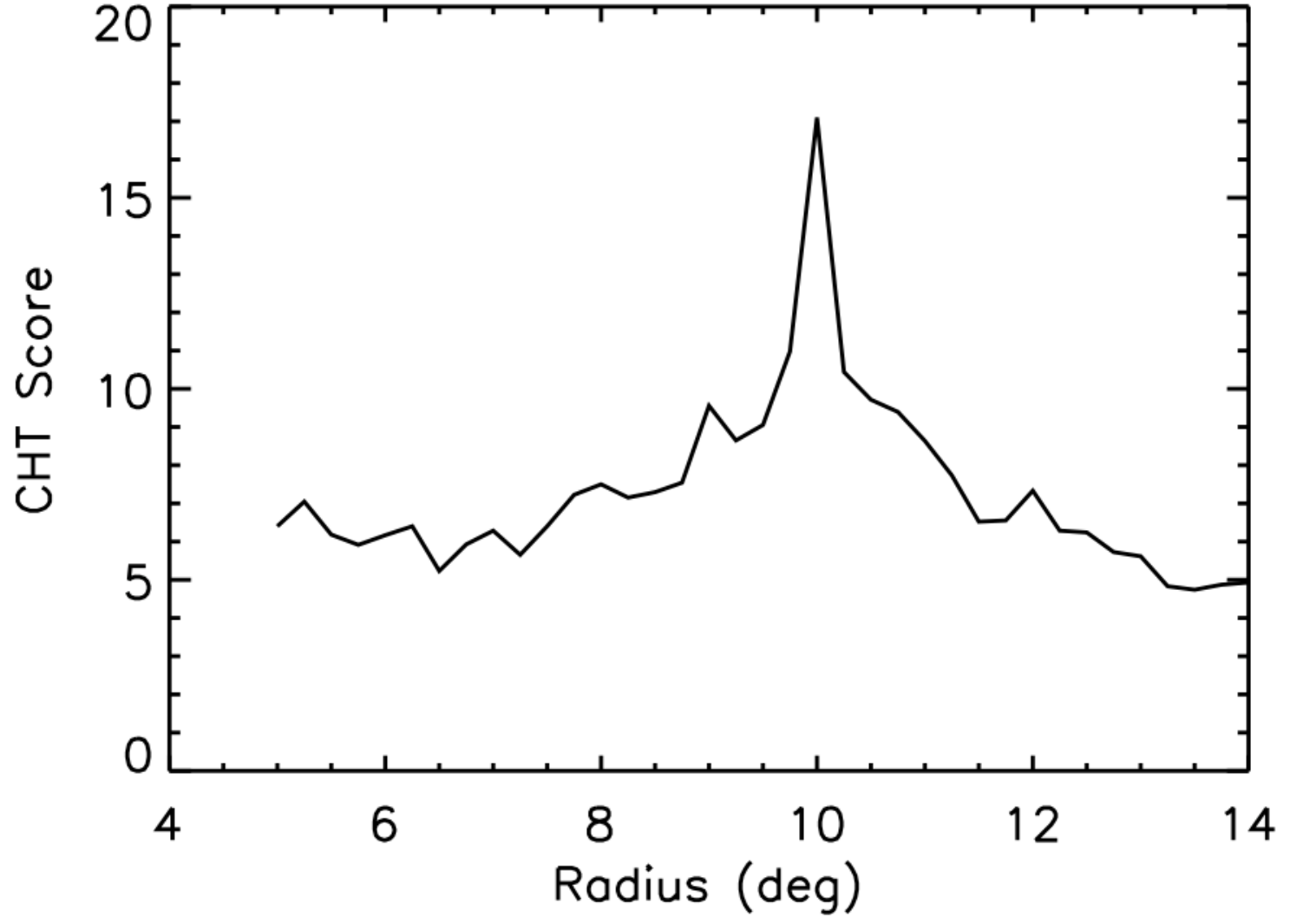}
\caption{The global maximum of the CHT score at each circle radius for the collision simulation shown in Fig.~\ref{fig-bubble_and_CHT}. The collision has a maximum response for standard needlets with $B=2.5$ at $j=3$, which from Table~\ref{tab:anglelookup} sets the search range to be $5^{\circ} \leq \theta_{\rm crit} \leq 14^{\circ}$. The peak of the CHT score at $10^{\circ}$ clearly identifies the correct angular scale of the simulated collision.}
\label{fig-needlet_score}
\end{figure}

In Fig.~\ref{fig-bubble_and_CHT} we show the output of the edge detection algorithm on our illustrative example bubble collision simulation (see Fig.~\ref{fig-needletbubble}). On the left is the portion of the temperature map containing the collision. On the right we plot the CHT score in the pixels that passed the needlet significance threshold for $\theta_{{\rm crit}, i} = 10^{\circ}$ (the input value). There is a clear peak at the location of the true center of the simulated bubble collision, which is $\sim 3$ times the average response at other pixels. Since this feature was flagged in the blob detection step for standard needlets with $B=2.5$ at $j=3$, the range of radii scanned during the CHT step is determined from Table~\ref{tab:anglelookup} to be $5^{\circ} \leq \theta_{\rm crit} \leq 14^{\circ}$. This range contains the true radius $\theta_{\rm crit} = 10^{\circ}$. In Fig.~\ref{fig-needlet_score} we plot the maximum CHT score found in the map for each circular radius, which contains a clear peak at the true radius of the causal boundary. This signal is a clear and unambiguous signature of a bubble collision. From visual inspection of the temperature map, it can be seen that we are able to clearly detect the edge even though the background fluctuations, noise and beam drastically reduce the sharpness of the observed temperature discontinuity.

\subsubsection{Analysis of the WMAP end-to-end simulation}
We expect strong circular edges to be extremely rare in a purely Gaussian CMB temperature map. However, it is possible that foregrounds, instrumental noise, the mask, and other experimental artifacts could lead to a spurious detection of a circular edge. To evaluate this, we have performed the edge detection step of our analysis pipeline on the features that passed the significance threshold in the WMAP end-to-end simulation (see Table~\ref{tab:endtoendfeatures}) with the KQ75 mask applied.

\begin{figure}[tb]
   \includegraphics[width=7 cm]{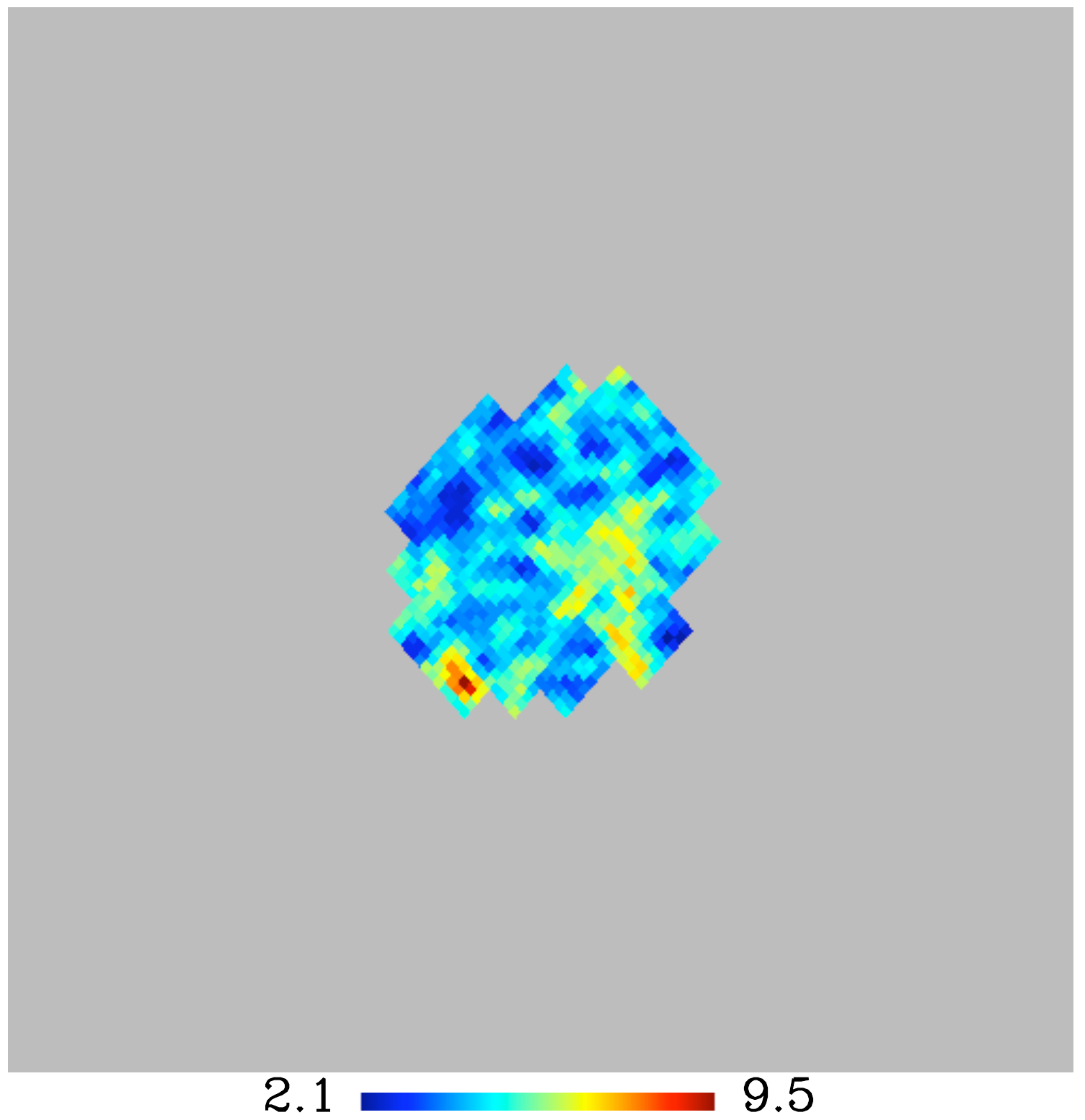}
   \includegraphics[width=9 cm]{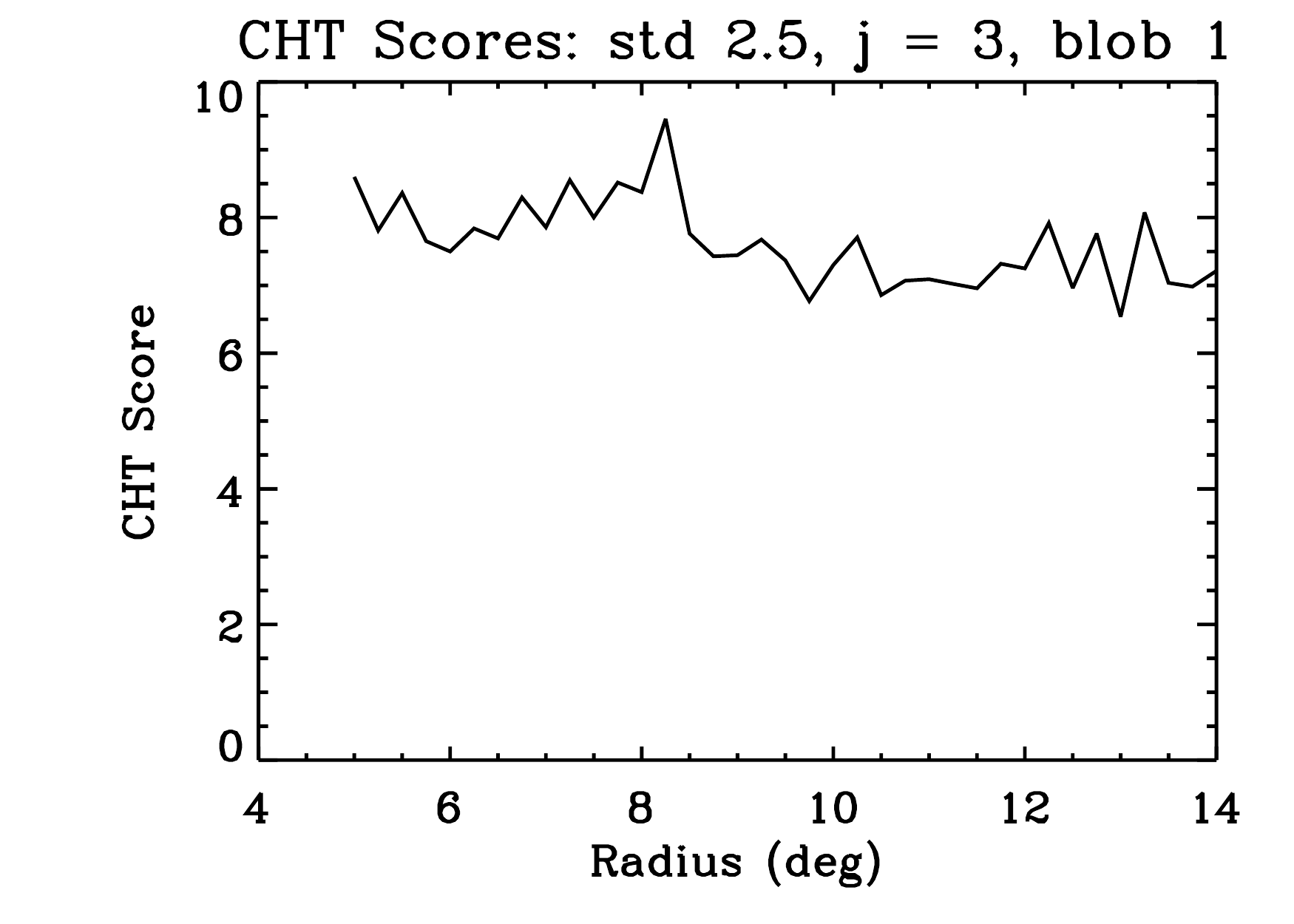}
\caption{The most edge-like feature in the WMAP end-to-end simulation. The contrast in scores as a function of position (left) and radius (right) is greatly reduced compared to the collision example (Figs.~\ref{fig-bubble_and_CHT}~and~\ref{fig-needlet_score}).}
\label{fig-WMAPendtoend}
\end{figure}

Comparing each feature in the end-to-end simulation with the bubble collision example studied above, the peak structure of the CHT score as a function of angle and morphology in pixel space are both drastically different. Examining the maximum CHT score as a function of circular radius, although there are several peaks, the clearest of which is shown in Fig.~\ref{fig-WMAPendtoend}, their amplitude relative to the average score is nowhere near that of the collision example shown in Fig.~\ref{fig-needlet_score}. In addition, from the plots of the CHT score at each pixel, there are typically a number of fairly broad local maxima at different locations with approximately the same score. This is in contrast to the collision example of Fig.~\ref{fig-bubble_and_CHT}, which yields a highly peaked score around a small number of pixels. 

\subsubsection{Analysis of bubble collision simulations}

\begin{figure}[tb]
   \includegraphics[width=6 cm]{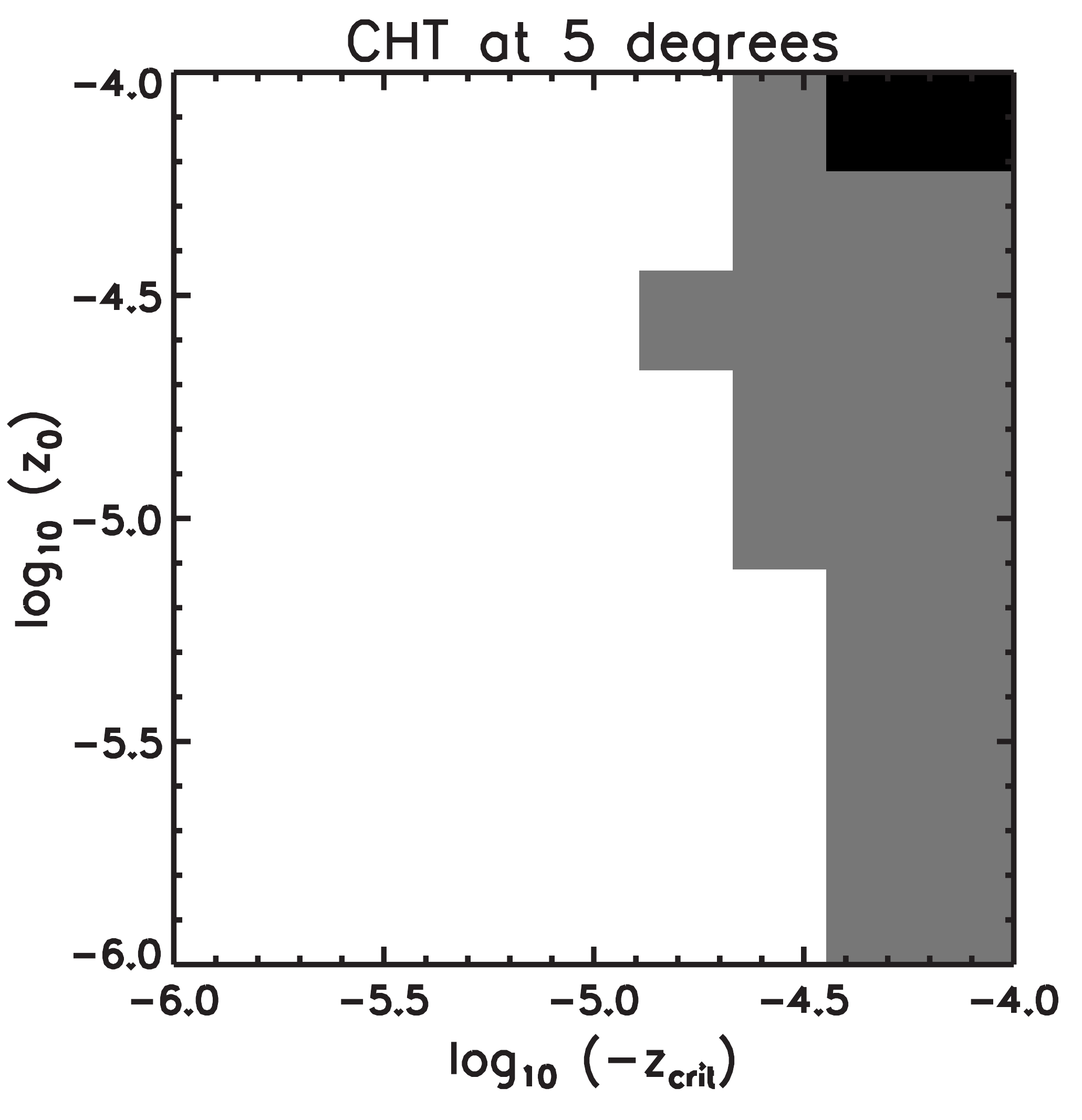}
    \includegraphics[width=6 cm]{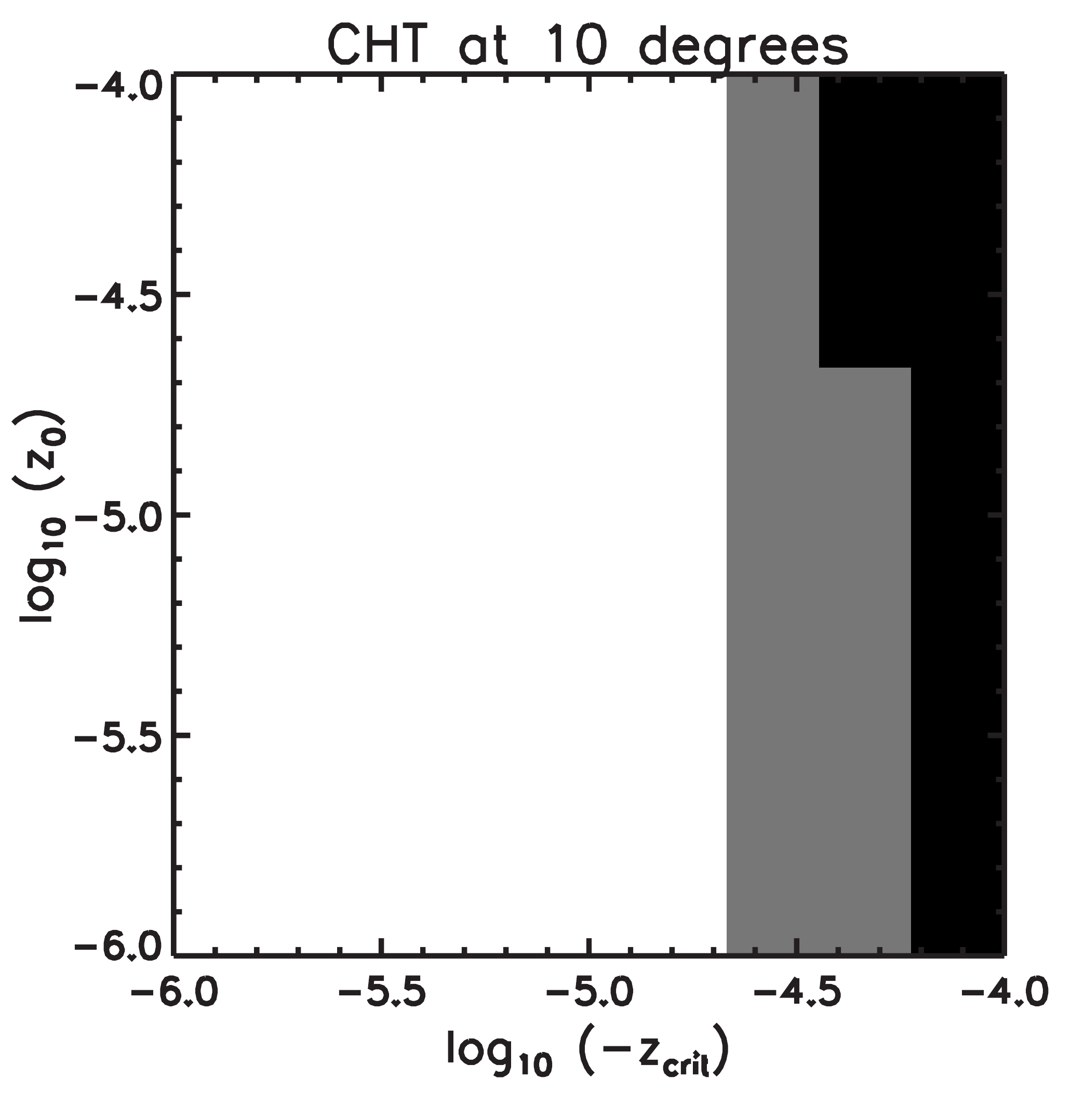}   
\caption{Exclusion (black) and sensitivity (grey) regions (see Fig.~\ref{fig-needletexclusion}) for the edge detection step of the analysis pipeline applied to a set of $5^{\circ}$ (left) and $10^{\circ}$ (right) simulated bubble collisions.}
\label{fig-chtexclusion}
\end{figure}

To better understand the response of our edge detection algorithm to the signal from a bubble collision in WMAP-quality data, we have analyzed the simulations described in Section~\ref{sec:simulatedmaps}. We use as inputs the blobs found using the first step of the pipeline, and search for circular edges over the range of angular scales appropriate to the needlet type and frequency for each blob (see Table~\ref{tab:anglelookup}). We conclude that a true causal edge has been detected if there is a global maximum for the CHT score at the radius of the true edge and the pixel with the highest score is within a typical CHT error ($0.5^{\circ}$) of the actual center. We again present our results in the form of a contour plot denoting exclusion and sensitivity regions in the parameter space of $z_{0}$ and $z_{\rm crit}$. This is shown in Fig.~\ref{fig-chtexclusion} for simulated bubbles with $\theta_{\rm crit} = 5^{\circ}$ and $10^{\circ}$. The plot for the $25^{\circ}$ collisions is identical to the plot for $10^{\circ}$ collisions.

Again, from the size of the sensitivity region, we conclude that our ability to make a detection is dependent on the location of the collision and the particular realization of the background CMB fluctuations. The exclusion region for the $10^{\circ}$ (and $25^{\circ}$) collisions is far larger than for the $5^{\circ}$ collisions. We attribute this to the proliferation of $\sim1^{\circ}$-sized features in the background fluctuations, which can both disrupt a significant fraction of the edge pixels in a small collision and swamp the collision signal with their own strong gradients. We therefore expect our sensitivity to edges at small angular scales $\theta_{\rm crit} \lesssim 5^{\circ}$ to be quite poor at WMAP resolution. As the performance of the edge detection algorithm for $10^{\circ}$ and $25^{\circ}$ collisions is identical, we conclude that the $10^{\circ}$ results are fairly representative of our sensitivities over a wide range of angular scales $\theta_{\rm crit} \gtrsim 10^{\circ}$. Most of the collisions we mark as a detection have a clear peak in the CHT score of the type seen in Fig.~\ref{fig-needlet_score}. If a collision has parameters in the exclusion region, it would be reliably detected.  Based on these results, the first two steps of our pipeline can detect bubble collisions with central modulations $|z_0| \agt 3 \times 10^{-5}$ and causal edges $|z_{\rm crit}| \agt  3 \times 10^{-5}$ at $\theta_{\rm crit} \agt 5^{\circ}$ in WMAP-quality data.

\subsection{Parameter estimation and model selection} \label{sec:bayes}

In many CMB anomaly analyses (but not all -- see, e.g., Refs.~\cite{Groeneboom:2009cb,Hoftuft:2009rq,2008MNRAS.390..913C}), the significance of a signal is quantified by calculating the frequentist {\em $P$-value} of some relevant statistic. This typically involves doing a large number of Monte Carlo realizations of the standard cosmological model (i.e., the ``null hypothesis''), calculating the above statistic for each, and finding the fraction for which the statistic has a ``more extreme'' value than was actually observed.  There are several problems with this approach. First, the calculated $P$-value is only a measure of how (un)likely the measured data were given the null hypothesis of the standard model; no comparison is made to an alternative model (which is what we are primarily interested in here). Second, the notion of ``more extreme'' is fundamentally ambiguous -- both ``more discrepant'' (i.e., values of the statistic which are further from some fiducial expected value than that which was measured) or ``less likely'' are common choices\footnote{For simple, single-peaked, likelihoods these two definitions are at least equivalent, but in some cases (e.g., a likelihood that is constant over some finite range) neither definition is satisfactory.}.  The heart of the problem is that all such $P$-values are integrals over the likelihood, whereas it is only the likelihood of the actual data that is relevant.  The fact that the likelihood and its integral generally have a similar qualitative dependence in the tail(s) of the distribution (i.e., both tend to zero for extreme values of the statistic) can mask this problem.  In particular, if the tails of the likelihood are Gaussian then the integral that gives the $P$-value falls off more rapidly than the likelihood itself, and so the resultant $P$-values are unreasonably harsh on the null hypothesis.  A related problem is that that many attempts to identify CMB anomalies using frequentist $P$-values are overly sensitive to {\em a posteriori} selection effects (see, e.g., Refs.~\cite{Bennett:2010jb,Pontzen:2010yw} for a discussion of this effect). Here the issue -- that the statistics being applied to the data are often chosen on the basis of interesting features initially identified in the same data -- is not intrinsic to frequentist methods (which, correctly, do not permit any data to be used more than once); but the need to invent a statistic from which to calculate a $P$-value can make it hard to avoid this trap.  For these reasons we do not use $P$-values in our analysis.

Instead, we adopt a Bayesian approach. Bayes' theorem provides a prescription for parameter estimation. In addition, given that we have two well-defined hypotheses, we can utilize Bayesian model comparison to make probabilistic statements about the degree to which the available data (and theoretical prior information) imply that bubble collisions have been observed. As shown by Ref.~\cite{Cox:1946}, Bayesian methods are the only self-consistent framework for such calculations. The optimal Bayesian calculation would be to evaluate the likelihood of the entire WMAP data-set under the two models; however this not computationally feasible at present. In Appendix~\ref{section:method}, we outline a set of simplifications that allow us to approximate the optimal Bayesian result. As outlined in Sec.~\ref{sec:summaryofpipeline}, we utilize the information on the location and scales of the most probable bubble collision sites obtained in the blob detection step of the pipeline to implement this procedure. Even this reduced problem is computationally demanding: analysis of the blobs detected in the WMAP 7-year data during the first steps of the pipeline requires three days' processing on 28 cores. Working at full resolution is necessary to ensure that any possible circular temperature discontinuities are examined.

These computational limitations also mean we are only able to process a limited number of simulated temperature maps with and without bubble collisions. The WMAP end-to-end simulation provides a great asset at this stage, giving the best possible measure of what false detections are to be expected from experimental effects and any systematic errors that are not included in our likelihood. We also analyze a small number of representative bubble collision simulations to obtain an estimate of the strength of signal we are looking for. 

We now describe our methods and the results from simulations in greater detail.

\subsubsection{Bayesian formalism}

A model of eternal inflation predicts the average number of collisions $\nsavge$ that are, in principle, detectable by our pipeline on the full sky~\footnote{The number of detectable sources $\nsavge$ is a subset the total number of sources on the sky $\bar{N}$ (Eq.~\ref{eq:collnum}).}. The ultimate goal of our Bayesian analysis is to evaluate the full posterior probability distribution for $\nsavge$, given a CMB data set ${\bf d}$ covering a sky fraction $\fsky$. Using Bayes' theorem, this can be written as
\begin{equation}
\label{eq:posterior_bayes}
\prob(\nsavge | \data, \fsky) 
  = \frac{\prob(\nsavge) 
    \, \prob(\data | \nsavge, \fsky)}{\prob(\data | \fsky)}.
\end{equation}
The form of the posterior depends on the model prior $\prob(\nsavge)$ and the {\em evidence} (also known as the model likelihood) $\prob(\data | \nsavge, \fsky)$. The evidence is defined by marginalizing the likelihood, $\prob(\data | \model, \nsavge, \fsky)$, over the $n$ parameters describing a collision, as specified by the model $\model$. Once the shape of the posterior has been determined, it is normalized using $\prob(\data | \fsky)$. The posterior leads directly to constraints on the values of $\nsavge$ consistent with a CMB data set. 

In a landscape scenario, $\nsavge$ can be considered as a continuous parameter  and the prior $\prob(\nsavge)$ would be determined from a measure over the possible values of $\nsavge$.  We can also view $\nsavge$ as a proxy for different models of eternal inflation (i.e., selecting a single value of $\nsavge$), as described further in Sec.~\ref{sec:collisionintro}. The standard cosmological model without bubble collisions is specified by the case $\nsavge = 0$. Using Eq.~\ref{eq:posterior_bayes}, the probability of a model which predicts $\nsavge$ collisions (on average) relative to that of the standard cosmological model is 
\begin{equation}
\label{eq:relative_prob}
\frac{\prob(\nsavge | \data, \fsky)}{\prob(0 | \data, \fsky)} = \frac{\prob(\nsavge) \prob(\data | \nsavge, \fsky) }{ \prob(0) \prob(\data | 0, \fsky) }.
\end{equation}
The model priors and the evidence values play an equal role in this relationship, but in the absence of a detailed understanding of the former, it is often useful to proceed under the assumption that the two models are equally probable {\em a priori}. A theory predicting an expected $\nsavge$ collisions is favoured over the standard model when the relative probability on the LHS of Eq.~\ref{eq:relative_prob} is greater than unity.

It is also useful to provide heuristic conversions between the Bayesian evidence ratio and other commonly used model comparison quantities. The number of ``sigma'' of an anomaly statistic, $N \sigma$, is often used to characterize the deviation from a null model, but it is unambiguously defined only in the case that the null distribution of the chosen statistic is Gaussian with zero mean. In such a case the probability of measuring an $N \sigma$ deviation is $P(N) \propto \exp(- N^2/2)$, which can be identified approximately with the inverse of the ratio in Eq.~\ref{eq:relative_prob}, so that, e.g., a $3\sigma$ detection is comparable to a ratio of approximately one hundred. However we emphasize that both the number of sigma and related statistics such as $\Delta \chi^2$ are of limited utility in the context of all but the most trivial model comparison problems.

Computing $\prob(\data | \nsavge, \fsky)$ by marginalizing over the likelihood for the full prior volume is an immense computational task, requiring the inversion of the full sky WMAP covariance matrix at full resolution and marginalizing over all possible numbers, locations, and sizes of collisions. However, taking advantage of the fact that bubble collisions produce discrete localized effects on the CMB sky, it is possible to approximate the full-sky Bayesian analysis by a patch-wise analysis if the most promising candidate signatures can be identified in advance. We describe in detail in Appendix~\ref{section:method} an algorithm to perform such a patch-wise approximation to this full multidimensional integral. 

The key ingredient is determining the regions of parameter space where the likelihood is significantly peaked, and hence gives the most  significant contributions to the evidence. If these regions can be identified, the integral need only be performed over the restricted ranges to obtain an estimate of the evidence at greatly-reduced computational cost. We use the results of the blob detection step of the analysis pipeline to identify the patches which are likely make the most significant contributions to the integral. Assuming that the bubble collision model likelihood is peaked in the $\nblob$ detected blobs, we show in Appendix~\ref{section:method} that the unnormalized posterior can be approximated as
\begin{equation}
\label{equation:posteriorfinal}
\prob(\nsavge | \data, \fsky) 
\mbox{}
  \propto \prob(\nsavge) \,  e^{-\fsky \nsavge}
  \sum_{\ns = 0}^\nblob
    \frac{(\fsky \nsavge)^\ns}{\ns!} 
  \sum_{\blob_1, \blob_2, \ldots, \blob_{\ns} = 1}^{\nblob}
    \left[
    \prod_{\src = 1}^{\ns} 
      \rho_{b_s}
    \prod_{i,j = 1}^{\ns} (1 - \delta_{\src_i, \src_j})
    \right],
\end{equation}
where the pre-factors reflect the fact that the number of collisions present on the observable sky, $\ns$, is the realization of a Poisson-like process (of mean $\fsky \nsavge$), and $\rho_b$ is the evidence ratio evaluated within a candidate collision region (with data sub-set $\data_\blob$) using a single bubble collision template
\begin{equation}
\rho_\blob = \frac{\prob(\data_\blob | 1)}{\prob(\data_\blob | 0)}.
\end{equation}
The posterior can therefore be built from local measures of how well the data are described by the standard model with and without a collision template. Once Eq.~\ref{equation:posteriorfinal} is obtained in a particular case, it can be normalized, although this is not strictly necessary to perform the parameter estimation and model selection analyses. 

\begin{figure}[tb]
   \includegraphics[width=6 cm]{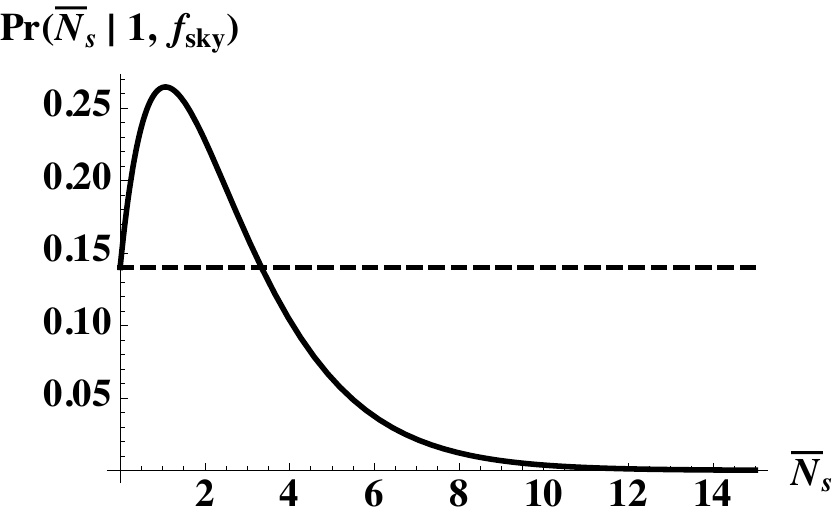}
    \includegraphics[width=6 cm]{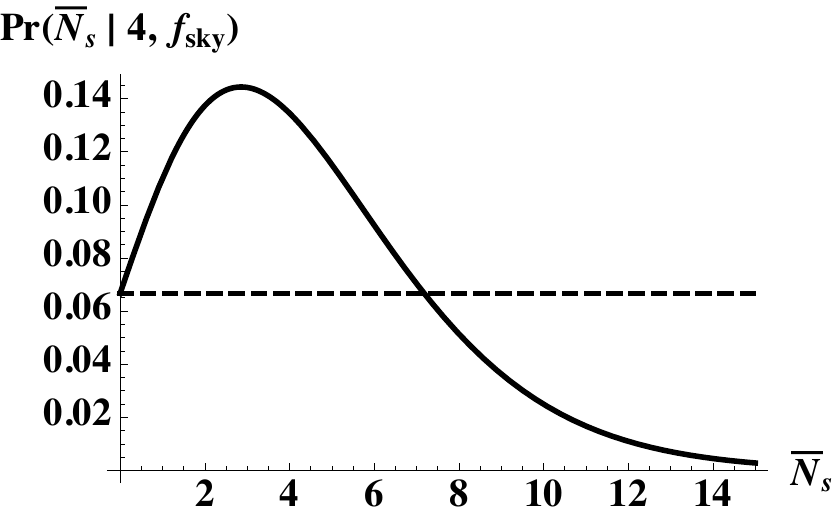}   
\caption{The normalized posterior $\prob(\nsavge | \nblob, \fsky)$ (see Eq.~\ref{eq:nblobposterior}) assuming $f_{\rm sky} = 0.7$. In the left panel, we show the posterior obtained for one blob $N_b = 1$ for a local evidence ratio $\rho_b = 4$. Comparing with the posterior at $\nsavge = 0$ (dashed line), we see that any theory predicting $\nsavge \alt 4$ will be preferred over the theory without bubble collisions. In the right panel, we show the posterior obtained for four blobs with identical local evidence ratios $\rho_b = 1/2$. Again, comparing with the posterior at $\nsavge = 0$, any theory with $\nsavge \alt 7$ will be preferred over the theory without bubble collisions. When there are multiple blobs, the bubble collision hypothesis can be supported even when the evidence ratio for each blob is less than one.}
\label{fig-postfull}
\end{figure}

To illustrate some possibilities, in Fig.~\ref{fig-postfull} we plot the normalized posterior assuming $\fsky=0.7$ (from the KQ75 mask) and a uniform prior on $\nsavge$, for the case where there is a single detected blob (left panel), and four detected blobs (right panel). A theory predicting a particular value of $\nsavge$ will be preferred to a theory without bubble collisions if the ratio in Eq.~\ref{eq:relative_prob} is larger than one. This amounts to comparing the posterior at some value of $\nsavge$ to the posterior at $\nsavge = 0$ (dashed line). To prefer {\em any} theory with bubble collisions, in the one-blob case it is necessary for the blob to yield a local evidence ratio larger than one (here, we plot the posterior assuming $\rho_b = 4$). This is not true when there are multiple blobs, as can be seen in the right panel of Fig.~\ref{fig-postfull}, where we plot the posterior assuming each blob has a local evidence ratio $\rho_b = 0.5$. The bubble collision hypothesis (for some values of $\nsavge$) is preferred even when the local evidence ratios are less than one: a number of marginal detections can be significant when considered together. We can also obtain any desired confidence intervals on $\nsavge$ by examining the shape of the posterior (although it is always the whole distribution that is the full answer to any parameter estimation problem).

\begin{figure}[tb]
   \includegraphics[width=6 cm]{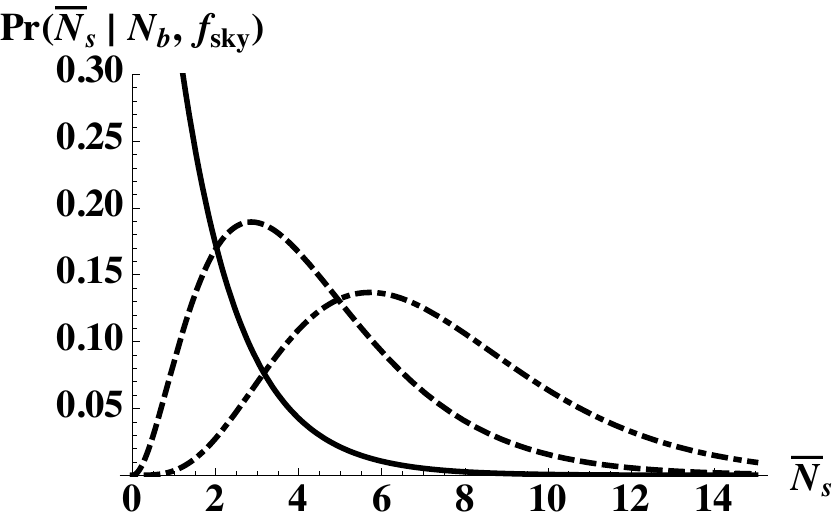}
\caption{The full posterior $\prob(\nsavge | \nblob, \fsky)$ (Eq.~\ref{eq:nblobposterior}) that would be obtained from a conclusive detection (i.e., $\rho_{\blob_s} \gg 1$) of $N_b = 0,2,4$ (solid, dashed, and dot-dashed curves) blobs containing bubble collisions assuming  $f_{\rm sky} = 0.7$. The presence of a sky cut skews the distribution towards $\nsavge > N_b$. Note that even when features are conclusively detected, there is an intrinsic uncertainty in $\nsavge$; this is a form of cosmic variance. }
\label{fig-PrNsNb}
\end{figure}

When the local evidence ratios are large, the posterior can be approximated by Eq.~\ref{eq:nblobposterior}, appropriately normalized. In Fig.~\ref{fig-PrNsNb}, we plot the posterior in the limit of large evidence ratios (again assuming $\fsky = 0.7$) for no blobs, two blobs, and four blobs. Even in the presence of large local evidence ratios, it can be seen that the posterior has a significant spread due to cosmic variance: we only have access to one realization of bubble collisions on the CMB sky. Note that this is true even when there are no detected blobs. When there are multiple decisively detected blobs, the posterior correctly assigns a very small probability to $\nsavge = 0$. 
 
Our analysis also provides constraints on the parameter values of each candidate collision. The constraints on the $n$ template parameters $\model$ are encoded in their joint posterior distribution
\begin{equation}\label{eq:modelposterior}
\prob(\model | \data_\blob, 1) 
  = \frac{\prob(\model) 
    \, \prob(\data_\blob | \model, 1)}{\prob(\data_\blob, 1)}.
\end{equation}
The marginal distribution of any subset of the parameters is given by integrating $\prob(\model | \data_\blob, 1)$ over the remaining parameters which are not of interest.  For the bubble collision model the parameters should include both those describing the collision and the global cosmological parameters; marginalizing over the latter would give constraints on the properties of a (putative) detected collision. We now discuss the analysis of the likelihood and evidence ratios for a patch in greater detail.

\subsubsection{Analysis of candidate bubble collision patches}

At the heart of the above formalism for assessing the full posterior for $\nsavge$ is the evaluation of the patch likelihood for a single collision, $\prob(\data_\blob | \model, 1)$.  Here the data, $\data_\blob$, are the measured temperature values of the pixels in the vicinity of the detected blob that are not in the sky cut.  The bubble collision model parameters, $\model$, should include both those that describe the collision, $\{ z_0, z_{\rm crit} , \theta_{\rm crit} , \theta_0 , \phi_0 \}$, as well as the cosmological parameters which determine the CMB power spectrum.  However any plausible bubble collision would be sufficiently localized that the cosmological parameters are essentially uncorrelated with them; moreover they are sufficiently tightly constrained from CMB measurements that their uncertainties are minimal in the context of a template-matching problem like this.  Hence we fix the cosmological parameters to their best-fit WMAP values \cite{Larson:2010gs} and only the bubble collision parameters are varied.  Hence $\model = \{ z_0, z_{\rm crit} , \theta_{\rm crit} , \theta_0 , \phi_0 \}$ for the bubble collision model, and there are no free parameters in the null model.  Indeed, the no-collision model can be treated as a special case of the collision model in which the collision has zero amplitude.

As both the CMB signal and the WMAP noise are Gaussian, the likelihood has the form
\begin{equation}
\label{eq:chisq}
\prob(\data_\blob | \model , 1) 
 \propto \exp\left( - \frac{1}{2} \chi^2 \right)
 = \exp\left\{ - \frac{1}{2} [\data_\blob - \template (\model)]^{\rm T}
 \matr{C}_\blob^{-1} [\data_\blob - \template (\model)] \right\},
\end{equation}
where $\template(\model)$ is the temperature modulation caused by the collision and $\matr{C}_\blob$ is the pixel-pixel covariance matrix. The temperature modulation of the $p^{\rm th}$ pixel is given from Eq.~\ref{eq:tempmod} as $t_p = 1 +  f(\skypos_p)$, where $\skypos$ is the position on the sky. The covariance matrix includes CMB cosmic variance, Gaussian smoothing approximating the WMAP W-band beam, and the pixel-dependent WMAP noise. The covariance between two pixels $p$ and $q$ with angular positions $\skypos_p$ and $\skypos_q$ is hence given by 
\begin{equation}
C_{p,q} = N_{p,q} + \sum_{\ell}\frac{2\ell + 1}{4\pi}\bar{C_{\ell}}P_{\ell}(\skypos_p \cdot \skypos_q),
\end{equation}
where $\bar{C_{\ell}}$ is the best-fit WMAP CMB power spectrum convolved with a Gaussian beam of FWHM $0.22^{\circ}$, $P_{\ell}(x)$ is the Legendre polynomial of degree $\ell$, and $N_{p,q}$ is the noise covariance between pixels.  This is taken to be 
\begin{equation}
N_{p,q} = \delta_{p,q} \frac{\sigma_{\rm W}^2}{N_{{\rm obs}, p}},
\end{equation}
where $\delta_{p,q}$ is the Kronecker delta function, $\sigma_{\rm W} = 6.549$\,mK is the RMS noise of the W-band detectors, and $N_{{\rm obs}, p}$ is the number of times WMAP has observed the $p^{\rm th}$ pixel. To preserve any edges, we must invert $\matr{C}_\blob$ at full resolution. Given available computational resources, the maximum area of the sky we can study at any one time is limited to patches of radius $\sim11^{\circ}$ surrounding the center of each detected blob.

The evaluation of the evidence integral Eq.~\ref{eq:sourcelikelihoodratio} and the full characterization of the posterior distribution of the parameters are both computationally challenging -- even when restricted to small patches -- as they require a large number of likelihood evaluations.  In all but the simplest of cases it is fatally inefficient to evaluate the likelihood over a multi-dimensional grid and so a variety of sampling algorithms have been developed in which the likelihood is only evaluated in the high posterior regions that are of most interest. For both parameter estimation and evidence calculations we use the nested sampling algorithm \cite{Skilling:2004} as implemented in the publicly available MultiNest package \cite{Feroz:2008xx}. MultiNest performs numerical integration in order to estimate the evidence values; the required convergence of the integration can be adjusted to balance computation speed with accuracy. At the settings we use, the evidence values returned by MultiNest are accurate to $\sim 10\%$. We use the {\tt getdist} routine included in CosmoMC~\cite{Lewis:2002ah} to extract parameter estimates and uncertainties.

The parameter prior $\prob(\model)$ in Eq.~\ref{eq:modelposterior} is derived from theory, previous experimental results, and the limitations of the data-set and pipeline: it encompasses the full prior understanding of what defines a detectable collision. Because we lack a detailed theoretical prediction for the amplitude parameters in each template (as discussed in Sec.~\ref{sec:collisionintro}), we assume a uniform prior on $z_0$ and $z_{\rm crit}$ over the ranges $-10^{-4} \leq z_0 \leq 10^{-4}$ and $-10^{-4} \leq z_{\rm crit} \leq 10^{-4}$, set by the observed temperature fluctuations in the CMB. Bubbles with larger values of these parameters would have been visible to the naked eye in any existing CMB data-set. Bubble collisions are expected to be distributed isotropically on the CMB sky, and so we assume uniform priors on the full ranges of \{$\cos \theta_0, \phi_0$\} to ensure that the probability of finding a bubble per unit area is constant across the sphere. Theory predicts that bubble collision radii should range from $0^{\circ}$ to half-sky, but our pipeline's sensitivity is restricted by CMB power at small scales and computational requirements at large scales. The non-zero prior range for {\em detectable} bubble collisions is accordingly restricted, and we assume uniform priors on $\theta_{\rm crit}$ values in the range $2^\circ \le \theta_{\rm crit} \le 11^\circ$.~\footnote{Eq.~\ref{eq-angdist} predicts that the angular scale distribution for all bubbles falling within our past light cone varies with $\sin \theta_{\rm crit}$. However, this is derived under the assumption that collisions do not affect our bubble interior, and a more careful treatment might lead to a correlation between the values of $z_0$, $z_{\rm crit}$, and $\theta_{\rm crit}$. To retain consistency with our uniform priors on $z_0$ and $z_{\rm crit}$, we assume a uniform prior on $\theta_{\rm crit}$. Regardless, both choices for the prior lead to identical conclusions for the WMAP 7-year data.}

To minimize computation time, the evidence integrals are only calculated over the parameter ranges within which the priors are non-zero and the likelihood is peaked. For each feature, the angular scale lookup tables (Table~\ref{tab:anglelookup}) indicate the range of interest for $\theta_{\rm crit}$. Merging all of the sets of significant pixels found for each feature yields the ranges for $\{ \theta_0, \phi_0 \}$. As the needlets are equally sensitive to cold and hot features with varying profiles, little information about the ranges of interest for $\{z_0, z_{\rm crit }\}$ can be extracted from the blob detection results, so the full prior volume must be considered. The accuracy of this procedure has been verified by performing the integral on the same patch of sky using different parameter ranges for $\{\theta_{\rm crit}, \theta_0, \phi_0\}$. As long as the likelihood peak is encompassed by the parameter ranges given to Multinest, the returned evidence values agree to within numerical accuracy.

For the fiducial collision example shown in Fig.~\ref{fig-needletbubble}, our analysis yields an evidence ratio of $\ln \rho_\blob = 119.8 \pm 0.3$: the collision model is a very good fit to the data. The full-sky posterior would favour any theory predicting bubble collisions over a large range of $\nsavge$. The marginalized bounds on the parameters are compared to the input parameters in the first row of Table~\ref{tab:evtestcases}: the agreement is excellent. However, to make a final judgment about a detection, we must ask what types of evidence ratios we get for false detections in the WMAP end-to-end simulation.

\subsubsection{Analysis of the WMAP end-to-end simulation}

We have performed the full Bayesian parameter estimation and model selection analysis on the blobs found in the WMAP end-to-end simulation (see Table~\ref{tab:endtoendfeatures}). The total processing time for the full pipeline to run on this single map is on the order of $12$ hours on 28 cores. Our results for the evidence ratios and marginalized parameter constraints for $\{ z_0, z_{\rm crit}, \theta_{\rm crit}, \theta_0, \phi_0 \}$ for each feature are recorded in Table~\ref{tab:end-to-endevidences}. 

The evidence ratios for the features identified in the blob-detection step of the pipeline are all significantly less than one. We can therefore approximate the full posterior for $\nsavge$ by Eq.~\ref{eq:noblobs}, and rule out $\nsavge \agt 1.6$ at the $68 \%$ confidence level. The posterior is maximized at $\nsavge = 0$, and we therefore correctly conclude that the data from the end-to-end simulation does {\em not} warrant augmenting $\Lambda$CDM with bubble collisions. 

These results from the end-to-end simulation yield quantitative information on the degree to which systematics and foregrounds could mimic the signal from a bubble collision. Reassuringly, no features yield evidence ratios greater than one. To be distinguishable from systematics and foregrounds, we require the evidence ratios that we find for any feature to at least exceed the evidence ratios found in the end-to-end simulation at similar needlet frequencies.

\begin{table*}
\begin{tabular}{c c c c c c c c}
\hline
\hline
 feature & $\theta_{\rm crit}$ range & \ \  $\ln \rho_\blob$ & \ \ $z_0$ & \ \ $z_{\rm crit}$ & \ \ $\theta_{\rm crit}$ & \ \ $\theta_0$ & \ \ $\phi_0$ \\
\hline
1 & $5-14$ & $-7.9 \pm 0.1$ & $-3.3^{+1.1}_{-1.0} \times 10^{-5}$ & $0.0^{+0.4}_{-0.4} \times 10^{-5}$ & $8.6^{+1.5}_{-1.5}$ & $120.4^{+1.5}_{-1.3}$ & $77.8^{+1.8}_{-1.4}$ \\
2 & $10-21$ & $-9.9 \pm 0.1$ & $-2.4^{+1.3}_{-1.2} \times 10^{-5}$ & $0.0^{+0.4}_{-0.4} \times 10^{-5}$ & $10.6^{+0.5}_{-0.6}$ & $46.9^{+3.2}_{-2.4}$ & $152.0^{+3.0}_{-2.8}$ \\
3 & $6-12$ & $-11.9 \pm 0.1$ & $-2.9^{+1.0}_{-1.0} \times 10^{-5}$ & $0.0^{+0.4}_{-0.5} \times 10^{-5}$ & $8.5^{+1.9}_{-1.9}$ & $13.9^{+0.3}_{-0.3}$ & $72.1^{+1.6}_{-1.5}$ \\
4 & $4-12$ & $-6.9 \pm 0.2$ & $4.6^{+1.2}_{-1.2} \times 10^{-5}$ & $-0.1^{+0.7}_{-0.6} \times 10^{-5}$ & $5.2^{+0.7}_{-1.2}$ & $50.3^{+0.6}_{-0.6}$ & $221.7^{+0.8}_{-0.9}$ \\
5 & $3-7$ & $-10.7 \pm 0.1$ & $4.1^{+1.3}_{-1.2} \times 10^{-5}$ & $0.0^{+0.7}_{-0.7} \times 10^{-5}$ & $4.4^{+0.5}_{-0.6}$ & $80.5^{+0.3}_{-0.3}$ & $218.3^{+0.3}_{-0.4}$ \\
6 & $2-4$ & $-11.3 \pm 0.1$ & $-4.2^{+1.5}_{-1.5} \times 10^{-5}$ & $0.1^{+1.0}_{-1.0} \times 10^{-5}$ & $2.7^{+0.4}_{-0.5}$ & $62.6^{+0.4}_{-0.3}$ & $146.1^{+0.3}_{-0.4}$ \\
7 & $2-4$ & $-6.6 \pm 0.1$ & $6.5^{+1.5}_{-1.4} \times 10^{-5}$ & $-0.1^{+0.9}_{-0.8} \times 10^{-5}$ & $3.0^{+0.3}_{-0.3}$ & $62.6^{+0.2}_{-0.2}$ & $142.0^{+0.3}_{-0.3}$ \\
8 & $2-4$ & $-8.2 \pm 0.1$ & $6.1^{+1.7}_{-1.7} \times 10^{-5}$ & $-0.2^{+1.1}_{-1.1} \times 10^{-5}$ & $2.5^{+0.2}_{-0.3}$ & $111.5^{+0.2}_{-0.2}$ & $69.7^{+0.2}_{-0.2}$ \\
9 & $3-6$ & $-9.7 \pm 0.1$ & $5.0^{+1.3}_{-1.3} \times 10^{-5}$ & $0.0^{+0.6}_{-0.6} \times 10^{-5}$ & $3.7^{+0.5}_{-0.4}$ & $36.4^{+0.3}_{-0.3}$ & $131.0^{+0.5}_{-0.5}$ \\
10 & $2-4$ & $-9.9 \pm 0.2$ & $5.6^{+1.6}_{-1.5} \times 10^{-5}$ & $-0.8^{+0.9}_{-0.9} \times 10^{-5}$ & $2.6^{+0.2}_{-0.4}$ & $160.2^{+0.2}_{-0.2}$ & $235.9^{+0.6}_{-0.7}$ \\
\hline
\hline
 \end{tabular}
 \begin{center}
 \caption{Results of the Bayesian parameter estimation and model selection analysis for the WMAP end-to-end simulation. The ranges of $\theta_{\rm crit}$ are determined from the needlet response (see Table~\ref{tab:anglelookup}). By computational necessity, the evidence integral is truncated at $11^{\circ}$. Reported error bars are at 68\% CL. Angular quantities are quoted in degrees.
   \label{tab:end-to-endevidences}}
 \end{center}
\end{table*}

\subsubsection{Analysis of bubble collision simulations}

The long processing time, even for a single map, prohibits us from running the Bayesian parameter estimation and model selection analysis on the full set of bubble collision simulations. We therefore choose a small number of representative examples from the set of simulated collisions passing the needlet significance threshold (drawn from the exclusion and sensitivity regions of Fig.~\ref{fig-needletexclusion}). Six 10$^\circ$ collision simulations were chosen to sample distinct areas of our parameter space, specifically collisions with:
\begin{enumerate}
\item a large central amplitude and edge;
\item a small central amplitude but large edge;
\item a large central amplitude but small edge;
\item a medium central amplitude and medium edge;
\item a small central amplitude and medium edge, and;
\item a small central amplitude and small edge.
\end{enumerate}
The first two collisions lie in the CHT exclusion zone, the third in the needlets exclusion zone, and the others in the sensitivity region. All collisions were placed at the low-noise location to maximize the chance of a detection. 

The results of the Bayesian analysis of the collision simulations are displayed in Table~\ref{tab:evtestcases}. The first example corresponds to the collision in Fig.~\ref{fig-needletbubble}, and is clearly a highly significant detection with an evidence ratio of $\ln \rho_\blob \simeq 120$. The second example is, again, an extremely clear detection, with $\ln \rho_\blob \simeq 136$. While the evidence for the third example is numerically lower than for the strongly discontinuous cases, at  $\ln \rho_\blob \simeq 29$, it is again a conclusive detection. In each of these examples, the full-sky posterior assuming $N_b = 1$ (which is well approximated by Eq.~\ref{equation:posterioroneblob}), would prefer models with bubble collisions over a wide range of $\nsavge$. 

For the collisions sampled in the sensitivity region, the maximum needlet significance recorded in each case was around $S \simeq 4$, which is on the upper end of the significances found in the end-to-end simulation: similar features in the data would be passed to the Bayesian analysis section of the pipeline. The evidence ratios were found to be $\ln \rho_\blob \simeq 9$ for the collision with a medium central amplitude and a medium edge, $\ln \rho_\blob \simeq -1$ for the collision with the medium edge but a smaller central amplitude, and $\ln \rho_\blob \simeq -7$ for the collision with a small edge. Since the latter two templates differ only by the value of $z_{\rm crit}$, this is further proof that the presence of a detectable causal boundary increases our ability to distinguish a collision. In addition, comparing examples 3 and 4, it can be seen that changing the central amplitude by a bit more than a factor of two yields an evidence ratio that is orders of magnitude larger. Apparently, there are rather sharply defined limits of detection. For these marginal cases, the parameter uncertainties are significantly underestimated due to the relative strength of the CMB and noise. Only in the case of the collision with a medium amplitude and medium edge could we conclude that models with bubble collisions are preferred over those without over a modest range in $\nsavge$.

In conclusion, for the simulated collisions in the needlet and CHT exclusion regions of parameter space, our pipeline can clearly determine that the bubble collision hypothesis is favoured for a variety of $\nsavge$. In the other cases we have studied, where the collision lies in the needlet sensitivity region, the conclusion is less clear. The evidence ratios are higher than most of those in the end-to-end simulation, but not much greater. They are also small in magnitude, and therefore do not yield full-sky posteriors that favor the bubble collision hypothesis. Thus, while we might rule these features out as being systematics or foregrounds, better data would be needed to definitively establish the bubble collision hypothesis. Furthermore, the bounds on parameter values in detections associated with the sensitivity regions of parameter space should be regarded as rough estimates only. Note also that since the data sets we consider for each blob are restricted to patches of the sky smaller than $11^{\circ}$, the gain in sensitivity that arises from the existence of a circular temperature discontinuity will not be present for modulations with $\theta_{\rm crit} \agt 11^{\circ}$. For large features with an edge, the evidence ratios we obtain would therefore be an {\em underestimate}. 

\begin{table*}
\begin{tabular}{c c c c c c c c c}
\hline
\hline
example & \ \ $z_0$ & \ \ $z_{\rm crit}$ & \ \ $\ln\rho_\blob$ & \ \ $\hat{z}_0$ & \ \ $\hat{z}_{\rm crit}$ & \ \ $\hat{\theta}_{\rm crit}$ & \ \ $\hat{\theta}_0$ & \ \ $\hat{\phi}_0$ \\
\hline
1 & $5.0 \times 10^{-5}$ & $-5.0 \times 10^{-5}$ & $119.8 \pm 0.3$ & $5.1^{+1.0}_{-1.0} \times 10^{-5}$ & $-5.0^{+0.3}_{-0.3} \times 10^{-5}$ & $10.0$ & $57.7$ & $99.2$ \\
2 & $1.0 \times 10^{-5}$ & $-5.6 \times 10^{-5}$ & $136.0 \pm 0.2$ & $2.3^{+1.1}_{-1.1} \times 10^{-5}$ & $-5.4^{+0.3}_{-0.3} \times 10^{-5}$ & $10.0$ & $57.7$ & $99.2$ \\
3 & $1.0 \times 10^{-4}$ & $-1.0 \times 10^{-6}$ & $28.9 \pm 0.3$ & $9.4^{+0.6}_{-0.4} \times 10^{-5}$ & $-0.2^{+0.8}_{-0.8} \times 10^{-5}$ & $10.0^{+0.9}_{-0.9}$ & $57.7^{+0.5}_{-0.7}$ & $99.8^{+0.5}_{-0.6}$ \\
4 & $3.2 \times 10^{-5}$ & $-3.2 \times 10^{-5}$ & $9.0 \pm 0.3$ & $6.4^{+1.0}_{-1.1} \times 10^{-5}$ & $-2.0^{+0.3}_{-0.3} \times 10^{-5}$ & $8.92^{+0.09}_{-0.02}$ & $57.01^{+0.03}_{-0.03}$ & $100.27^{+0.04}_{-0.04}$ \\
5 & $1.0 \times 10^{-5}$ & $-3.2 \times 10^{-5}$ & $-1.0 \pm 0.3$ & $4.2^{+0.6}_{-0.6} \times 10^{-5}$ & $-1.2^{+0.3}_{-0.4} \times 10^{-5}$ & $8.6^{+0.4}_{-0.9}$ & $57.3^{+0.1}_{-0.3}$ & $100.1^{+0.3}_{-0.1}$ \\
6 & $1.0 \times 10^{-5}$ & $-1.8 \times 10^{-5}$ & $-7.2 \pm 0.2$ & $4.8^{+1.2}_{-1.1} \times 10^{-5}$ & $-0.1^{+0.7}_{-0.7} \times 10^{-5}$ & $6.5^{+0.4}_{-0.7}$ & $58.2^{+0.3}_{-0.3}$ & $99.9^{+0.3}_{-0.3}$ \\
\hline
\hline
 \end{tabular}
 \begin{center}
 \caption{The input and marginalized 68\% CL parameter bounds for the representative sample of simulated $10^{\circ}$ collisions. All simulated collisions are located at $\theta_0 = 57.7^{\circ}$, $\phi_0 = 99.2^{\circ}$. Hatted quantities are estimates, and un-hatted quantities are inputs. No errors are quoted for the estimated central positions and radii for the cases where there was an extremely strong detection. This is due to the pixelization of the map: variations in collision-centre coordinates or radius of much less than a pixel's width will not affect the pixelated template, and hence will not affect the likelihood.  Angular quantities are quoted in degrees.
   \label{tab:evtestcases}}
 \end{center}
\end{table*}

\subsection{Summary of the analysis pipeline and conditions for claiming a detection} \label{sec:det_conditions}

We now summarize the analysis pipeline and the interpretation of its outputs. First, the analysis pipeline segments the sky into ``blobs," each of which corresponds to a region which, for some needlet type and frequency, passes our needlet significance threshold. A specific region of the temperature map can be covered by multiple blobs if there is a response for multiple needlet types/frequencies at the same location. The output of this first step in our pipeline is the location, size, and maximum significance associated with each blob. The edge detection step of our pipeline finds the CHT score as a function of assumed circle size and pixel. If there is a clearly peaked global maximum for the CHT score, this can be processed into the most likely circle center and angular scale. In parallel, we calculate the marginalized constraints on the parameters $\{ z_0, z_{\rm crit}, \theta_{\rm crit}, \theta_0, \phi_0\}$ and Bayesian evidence ratio $\rho_\blob$ for each feature. These evidence ratios are then used to construct the full-sky posterior $\prob(\nsavge | \nblob, \fsky)$ (Eq.~\ref{equation:posteriorfinal})  which is a function of $\nsavge$.

The posterior allows us to put constraints on the possible values of $\nsavge$ that are consistent with the data. Comparing the value of the posterior at $\nsavge=0$ and some particular value of $\nsavge$ specifies whether or not the $\Lambda$CDM model should be superseded by a model that also predicts on average $\nsavge$ bubble collisions. If a large ratio of the posteriors is obtained,  a conclusive detection of the bubble collision hypothesis can be claimed (provided a model that predicts an appropriate value of $\nsavge$ exists). A clear peak in the CHT score would indicate the presence of a circular temperature discontinuity in the CMB. This is a clear signature of  bubble collisions, and would be nearly conclusive evidence for the eternal inflation scenario. We have found using simulations that a clear edge also yields large evidence ratios, indicating that these two tests are complementary. However, an edge is not necessary to verify the bubble collision hypothesis. There is a clear expectation obtained from the end-to-end simulation for the contribution from false detections due to systematics and foregrounds: the absence of a clear peak in the CHT score, and evidence ratios for each blob not exceeding $\ln \rho_\blob \sim -6.6$ at detectable scales.

\section{Analysis of the WMAP 7-year data}\label{sec:WMAP7}

Our analysis of the W-band WMAP 7-year foreground-reduced temperature map with the KQ75 mask produces a total of 38 blobs passing our needlet sensitivity thresholds. These blobs can be grouped into 15 distinct features, four of which either intersect or are within a few pixels of the main Galactic cut; these features are assumed to be responses to the mask, and we do not consider them further. The properties of the blobs belonging to the 11 remaining features are given in Table~\ref{tab:wmapfeatures}.

A number of these features have been noted previously. Feature 2 is at the same position as the famous Cold Spot~\cite{Cruz:2004ce}. In addition, features 1 and 3 are coincident with the most significant hot spots identified in the needlet analysis of Ref.~\cite{Pietrobon:2008rf}. The number of features we have found is consistent with the results from the WMAP end-to-end simulation, although the simulation does not contain as many high-significance features at low $j$. In addition, the most significant features in the WMAP 7-year data generate responses from multiple needlet types at multiple frequencies (e.g., the Cold Spot is picked out by seven needlet frequencies), whereas features in the end-to-end simulation tend to be highlighted only by a single needlet. Interestingly, 9 of the 11 features identified as significant are in the Southern Galactic hemisphere.

\begin{table*}
\begin{tabular}{c c c c c c c c}
\hline
\hline
 feature & \ \ blob & \ \ $B$ & \ \ $j$ & \ \ $\theta_0$ & \ \ $\phi_0$ & \ \ blob radius & \ \ $S$ \\
\hline
1 & 1 & 2.5 & 2 & 140.1 & 173.7 & 4.5 & 3.76 \\
1 & 2 & 1.8 & 3 & 140.9 & 174.4 & 5.3 & 3.48 \\
2 & 1 & 2.5 & 3 & 147.8 & 209.5 & 4.1 & 4.49 \\
2 & 2 & 1.8 & 4 & 148.2 & 207.7 & 5.5 & 4.55 \\
2 & 3 & 1.8 & 5 & 148.5 & 210.2 & 1.4 & 3.37 \\
2 & 4 & 1.4 & 7 & 147.8 & 209.5 & 2.5 & 3.81 \\
2 & 5 & 1.4 & 8 & 147.8 & 209.5 & 4.1 & 4.58 \\
2 & 6 & 1.4 & 9 & 147.4 & 208.1 & 2.8 & 4.30 \\
2 & 7 & 1.4 & 10 & 146.6 & 207.5 & 1.3 & 3.74 \\
3 & 1 & 2.5 & 3 & 123.2 & 321.3 & 2.5 &  4.09 \\
3 & 2 & 1.8 & 4 & 122.8 & 322.4 & 4.9 & 3.82 \\
3 & 3 & 1.4 & 7 & 122.8 & 321.0 & 1.9 & 3.59 \\
3 & 4 & 1.4 & 8 & 122.8 & 321.0 & 3.2 & 4.01 \\
3 & 5 & 1.4 & 9 & 122.8 & 321.0 & 2.7 & 4.30 \\
3 & 6 & 1.4 & 10 & 122.4 & 320.6 & 1.5 & 3.78 \\
4 & 1 & 2.5 & 4 & 145.1 & 33.0 & 0.9 &  4.20 \\
4 & 2 & 1.8 & 6 & 145.5 & 32.4 & 0.7 & 3.72 \\
4 & 3 & 1.4 & 11 & 145.1 & 33.0 & 0.9 & 3.95 \\
5 & 1 & 1.8 & 5 & 32.2 & 74.0 & 1.2 & 3.41 \\
6 & 1 & 1.8 & 5 & 128.7 & 91.1 & 1.2 & 3.37 \\
7 & 1 & 1.8 & 5 & 169.8 & 181.6 & 2.3 & 3.82 \\
7 & 2 & 1.8 & 6 & 169.0 & 187.5 & 0.8 & 3.76 \\
7 & 3 & 1.4 & 10 & 169.4 & 184.7 & 1.5 & 4.12 \\
7 & 4 & 1.4 & 11 & 168.7 & 187.3 & 1.1 & 4.07 \\
8 & 1 & 1.8 & 6 & 57.9 & 115.7 & 0.7 & 3.78 \\
9 & 1 & 1.8 & 7 & 152.3 & 241.8 & 0.6 & 4.12 \\
10 & 1 & 1.4 & 10 & 167.2 & 268.7 & 1.0 & 3.99 \\
10 & 2 & 1.4 & 11 & 166.8 & 271.3 & 1.0 & 4.09 \\
11 & 1 & 1.4 & 11 & 115.0 & 22.5 & 0.5 & 3.80 \\
11 & 2 & 1.4 & 12 & 114.6 & 22.1 & 0.5 & 4.32 \\
\hline
\hline
 \end{tabular}
 \begin{center}
 \caption{Features found by the needlet transform in the WMAP 7-year data. Features 1 and 3 correspond to the hot spots found in Ref.~\cite{Pietrobon:2008rf}; feature 2 is the Cold Spot~\cite{Cruz:2004ce}. Angular quantities are reported in degrees.}
   \label{tab:wmapfeatures}
 \end{center}
\end{table*}

\begin{figure}[tb]
   \includegraphics[width=7 cm]{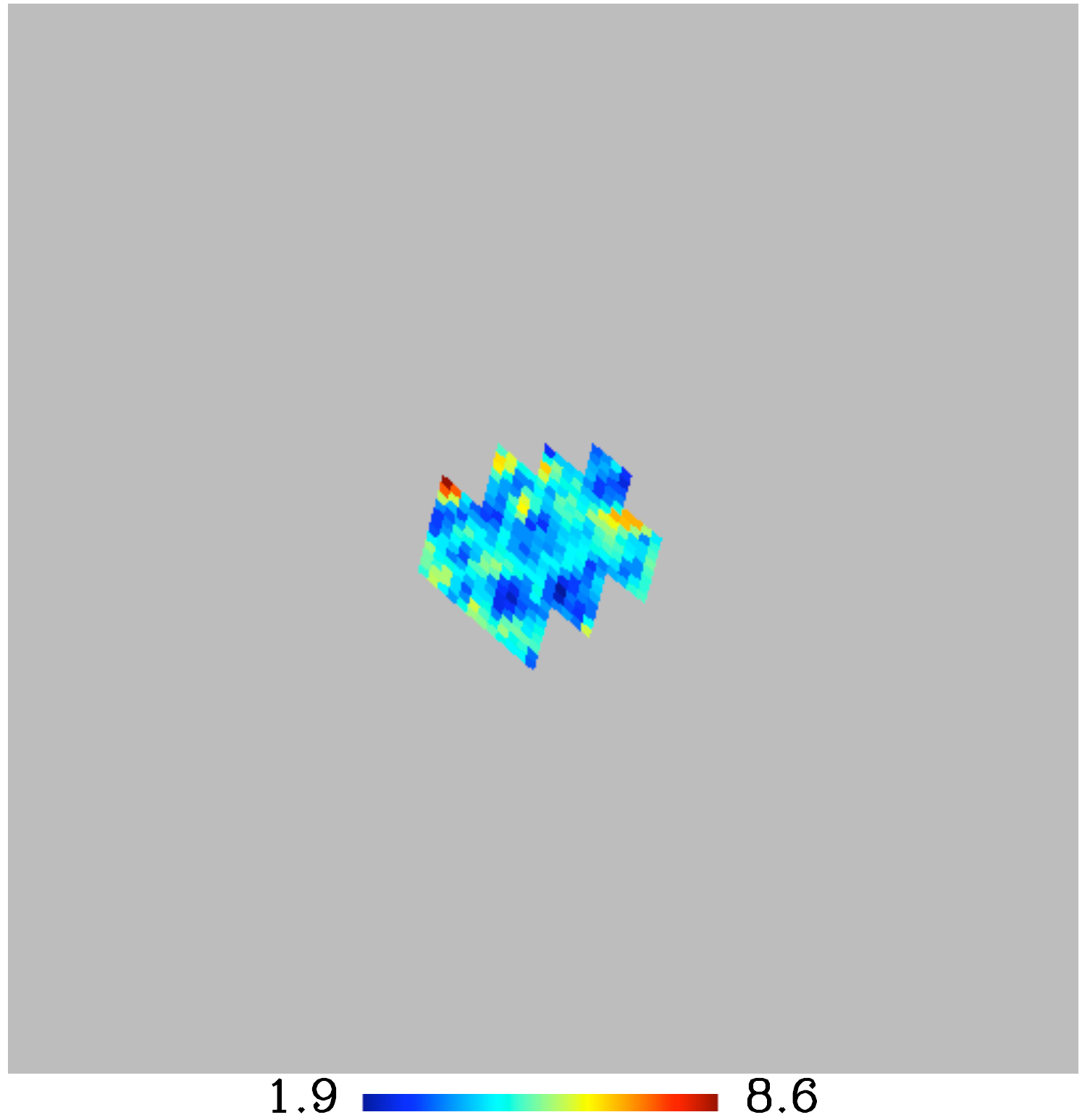}
   \includegraphics[width=9 cm]{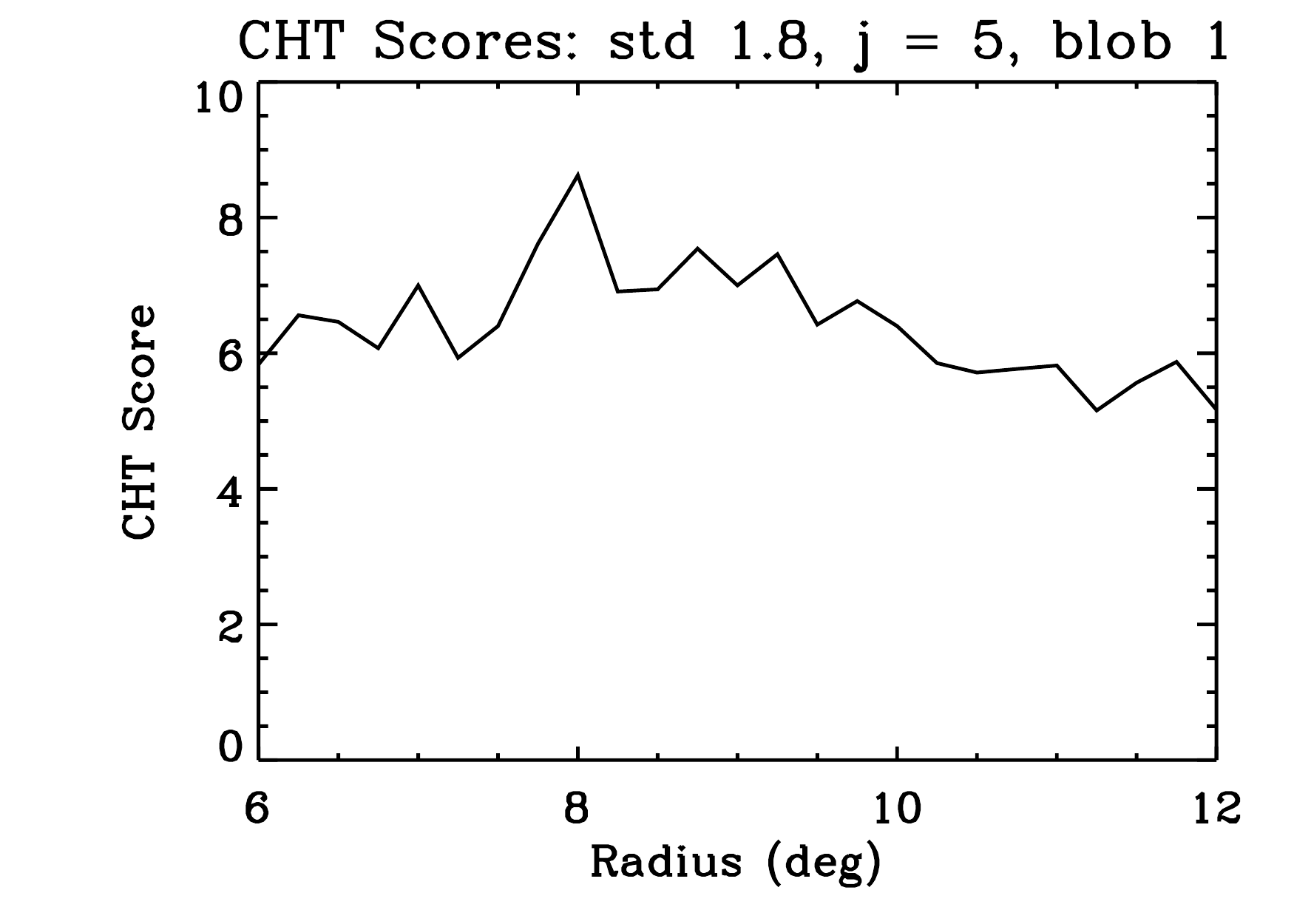}
\caption{The clearest peak found during the edge-detection analysis of the WMAP data. The contrasts in scores as a function of position (left) and radius (right) are comparable to those obtained in the analysis of the end-to-end simulation (Fig.~\ref{fig-WMAPendtoend}), and greatly reduced compared to the collision example (Figs.~\ref{fig-bubble_and_CHT}~and~\ref{fig-needlet_score}).}
\label{fig-WMAP7yearedges}
\end{figure}

The CHT scores do not have a clear peak at any angular scale or location for any of the detected features. Indeed, the detailed outputs for the data are completely consistent with those obtained for the end-to-end simulation. The largest CHT peak found in the data is shown in Fig.~\ref{fig-WMAP7yearedges} (which should be compared to the most peak-like feature found in the end-to-end simulation, shown in Fig.~\ref{fig-WMAPendtoend}). We therefore find no evidence for circular temperature discontinuities in the WMAP 7-year data, and can rule out bubble collisions in the CHT exclusion region defined by simulated collisions shown in Fig.~\ref{fig-chtexclusion}.

The marginalized parameter constraints and local evidence ratio for each of the features is recorded in Table~\ref{tab:wmapevidences}.~\footnote{Since we are limited to patches of the sky $11^{\circ}$ in radius, the evidence ratios we have obtained for features whose $\theta_{\rm crit}$ priors extend beyond $\sim 11^{\circ}$ will be underestimated if a weak edge exists outside the patch of sky considered.} 
Features 2, 3, 7, and 10 have evidence ratios significantly larger than those found in the collision-free end-to-end simulation ($\ln \rho_\blob \sim -6.6$), specifically $-4.6$, $-4.1$, $-5.4$ and $-3.8$ respectively. Assuming $\nblob = 4$, and using these values for the local evidence ratios in Eq.~\ref{equation:posteriorfinal}, we find that the posterior is maximized at $\nsavge=0$, and we can constrain $\nsavge < 1.6$ at $68 \%$ CL. One would need roughly $\ln \rho_\blob \sim -1$ for each of the four features to prefer the bubble collision hypothesis for any value of $\nsavge$. Therefore, the WMAP 7-year data does not warrant adding bubble collisions to $\Lambda$CDM. 

\begin{figure}[tb]
   \includegraphics[width=10 cm]{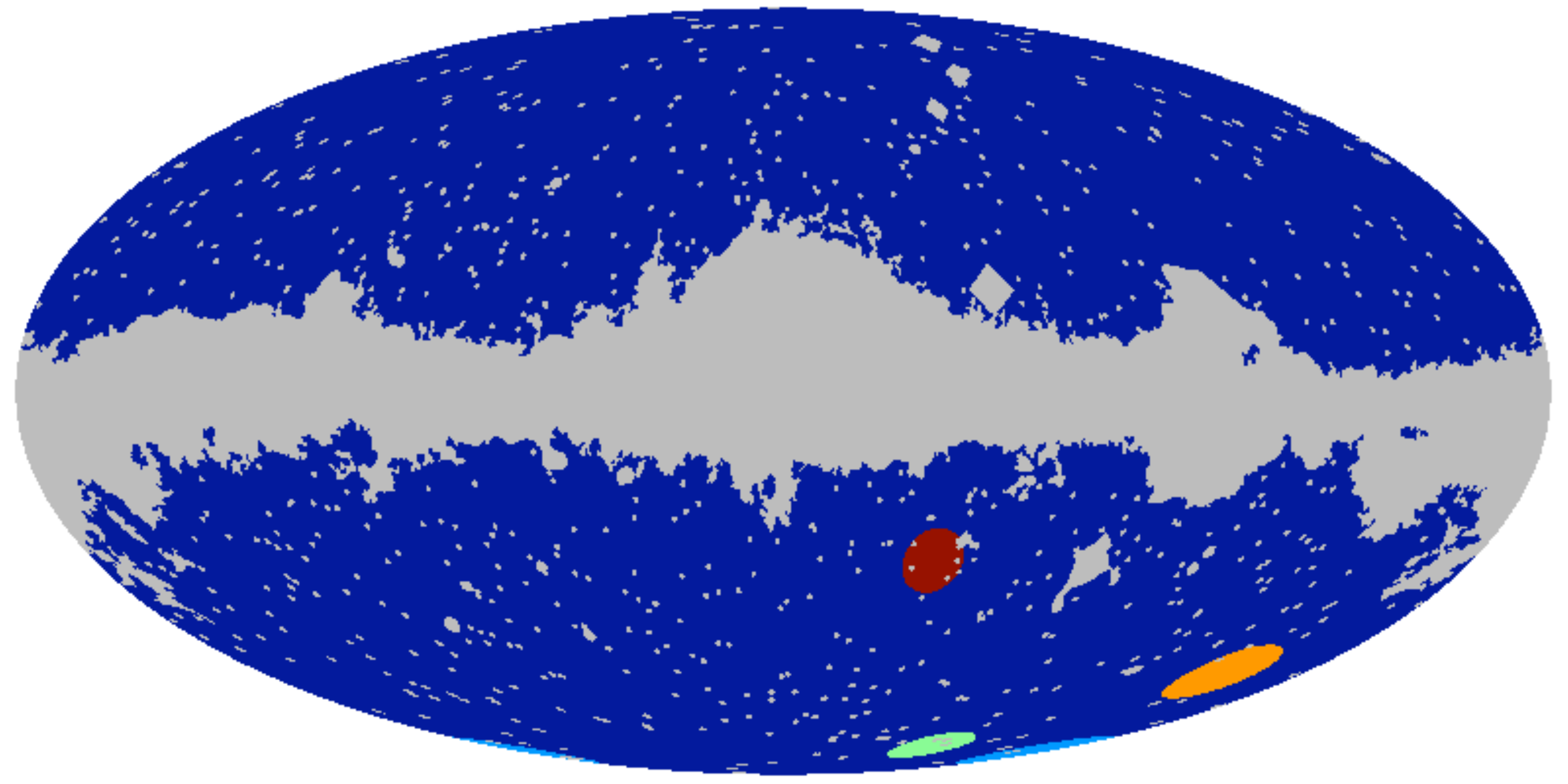}
\caption{Full-sky map showing the positions and sizes of the four features with largest evidence ratios, alongside the 7-year KQ75 sky cut. Feature 2 is plotted in orange, feature 3 in red, feature 7 in light blue and feature 10 in light green.}
\label{fig-detection_locations}
\end{figure}

\begin{figure}[tb]
   \includegraphics[width=17.5 cm]{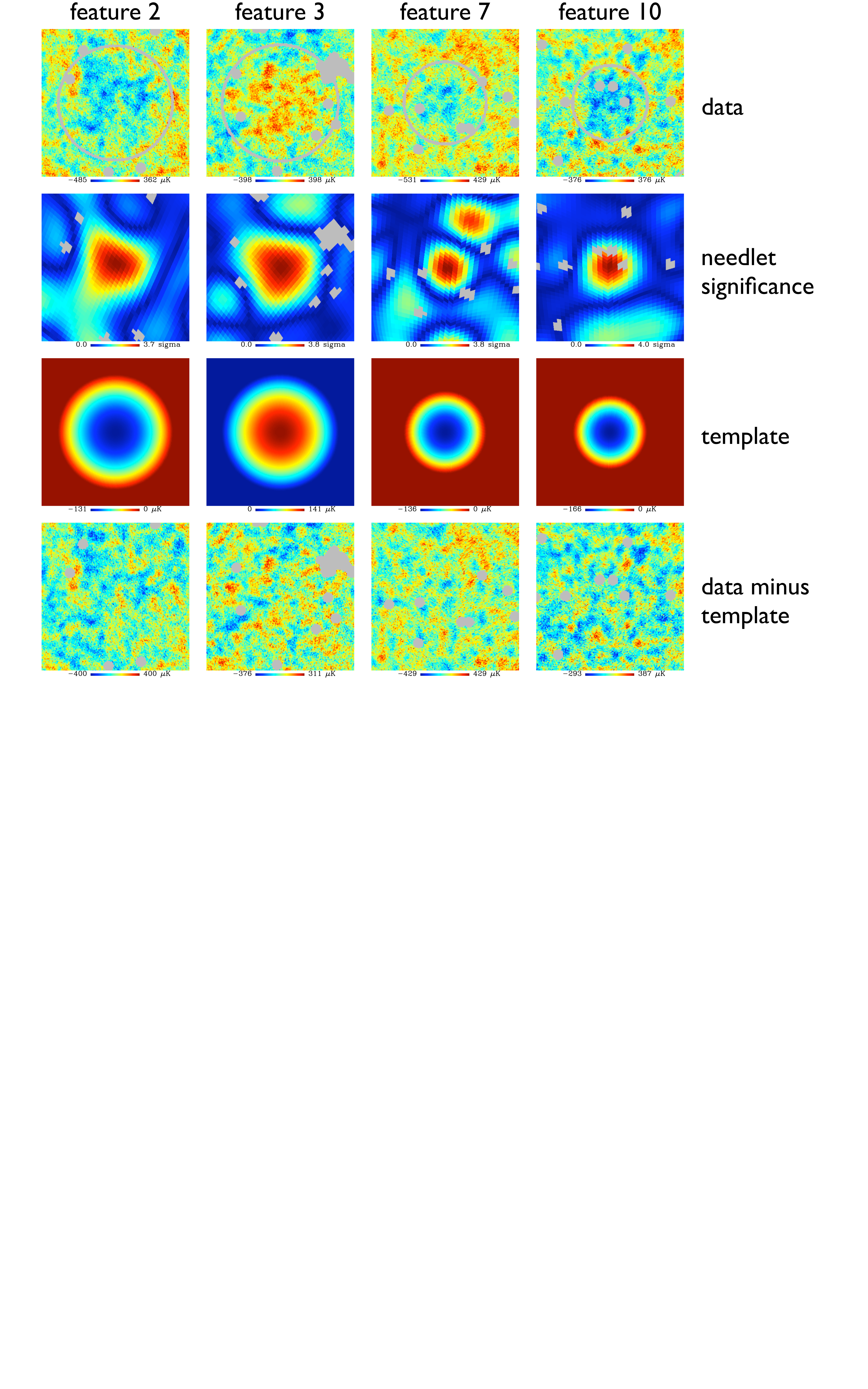}
\caption{Maps of the four features with largest evidence ratios. The top row shows the W-band temperature map in the locality of the four features, masked with the KQ75 mask. Overlaid are circles indicating the estimated position and angular scale found in each case. The second row contains plots of the masked needlet significances for the needlets whose $\theta_{\rm crit}$ priors produced the largest evidence ratios. These plots appear pixelated as the blob detection step is carried out at reduced resolution. The third row shows the bubble collision templates corresponding to the estimated model parameters; these templates are subtracted from the W-band data in the fourth row. The width of each plot is $\sim 16.7^{\circ}$.}
\label{fig-detection_closeups}
\end{figure}

Although the local evidence ratios found for the WMAP 7-year data were not large enough to yield support for the bubble collision hypothesis, they are about an order of magnitude larger than what was expected from systematics based on the end-to-end simulation. The analysis of future data sets may increase the significance of these blobs if they are indications of bubble collisions, or else they will decrease in significance if they are not; in any case they are the most significant features on our sky, and thus take priority in being further investigated with better data. Thus, we now examine these four most significant features in more detail.
The location of each of the four features on the sky is shown in Fig.~\ref{fig-detection_locations}. A closer view of each feature is shown in Fig.~\ref{fig-detection_closeups}, along with plots of the needlet significances $S$ triggering the Bayesian analysis step, collision templates for the marginalized parameter constraints found in each case, and the CMB sky as it would appear with these template contributions removed.

\begin{figure}[tb]
   \includegraphics[width=8 cm]{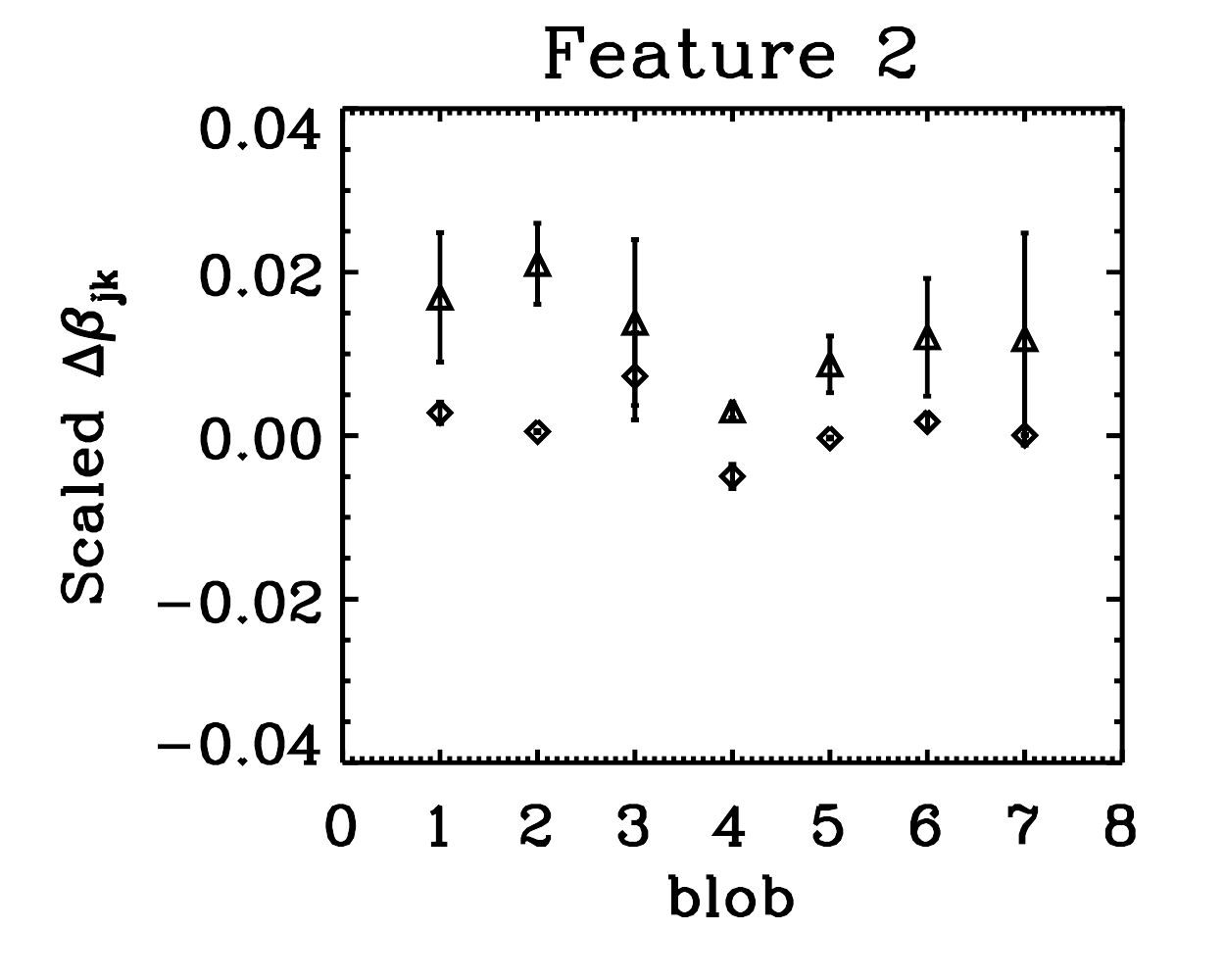}
   \includegraphics[width=8 cm]{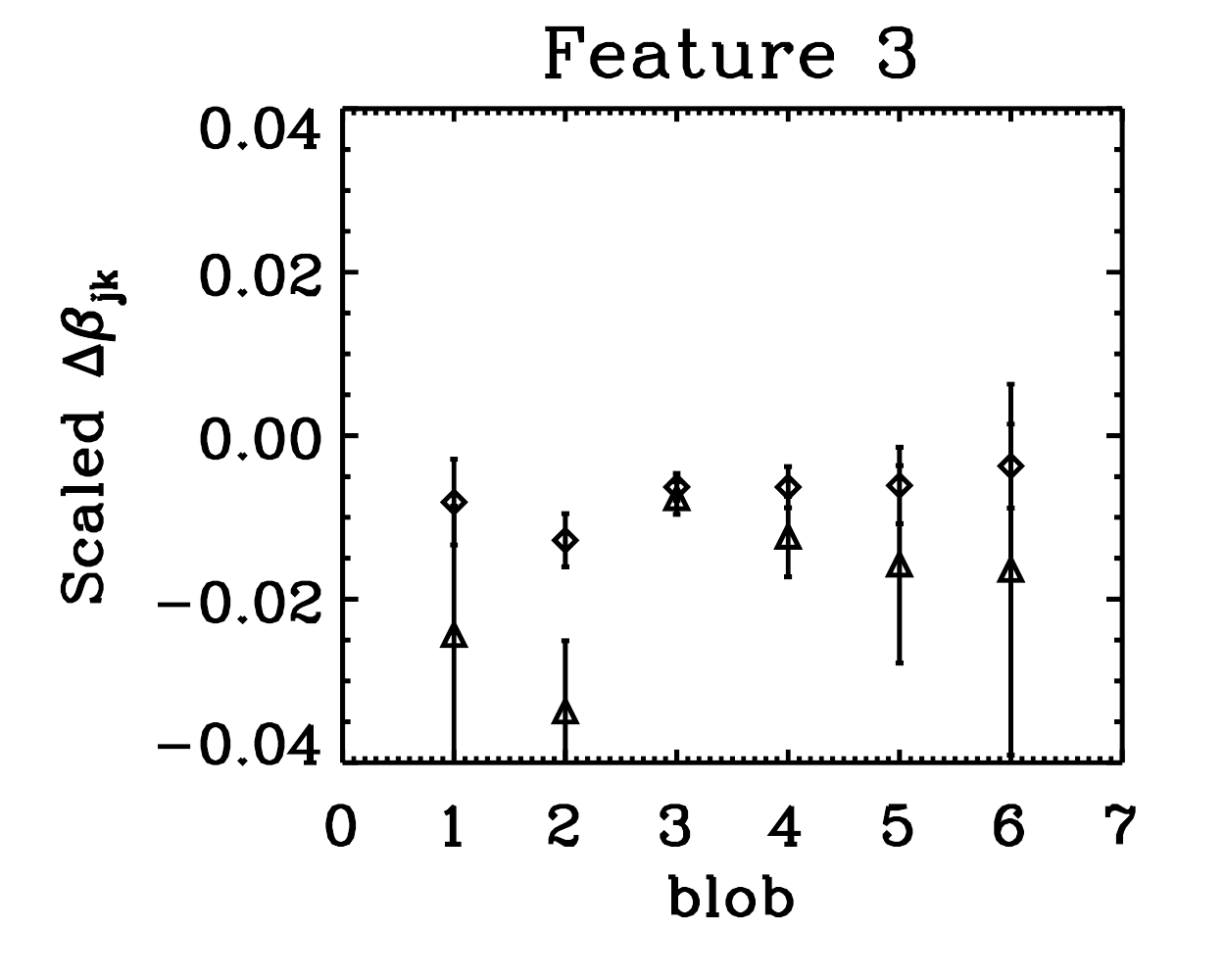}
   \includegraphics[width=8 cm]{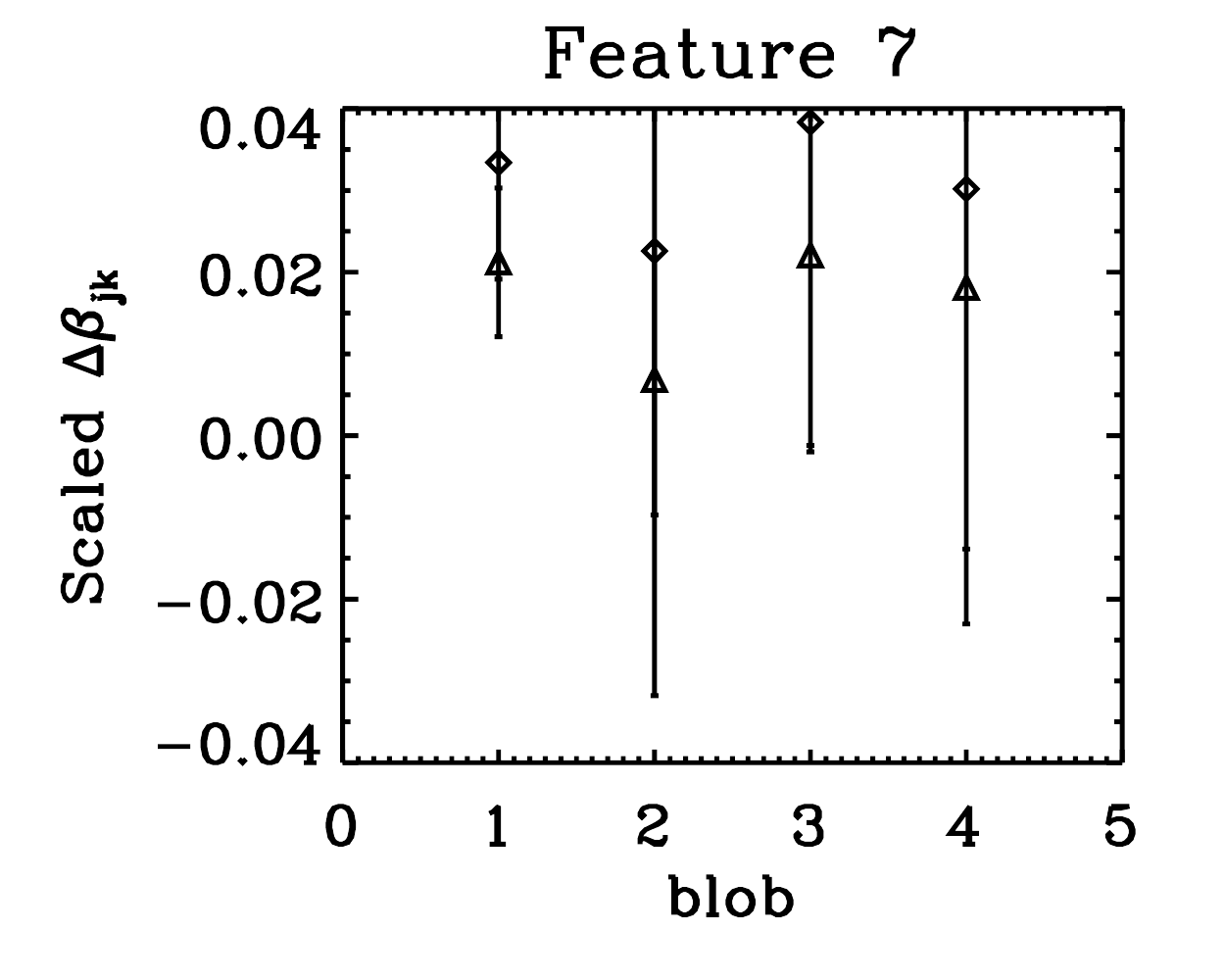}
   \includegraphics[width=8 cm]{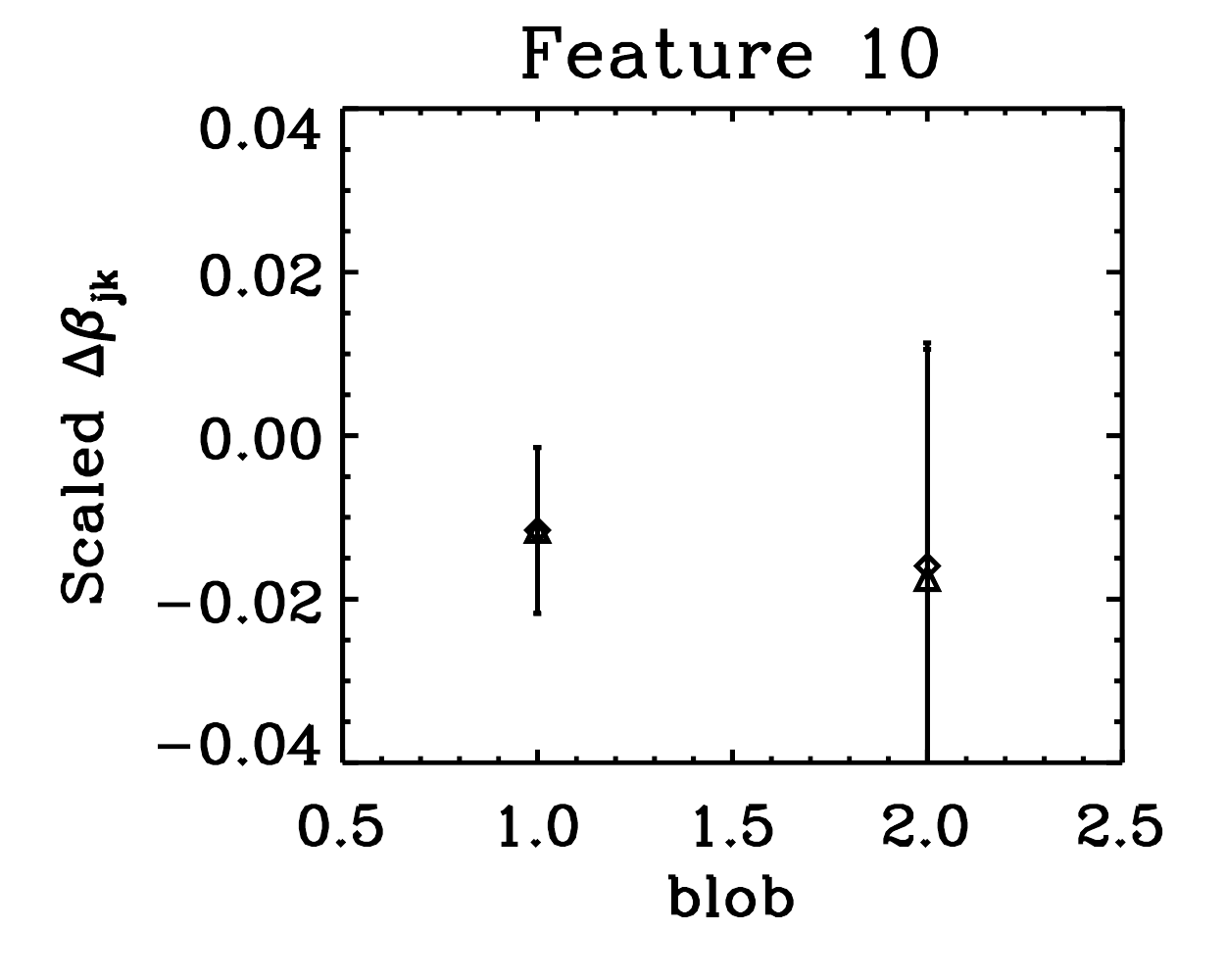}
\caption{WMAP channel frequency dependence of the features highlighted by our pipeline. The W-band-normalized difference in pixel-averaged needlet coefficients between the Q and W bands (triangles) and V and W bands (diamonds) are plotted for each blob making up a feature, as highlighted by the full suite of needlet transforms.}
\label{fig-detection_freqdep}
\end{figure}

To confirm that these features are not due to residual foregrounds, we have also applied our suite of needlet transforms to the WMAP 7-year Q (41 GHz) and V (61 GHz) band foreground-reduced maps. Taking all of the needlets which generate a significant response for the four most significant features, we calculate the average of the needlet coefficients within the regions described by the estimated bubble templates. The results are plotted in Fig.~\ref{fig-detection_freqdep}. We show, for each blob forming part of a feature, the W-band-normalized needlet coefficient averages given by
\begin{equation}
\Delta\beta_{jk, {\rm Q / V}} = \frac{\bar{\beta}_{jk, {\rm Q / V}} - \bar{\beta}_{jk, {\rm W}}}{\bar{\beta}_{jk, {\rm W}}},
\end{equation}
where $\bar{\beta}_{jk, {\rm Q / V / W}}$ is the pixel-averaged needlet coefficient value in a given WMAP frequency band. The plots are consistent with no change in the strength of the signal with frequency, suggesting that the features are not due to foreground contamination.

\begin{table*}
\begin{tabular}{c c c c c c c c}
\hline
\hline
 feature & $\theta_{\rm crit}$ range & \ \  $\ln \rho_\blob$ & \ \ $z_0$ & \ \ $z_{\rm crit}$ & \ \ $\theta_{\rm crit}$ & \ \ $\theta_0$ & \ \ $\phi_0$ \\
\hline
1 & $12-38$ & N/A & N/A & N/A & N/A & N/A & N/A \\
2 & $4-21$ & $-4.6 \pm 0.2$ & $-4.8^{+1.4}_{-1.4} \times 10^{-5}$ & $-0.1^{+0.7}_{-0.7} \times 10^{-5}$ & $6.4^{+1.7}_{-1.1}$ & $147.3^{+0.7}_{-0.6}$ & $208.0^{+1.5}_{-1.4}$ \\
3 & $4-21$ & $-4.1 \pm 0.2$ & $5.2^{+1.2}_{-1.2} \times 10^{-5}$ & $0.1^{+0.5}_{-0.5} \times 10^{-5}$ & $6.5^{+0.7}_{-0.7}$ & $123.0^{+0.7}_{-0.7}$ & $320.8^{+1.0}_{-1.0}$ \\
4 & $2-7$ & $-7.3 \pm 0.1$ & $-5.2^{+1.3}_{-1.3} \times 10^{-5}$ & $0.0^{+0.9}_{-0.9} \times 10^{-5}$ & $3.2^{+0.8}_{-0.8}$ & $145.3^{+0.5}_{-0.4}$ & $32.8^{+0.9}_{-1.1}$ \\
5 & $6-12$ & $-9.2 \pm 0.2$ & $3.7^{+1.2}_{-1.1} \times 10^{-5}$ & $-0.1^{+0.4}_{-0.4} \times 10^{-5}$ & $6.9^{+0.5}_{-0.9}$ & $32.6^{+0.6}_{-0.6}$ & $74.3^{+1.3}_{-1.2}$ \\
6 & $6-12$ & $-9.7 \pm 0.1$ & $3.0^{+1.1}_{-1.2} \times 10^{-5}$ & $-0.1^{+0.4}_{-0.4} \times 10^{-5}$ & $7.9^{+2.2}_{-1.9}$ & $128.6^{+0.7}_{-0.7}$ & $91.8^{+1.2}_{-1.3}$ \\
7 & $3-12$ & $-5.4 \pm 0.2$ & $-5.0^{+1.3}_{-1.4} \times 10^{-5}$ & $0.0^{+0.6}_{-0.6} \times 10^{-5}$ & $4.6^{+0.9}_{-1.1}$ & $169.0^{+0.6}_{-0.6}$ & $185.7^{+3.9}_{-3.9}$ \\
8 & $3-7$ & $-8.6 \pm 0.1$ & $4.4^{+1.3}_{-1.2} \times 10^{-5}$ & $0.0^{+0.7}_{-0.7} \times 10^{-5}$ & $4.4^{+0.6}_{-0.6}$ & $58.0^{+0.4}_{-0.4}$ & $115.7^{+0.5}_{-0.6}$ \\
9 & $2-4$ & $-9.0 \pm 0.2$ & $-6.0^{+1.6}_{-1.6} \times 10^{-5}$ & $0.0^{+1.0}_{-1.0} \times 10^{-5}$ & $2.3^{+0.2}_{-0.3}$ & $152.1^{+0.2}_{-0.2}$ & $241.9^{+0.4}_{-0.4}$ \\
10 & $3-8$ & $-3.8 \pm 0.2$ & $-6.1^{+1.3}_{-1.3} \times 10^{-5}$ & $0.1^{+0.9}_{-0.8} \times 10^{-5}$ & $4.2^{+0.4}_{-0.4}$ & $167.2^{+0.3}_{-0.3}$ & $269.1^{+1.6}_{-1.4}$ \\
11 & $2-6$ & $-8.1 \pm 0.1$ & $-5.9^{+1.5}_{-1.6} \times 10^{-5}$ & $0.2^{+0.8}_{-0.8} \times 10^{-5}$ & $2.5^{+0.4}_{-0.5}$ & $114.9^{+0.4}_{-0.4}$ & $22.4^{+0.4}_{-0.4}$ \\
\hline
\hline
\end{tabular}
\begin{center}
\caption{Results of the Bayesian parameter estimation and model selection analysis for the WMAP 7-year data. Reported error bars are at 68\% CL. Angular quantities are reported in degrees. The ranges of $\theta_{\rm crit}$ are determined from the needlet response (see Table~\ref{tab:anglelookup}). By computational necessity, the evidence integral is truncated at $11^{\circ}$. Hence, an evidence ratio for feature 1 could not be calculated as its $\theta_{\rm crit}$ range lies entirely beyond this upper bound. The angular positions $\{\theta_0, \phi_0\}$ can be related to Galactic coordinates through longitude $l_0 = \phi_0$ and latitude $b_0 = 90^{\circ} - \theta_0$. 
\label{tab:wmapevidences}}
\end{center}
\end{table*}

\section{Conclusions and outlook}\label{sec:conclusions}

An exciting opportunity to confront the eternal inflation scenario with experiment lies in the observation of collisions between other bubble universes and our own. In this paper, we have described an algorithm to search for the imprint of bubble collisions on the cosmic microwave background, and applied it to the WMAP 7-year data. Our search algorithm targets the generic signatures expected from bubble collisions: azimuthal symmetry, long-wavelength modulation of the temperature confined to discs on the sky, and circular temperature discontinuities. For this reason, we expect our analysis to be fairly robust under changing assumptions about the underlying theory, which is presently rather poorly understood.

The analysis pipeline we have developed takes a two-pronged approach, applied in parallel. The first uses heuristic techniques to test for the presence of features specific to bubble collisions. The second is a fully Bayesian algorithm for the general problem of non-Gaussian source detection, implemented as a patch-wise approximation to the full-sky model selection and parameter estimation problem. The data set is segmented in a completely automated way, allowing us to avoid {\em a posteriori} selection effects associated with choosing the most ``interesting" features on the CMB sky by hand. The algorithm is tested and thresholds at each step are calibrated using extensive simulations, and then frozen before ever looking at the data, to follow as much as possible the philosophy of a blind analysis. Candidate collisions are identified from an input temperature map based on the response to a suite of needlet transforms (calibrated using simulations with and without bubble collisions), and grouped into ``blobs." These blobs are scrutinized for circular temperature discontinuities using an edge detection algorithm. The quantitative significance of an edge is characterized using the Circular Hough Transform (CHT). The blobs are also used to construct an approximation to the full-sky Bayesian parameter estimation and model selection problem for bubble collisions. The posterior probability distribution over the expectation value for the number of detectable collisions, $\nsavge$, is then obtained. This allows us to quantify which of the two models -- a theory which predicts on average $\nsavge$ bubble collision signatures described by temperature modulations of the form given in Eq.~\ref{eq:collfluct}, or else the standard model (specified by $\nsavge=0$) with CMB plus realistic noise and beam effects -- better explains the data.

Applying our analysis pipeline to simulations, we have found that a circular temperature discontinuity at the causal boundary is a clear  signature of bubble collisions.\footnote{The observational detection of a circular temperature discontinuity is so unlikely to arise spuriously that it provides conclusive evidence of a detection.} Although our analysis can identify collisions without temperature discontinuities, their presence greatly increases our ability to make a conclusive detection. Both the edge-detection and Bayesian model selection steps have the ability to identify a causal boundary in the patches of the sky that are highlighted as candidate collisions by the blob detection step of our analysis pipeline. We have found no evidence for  circular temperature discontinuities in the WMAP 7-year data using either method. Based on our analysis of simulations, this allows us to rule out the presence of collisions in the exclusion region of Fig.~\ref{fig-chtexclusion}. For collisions larger than $\theta_{\rm crit} \agt 10^{\circ}$, this corresponds to $10^5 |z_{\rm crit}| \alt 3$--$6$ for the amplitude of the circular temperature discontinuity defined in Eq.~\ref{eq:collfluct}. For collisions on smaller scales, the CHT step loses sensitivity due to the proliferation of degree-scale blobs in the background CMB fluctuations.

The posterior evaluated using the WMAP 7-year data is maximized at $\nsavge=0$, and constrains $\nsavge < 1.6$ at $68 \%$ confidence. We therefore conclude that this data set does not favor the bubble collision hypothesis for any value of $\nsavge$. In light of this null detection, comparing with the simulated bubble collisions, we can constrain the central amplitude of the temperature modulation caused by the collision (defined in Eq.~\ref{eq:collfluct}) to be $z_0 \alt 1 \times 10^{-4}$ over the range of scales $\theta_{\rm crit} \agt 5^{\circ}$ we have simulated. If the collision is described by a single super-Hubble wavelength mode confined to a disc on the sky, from Eq.~\ref{eq:griszeleffect} we can use these bounds (with the largest collision size we have simulated: $\theta_{\rm crit}=25^{\circ}$) to constrain $\Omega_k^{1/2} \Phi(0) \alt 7 \times 10^{-4}$ (where $\Omega_k$ is the present component in curvature and $\Phi(0)$ is the initial magnitude of the Newtonian potential caused by the collision). More generally, Eq.~\ref{eq:collbound} bounds the nucleation rate of bubbles in our parent vacuum, provided gravitational waves and negative curvature are observed with future experiments.

Although we have obtained a null result, our analysis pipeline has identified four features in the WMAP 7-year data that have Bayesian evidence ratios that are significantly larger than expected for false detections from an end-to-end simulation of the WMAP experiment. Two of these features (features 2 and 3) have been noted in previous literature. Feature 2 corresponds to the WMAP Cold Spot~\cite{Cruz:2004ce} (see Ref.~\cite{Cruz:2009nd} for a review of its properties), and feature 3 was identified using standard needlets in Ref.~\cite{Pietrobon:2008rf}. All four features are far from the Galactic cut of the KQ75 7-year mask (see Fig.~\ref{fig-detection_locations}), and none appear to be responses to the point source components of the mask (see Fig.~\ref{fig-detection_closeups}). We have confirmed that the signal in each case is not strongly dependent on the frequency band used (see Fig.~\ref{fig-detection_freqdep}), providing evidence that these features are not due to astrophysical foregrounds. A number of analyses, most recently the redshift analysis of Ref.~\cite{2010arXiv1004.1178B}, suggest that the Cold Spot is primordial and not associated with the integrated Sachs-Wolfe effect of a large void. Further studies of the other three features would be needed to confirm that they are truly primordial.

Our ability to detect bubble collisions will improve greatly with data from the {\em Planck} satellite. Decreased instrumental noise will enlarge the exclusion and sensitivity regions in parameter space for the needlet step of the analysis, as evidenced by our ability to detect more simulated collisions in low-noise regions of the WMAP data. The threefold increase in resolution will greatly improve our ability to detect circular edges. In addition, the polarization data from {\em Planck} will be of sufficient resolution to look for complementary signatures of bubble collisions~\cite{Czech:2010rg,Dvorkin:2007jp}. Such an analysis should be able to confirm if the features we have identified are in fact bubble collisions.

It is also important to determine if other theories predicting azimuthally symmetric features in the CMB~\cite{Cornish:1997ab,2008MNRAS.390..913C,Afshordi:2010wn,Kovetz:2010kv} are better fits to the data. The blob and edge detection steps in our analysis pipeline are sensitive to a variety of possible signatures, and given a model, the Bayesian model comparison step could be easily tailored to accommodate different forms of the temperature modulation. Because our pipeline is automated, we can compare the evidence ratios obtained for different models to decide which is a better fit, without recourse to {\em a posteriori} choices of which features to analyze. 

In conclusion, we have presented a powerful algorithm for analyzing CMB data for signatures of bubble collisions. Applying this pipeline to the WMAP 7-year data, we have constrained the possible parameter space of bubble collisions, as well as identifying interesting candidate signatures in the data for further investigation. Future data from the {\em Planck} experiment will allow us to greatly improve on these results. If confirmed, the presence of bubble collisions in the CMB would be an extraordinary insight into the origins of our universe. 

\acknowledgements
We are very grateful to Eiichiro Komatsu and the WMAP Science Team for supplying the end-to-end WMAP simulations used in our null tests. SMF is supported by the Perren Fund. MCJ acknowledges support from the Moore Foundation. HVP is supported by Marie Curie grant MIRG-CT-2007-203314 from the European Commission, and by STFC and the Leverhulme Trust. MCJ and HVP thank the Aspen Center for Physics, where this project was initiated, for hospitality. HVP and DJM acknowledge the hospitality of the Statistical Frontiers of Astrophysics workshop at IPMU, Tokyo. SMF thanks Filipe Abdalla, Ingo Waldmann and Michael Hirsch for useful discussions, and Jenny Feeney for proofreading the manuscript. MCJ thanks Rebecca Danos for discussions regarding the edge detection algorithm. HVP thanks Andrew Pontzen for interesting discussions regarding the general problem of CMB anomaly hunting. We acknowledge use of the HEALPix package and the Legacy Archive for  Microwave Background Data Analysis (LAMBDA).  Support for LAMBDA is provided by the NASA Office of Space Science. Research at Perimeter Institute is supported by the Government of Canada through Industry Canada and by the Province of Ontario through the Ministry of Research and  Innovation. A preprint version of this paper presented only evidence ratios confined to patches. We thank an anonymous referee who encouraged us to develop this algorithm into a full-sky formalism. This calculation is now presented in Appendix A and incorporated into our analysis pipeline and results.

\bibliography{cmbbubbles}

\begin{appendix}

\section{Statistical formalism}
\label{section:method}

\subsection{Posterior}

In this appendix, we discuss how Bayesian parameter estimation and model selection for theories which predict localized sources can be approximated by a patch-wise analysis.  Consider astronomical observations covering solid angle $\aobs = 4 \pi \fsky$ that are of sufficient depth/resolution to identify sources with a particular range of properties (which can then be deemed ``detectable''). Given a theory that predicts an expectation value of $\nsavge$ sources over the whole sky, we want to know both: what constraints the available data place on $\nsavge$; and whether the data favour a model which predicts one value of $\nsavge$ over another. all the relevant information is encoded in the posterior distribution $\prob(\nsavge | \data, \fsky)$, where $\data$ are the pixelized flux or temperature measurements (and, optionally, any statistics derived from them). Bayes' theorem allows the posterior to be written as 
\begin{equation}
\prob(\nsavge | \data, \fsky) 
  = \frac{\prob(\nsavge) 
    \, \prob(\data | \nsavge, \fsky)}{\prob(\data | \fsky)},
\end{equation}
where $\prob(\nsavge)$ is the prior distribution on $\nsavge$, $\prob(\data | \nsavge, \fsky)$ is the likelihood of getting the observed data given the area of observation and the expected number of sources, and $\prob(\data | \fsky)$ ensures that the posterior is normalized over $\nsavge$. Constraints on $\nsavge$ can be drawn directly from this normalized posterior; the relative probability of models predicting different values of $\nsavge$ can be found by picking out the posterior at two values of $\nsavge$:
\begin{equation}
\frac{\prob(\bar{N}_{\rm s, 1} | \data, \fsky)}{\prob(\bar{N}_{\rm s, 2} | \data, \fsky)} = \frac{\prob(\bar{N}_{\rm s, 1}) \prob(\data | \bar{N}_{\rm s, 1}, \fsky) }{ \prob(\bar{N}_{\rm s, 2}) \prob(\data | \bar{N}_{\rm s, 2}, \fsky) }.
\end{equation}

In the absence of a prescriptive theory,  it is useful to emphasize the role of the data, which can be done by adopting a flat prior on $\nsavge$; further assuming that the data will give an upper limit on $\nsavge$, it is possible to adopt an improper uniform prior $\prob(\nsavge) = \step (\nsavge)$ without any high-$\nsavge$ cut-off. The resultant posterior has the form
\begin{equation}
\label{equation:posterior_general}
\prob(\nsavge | \data, \fsky) \propto
  \step(\nsavge) \, \prob(\data | \nsavge, \fsky),
\end{equation}
up to a normalization constant that depends on the data and $\fsky$ but not on $\nsavge$. 

In general $\nsavge$ is not directly measurable, even for perfect data, because the number of sources present in the observable sky, $\ns$, 
is the realization of a Poisson-like process (of mean $\fsky \nsavge$). The possibility that $\ns$ is itself subject to some uncertainty 
(\eg, due to noisy data or confusion problems) can be incorporated by marginalizing over $\ns$ to give
\begin{eqnarray}
\label{equation:likelihood}
\prob(\data | \nsavge, \fsky)
  & = & \sum_{\ns = 0}^\infty
    \prob(\ns | \nsavge, \fsky) \,
    \prob(\data | \ns, \fsky) \nonumber \\
  & = & \sum_{\ns = 0}^\infty
   \frac{(\fsky \nsavge)^\ns e^{-\fsky \nsavge}}{\ns!} 
   \, \prob(\data | \ns, \fsky),
\end{eqnarray}
where the second formula explicitly assumes that the number of observable sources is drawn from a Poisson process. Inserting this second expression into Eq.~\ref{equation:posterior_general} then gives
\begin{equation}
\label{equation:posterior_poisson}
\prob(\nsavge | \data, \fsky) \propto
  \step(\nsavge) \, e^{- \fsky \nsavge}
  \sum_{\ns = 0}^\infty
  \frac{(\fsky \nsavge)^\ns}{\ns!} 
   \, \prob(\data | \ns, \fsky).
\end{equation}

The form of the likelihood $\prob(\data | \ns, \fsky)$ is treated largely in abstract here, with the specific details of the likelihood calculation for the bubble collision hypothesis given WMAP 7-year data described in Sec.~\ref{sec:bayes}. Assuming the measurements take the form of flux/counts at different positions on the sky (as for a CMB experiment) and that they are subject to (possibly correlated) Gaussian noise, the likelihood would have the form
\begin{equation}
\label{eq:pdatans} 
\prob(\data | \ns, \fsky)
 = \int 
    \diff \model_1 \ldots \diff \model_\ns \,
    \prod_{\src = 1}^{\ns} \prob(\model_\src) 
    \frac{1}{(2 \pi)^{\npix / 2} |\matr{C}|}
    e^{- [\data - \template(\model_1) \ldots
      - \template(\model_{\ns})] 
      \matr{C}^{-1} 
    [\data - \template(\model_1) \ldots
      - \template(\model_{\ns})]^{\rm{T}} / 2},
\end{equation}
where $\template(\model)$ is the data template that would result from a source whose position and profile/scale are defined by the model parameters $\model$, $\prob(\model_\src)$ is the prior distribution of source parameters for ``detectable'' sources, and $\matr{C}$ is the pixel-pixel covariance matrix of the  non-source noise (which could include contributions that are considered signal in other contexts, such as the CMB). 

Evaluating the full sum in Eq.~\ref{equation:posterior_poisson} is not always practical or even possible. In addition, the evaluation of individual terms in this sum will be computationally limited by the size of the covariance matrix $\mathbf{C}$ and the cost of performing the integral over model parameters for each template. However, it is possible to circumvent these problems, and estimate the posterior Eq.~\ref{equation:posterior_poisson} if one knows in advance some of the properties of the integrand in Eq.~\ref{eq:pdatans}. 

To see how this works, assume that one has located a set of $\nblob$ ``blobs" on the sky that are candidate sources. Segment the sky into $\nregion = \nblob+1$ regions consisting of those containing blobs, and the rest of the sky. Given $\nblob$, we can now evaluate Eq.~\ref{equation:posterior_poisson} term-by-term. The likelihood in the first term, for $\ns =0$, is simply given by:
\begin{equation}
\prob(\data | 0, \fsky) = \frac{1}{(2 \pi)^{\npix / 2} |\matr{C}|} e^{- \data \matr{C}^{-1}  \data^{\rm{T}} /2},
\end{equation}
which is the likelihood for the null-hypothesis with no sources. Moving on to the $\ns = 1$ term, we first expand the integral over source positions to cover each of the $\nregion$ regions:
\begin{equation}
\prob( \data | 1, \fsky) = \sum_{r=1}^{\nregion} \int_{\rm region \ r} \diff \model \, \prob(\model)  \frac{1}{(2 \pi)^{\npix / 2} |\matr{C}|}
    e^{- [\data - \template(\model)]
      \matr{C}^{-1}
    [\data - \template(\model)]^{\rm{T}} /2},
\end{equation}
We now assume that the blobs containing candidate sources include all of the significant contributions to the integral, and replace $\nregion$ in the sum by $\nblob$. This will give us a {\em lower bound} on the likelihood, even if a number of actual sources are not contained within the blobs defined by the candidate sources. We further assume that sources do not overlap. If the covariance matrix is small enough to invert, we could stop here. However, in cases where the covariance matrix is too large to feasibly invert (as is the case for the WMAP 7 year data), we can make one further approximation:
\begin{equation}
 \int_{\rm region \ b} \diff \model \, \prob(\model)  \frac{1}{(2 \pi)^{\npix / 2} |\matr{C}|}
    e^{- [\data - \template(\model)]
      \matr{C}^{-1}
    [\data - \template(\model)]^{\rm{T}} /2} \simeq 
  \int_{\rm region \ b}  \diff \model \, \prob(\model)
  \prod_{\region = 1}^{\nregion} 
  \likelihood_\region(\model ),
\end{equation}
where the product is over $\nregion$ disjoint regions on the sky and
\begin{equation}
\likelihood_\region(\model) 
  = \frac{1}{(2 \pi)^{\npix / 2} |\matr{C}_\region|}
    e^{- [\data_\region - \template_\region(\model)]
      \matr{C}^{-1}_\region 
    [\data_\region - \template_\region(\model)]^{\rm{T}} /2}.
\end{equation}
is the contribution to the likelihood of data in region $\region$, defined in terms of the covariance of the pixels in this region, $\matr{C}_\region$, the data in this region, $\data_\region$, and the source template in this region, $\template_\region(\model)$. This is exact for a diagonal covariance, but is only approximate in the case where there are off-diagonal elements. Using the assumption that the integral has a significant contribution only inside the blobs, in the rest of the sky we can replace $\template = \mathbf{0}$. The one-source model likelihood then becomes
\begin{equation}
\prob(\data | 1, \fsky)
  \simeq 
   \sum_{\blob = 1}^\nblob \prod_{\region = 1}^\nregion \likelihood_\region(\vect{0}) \rho_{b},
\end{equation}
where 
\begin{equation}\label{eq:sourcelikelihoodratio}
\rho_\blob = \frac{\int_{{\rm{region}} \ b} 
  \diff \model \, \prob(\model) 
  \likelihood_\blob(\model)}{\likelihood_\blob(\vect{0})}.
\end{equation}
This is the evidence ratio for a single source template 
centered in region $b$.

For a general number of blobs and sources, the model likelihood is
\begin{eqnarray}
\label{equation:likelihoodfinal}
\prob(\data | \ns, \fsky) 
  = 
  \left\{
    \begin{array}{lll}
    0, & {\rm if} & \ns > \nblob , \\
    & & \\
\sum_{\blob_1, \blob_2, \ldots, \blob_{\ns} = 1}^{\nblob}
    \left[
    \prod_{\src = 1}^{\ns} 
      \rho_{b_s}
    \prod_{i,j = 1}^{\ns} (1 - \delta_{\src_i, \src_j})
    \prod_{\region = 1}^{\nregion} \likelihood_\region(\vect{0})
    \right],
    & {\rm if}
        & \ns \leq \nblob,
        \end{array}
        \right.
\end{eqnarray}
where the combinatorics require some explanation. If there are fewer blobs on the sky than proposed sources then the likelihood is very small: by assumption, the likelihood evaluated outside of a blob is small. If there are at least as many blobs as proposed sources, then the likelihood takes the form of a sum that includes every possible association of the $\ns$ sources with the $\nblob$ blobs, provided that no two sources are matched to the same blob. Hence the multiple sum generates all possible combinations of source-blob associations and the product over evidence ratios gives the relevant weightings; the product over delta functions removes the terms in which any two sources are attached to the same blob.

Inserting the likelihood given in Eq.~\ref{equation:likelihoodfinal} into Eq.~\ref{equation:posterior_poisson} yields the unnormalized posterior on $\nsavge$:
\begin{equation}
\label{equation:posteriorfinalapp}
\prob(\nsavge | \data, \fsky) 
\mbox{}
  \propto \step(\nsavge) \,  e^{-\fsky \nsavge}
  \sum_{\ns = 0}^\nblob
    \frac{(\fsky \nsavge)^\ns}{\ns!} 
  \sum_{\blob_1, \blob_2, \ldots, \blob_{\ns} = 1}^{\nblob}
    \left[
    \prod_{\src = 1}^{\ns} 
      \rho_{b_s}
    \prod_{i,j = 1}^{\ns} (1 - \delta_{\src_i, \src_j})
    \right],
\end{equation}
under the assumption that
\begin{equation}
\prob(\data | 0, \fsky) = \prod_{r = 1}^{\nregion} \likelihood_r (\vect{0}),
\end{equation}
in which case regions that do not contain a blob are irrelevant for determining the posterior. Eq.~\ref{equation:posteriorfinalapp} is the main result of this calculation, from which all following results can be derived.  In the limit of a single isolated observation Eq.~\ref{equation:posteriorfinalapp} reproduces the  Bayesian source detection formalism developed in \cite{Hobson_McLachlan:2003,Hobson_etal:2010}.

\subsection{Special cases}

\subsubsection{Perfect data}

With infinite, perfect data the number of sources on the sky
would be directly determined by counting the $\nblob$ ``blobs''
in the data and so
$\prob(\data | \ns, \fsky) = \delta_{\ns,\nblob}$.
The posterior in Eq.~\ref{equation:posteriorfinalapp} would become
\begin{equation}\label{eq:nblobposterior}
\prob(\nsavge | \nblob, \fsky)
  \propto \step(\nsavge) \nsavge^\nblob e^{- \fsky \nsavge},
\end{equation}
the standard result for constraining a rate variable from a single
measurement, 
modified slightly to account for the fact that the constraint
on $\nsavge$ is weakened if $\fsky \ll 1$.
In the even more particular case that no blobs were detected in 
perfect data, the posterior would be 
\begin{equation}
\label{equation:posteriornoblob}
\prob(\nsavge | 0, \fsky) = \step(\nsavge) \, \fsky e^{- \fsky \nsavge}.
\end{equation}
If a single blob was detected unequivocally then the posterior would be 
\begin{equation}
\label{equation:posterioroneblob}
\prob(\nsavge | 1, \fsky) = \step(\nsavge) \, 
  \fsky (\fsky \nsavge) e^{- \fsky \nsavge}.
\end{equation}
If two blobs were detected unequivocally then the posterior would be
\begin{equation}
\label{equation:posteriortwoblob}
\prob(\nsavge | 2, \fsky) = \step(\nsavge) \, 
  \frac{\fsky}{2} (\fsky \nsavge)^2 e^{- \fsky \nsavge}.
\end{equation}

\subsubsection{No blobs}

If there are no identified blobs then $\nblob = 0$ and there is 
no evidence for any sources at all.  This is really a weaker constraint
than the above situation if the data are perfect, but in the 
approximation used here the final result is the same.
In Eq.~\ref{equation:posteriorfinalapp} the first sum is truncated at the first term
and so
\begin{equation}\label{eq:noblobs}
\prob(\nsavge | 0, \data, \fsky) 
  = \step(\nsavge) \, \fsky e^{- \fsky \nsavge},
\end{equation}
matching Eq.~\ref{equation:posteriornoblob}.

Adopting a flat prior on $\nsavge$, the posterior probability ratio for a model predicting a generic $\nsavge > 0$ versus one predicting no collisions in this case is given by
\begin{equation}\label{eq:noblobsrho}
\frac{ \prob(\nsavge \, | \, 0, \data, \fsky) }{\prob(0 \, | \, 0, \data, \fsky)} = e^{- \fsky \nsavge}.
\end{equation}
This is always less than one, and so as expected, a theory which predicts $\nsavge$ sources on the sky is always disfavoured when compared to a theory that predicts no sources on the sky.

\subsubsection{One blob}

Probably the most important simple case is where there is a single
identified blob, which might represent a first detection of 
this class of source.  
Inserting $\nblob = 1$ into Eq.~\ref{equation:posteriorfinalapp},
the sum includes the possibilities of either one source on the 
(observed) sky or no sources;
the posterior evaluates to
\begin{equation}
\prob(\nsavge | 1, \data, \fsky)
  = \step(\nsavge) \,
   \fsky e^{- \fsky \nsavge}
   \frac{1 + \fsky \nsavge \rho_\blob}
     {1 + \rho_\blob},
\end{equation}
In the limit that the data in this region are much better fit by a source then $\rho_\blob \gg 1$, and the posterior becomes
\begin{equation}
\prob(\nsavge | 1, \data, \fsky)
  = \step(\nsavge) \,
   \fsky (\fsky \nsavge) e^{- \fsky \nsavge},
\end{equation}
which matches \ref{equation:posterioroneblob} above. Conversely, in the limit that the source is a worse fit to the data (possible given that the source has been forced to be detectable), then $\rho_\blob \ll1$ and 
\begin{equation}
\prob(\nsavge | 1, \data, \fsky)
  = \step(\nsavge) \,
   \fsky e^{- \fsky \nsavge},
\end{equation}
matching Eq.~\ref{equation:posteriornoblob} which was obtained under the assumption that there was no blob in the first place.

Adopting again a flat prior on $\nsavge$, the posterior probability ratio for a model predicting a generic $\nsavge > 0$ versus the no-bubble case is given by
\begin{equation}\label{eq:oneblobev}
\frac{ \prob(\nsavge | \, 1, \data, \fsky) }{\prob(0 | 1, \data, \fsky)} =  e^{- \fsky \nsavge}
\left( 1 + \fsky \nsavge \rho_\blob \right).
\end{equation}
Here, it can be seen that two things are necessary to favour the theory with sources given one detection:  $\nsavge \sim \mathcal{O}(1)$ {\em and} $\rho_\blob \gg 1$.

\subsubsection{Two blobs}

If two blobs are identified then the sum in \ref{equation:likelihoodfinal}
has three terms, for which the likelihoods are:
\begin{equation}
\prob(2, \data | 0, \fsky) = \prod_{\region = 1}^{\nregion} 
  \likelihood_\region(\vect{0}),
\end{equation}
\begin{equation}
\prob(2, \data | 1, \fsky)
  = 
   \left[
   \rho_{\blob_1}
+ 
   \rho_{\blob_2}
   \right]
   \prod_{\region = 1}^{\nregion} \likelihood_\region(\vect{0})
\end{equation}
and
\begin{equation}
\prob(2, \data | 2, \fsky)
  =
    \rho_{\blob_1} \rho_{\blob_2}
    \prod_{\region = 1}^{\nregion} \likelihood_\region(\vect{0}).
\end{equation}
Hence the (unnormalized) posterior is
\begin{equation}
\prob(\nsavge | 2, \data, \fsky)
  \propto \step(\nsavge)
  e^{-\fsky \nsavge}
  \left\{
   1
   + \fsky \nsavge  \left[
   \rho_{\blob_1}
   + 
   \rho_{\blob_2}
   \right]
   + (\fsky \nsavge)^2 \rho_{\blob_1} 
   \rho_{\blob_2}
  \right\}.
\end{equation}
In the limit that the evidence for both sources is strong (i.e., $ \rho_{\blob_1}  \gg1$ and $\rho_{\blob_2} \gg 1$) then the third term in the curly braces dominates and 
\begin{equation}
\prob(\nsavge | 2, \data, \fsky) 
  = \step(\nsavge) \frac{\fsky}{2} (\fsky \nsavge)^2 e^{- \fsky \nsavge},
\end{equation}
which matches the perfect data case with $\nblob = 2$, as expected. In the limit where one blob is a false candidate, but the other yields a strong evidence (e.g., $ \rho_{\blob_1}  \gg1$ and $\rho_{\blob_2} \ll 1$), then we recover the perfect data case with $\nblob=1$.

\end{appendix}

\end{document}